\documentclass[
reprint,
nofootinbib,
amsmath,amssymb, aps
]{revtex4-2}

\usepackage[font=small, labelfont=bf]{caption}
\usepackage{booktabs}
\usepackage{graphicx}
\usepackage{bm}
\usepackage{physics}
\usepackage{mathtools}
\usepackage{mathrsfs}
\usepackage{float}
\usepackage{hyperref}

\newcommand\numeqarrow[1]%
{\stackrel{\scriptscriptstyle(\mkern-1.5mu#1\mkern-1.5mu)}{\rightarrow}}
\newcommand\numeqequal[1]%
{\stackrel{\scriptscriptstyle(\mkern-1.5mu#1\mkern-1.5mu)}{=}}

\begin{document}
\title{Coarse-graining in time with the Functional Renormalisation Group:\\ Relaxation in Brownian Motion}

\author{Ashley Wilkins}
\email{a.wilkins3@newcastle.ac.uk}
\author{Gerasimos Rigopoulos}%
\email{gerasimos.rigopoulos@newcastle.ac.uk}
\affiliation{School of Mathematics, Statistics and Physics, Newcastle University,\\Newcastle upon Tyne, NE1 7RU, United Kingdom
}%
\author{Enrico Masoero}
\email{masoeroe@cardiff.ac.uk}
\affiliation{Cardiff School of Engineering, Cardiff University\\ Queens Buildings, The Parade, Cardiff, CF24 3AA, United Kingdom}

\date{\today}

\begin{abstract}
\noindent We apply the functional Renormalisation Group (fRG) to study relaxation in a stochastic process governed by an overdamped Langevin equation with one degree of freedom, exploiting the connection with supersymmetric quantum mechanics in imaginary time. After reviewing the functional integral formulation of the system and its underlying symmetries, including the resulting Ward-Takahashi identities for arbitrary initial conditions, we compute the effective action $\Gamma$ from the fRG, approximated in terms of the leading and subleading terms in the gradient expansion: the Local Potential Approximation and Wavefunction Renormalisation respectively. This is achieved by coarse-graining the thermal fluctuations in time resulting in e.g. an effective potential incorporating fluctuations at all timescales. We then use the resulting effective equations of motion to describe the decay of the covariance, and the relaxation of the average position and variance towards their equilibrium values at different temperatures. We use as examples a simple polynomial potential, an unequal Lennard-Jones type potential and a more complex potential with multiple trapping wells and barriers. We find that these are all handled well, with the accuracy of the approximations improving as the relaxation's spectral representation shifts to lower eigenvalues, in line with expectations about the validity of the gradient expansion. The spectral representation's range also correlates with temperature, leading to the conclusion that the gradient expansion works better for higher temperatures than lower ones. This work demonstrates the ability of the fRG to expedite the computation of statistical objects in otherwise long-timescale simulations, acting as a first step to more complicated systems. 

\end{abstract}

\maketitle


\section{\label{sec:Intro}Introduction}

Stochastic processes appear in all kinds of contexts in physics: from the Brownian motion of small particles in a thermal bath \cite{VanKampen2007, gardiner2009stochastic} to exotic scalar fields experiencing quantum fluctuations in the early inflationary universe \cite{Starobinsky1994}, many problems of interest can be described by the overdamped Langevin equation (\ref{eq:langevin}). In this work  we employ an effective description of the stochastic dynamics that captures the aggregate effect of fluctuations embodied in an \emph{effective action} $\Gamma[\chi(t)]$, a functional of the average position $\chi(t) \equiv \langle x(t) \rangle$  which can be thought of as an analogue to the statistical free energy and can be derived from the partition function or generating functional via a Legendre transform. Once obtained, the effective action can be used to compute $n$-point correlation functions of the particle's position $\langle x(t_1) x(t_2)\ldots x(t_n) \rangle $, characterizing the system's statistical properties. To obtain this effective action we will be coarse-graining the system in time such that we obtain e.g. an effective potential that incorporates thermal fluctuations on all timescales. To achieve this we will use a technique known as the functional, or exact, or non-perturbative Renormalisation Group \cite{Wetterich1993, Morris1994} -- see \cite{Berges2002} for a review and an entry point to the literature on the subject, \cite{Dupuis2020} for a comprehensive overview of applications as well as e.g. \cite{Gies2012, Delamotte2012} for more elementary introductions. 

The renormalisation group (RG) was brought to full force through the work of K. Wilson \cite{Wilson1983} who used it to understand phase transitions and since then the RG has become a widely used technique in modern physics with many applications in both particle physics \cite{Peskin:1995ev} and condensed matter physics \cite{Chaikin1995,Vasiliev2004}. 
The RG is relevant whenever fluctuations significantly influence the state (static or dynamical) of a physical system. A recent popular incarnation of this programme is the functional RG (fRG) formulation. Wetterich showed \cite{Wetterich1993} - see also \cite{Morris1994} - how one can define $\Gamma$ at some particular energy or momentum scale $\Lambda$ in the UV (for us this will correspond to small timestep/high frequency) where the theory is known and then create an RG flow that interpolates through all energy (frequency/momentum) scales down to the IR (i.e. decreasing fluctuation frequency/increasing characteristic timestep of fluctuation here). This change of the effective theory at different scales, the fundamental idea behind the RG, can be formulated in an integro-differential equation known as the Wetterich equation:
\begin{equation}\label{Wetterich eqn}
	k\partial_{k}\Gamma_{k} = \dfrac{1}{2}\textbf{Tr}\left[k\partial_{k}R_{k}\left(\Gamma_{k}^{(2)} + R_k\right)^{-1} \right]
\end{equation}
where $\Gamma_{k}$ is the effective action at scale $k$, $\textbf{Tr}$ denotes a trace over spatio-temporal points (an integral over spacetime) and a trace over all other relevant indices, $R_{k}$ is an IR regulator that acts as a cut-off for fluctuations below (momentum/frequency) scale $k$, and  $\Gamma_{k}^{(2)}$ is the second functional derivative of $\Gamma_{k}$. A simple zero-dimensional manifestation of this flow equation in the context of the Boltzmann equilibrium distribution is illuminating and is reviewed in Appendix \ref{app:equil-flow}.  At the end of the flow, as $k \rightarrow 0$, \emph{all} fluctuations are included and the full effective action $\Gamma$ is obtained. 

In this work we solve equation (\ref{Wetterich eqn}) in the context of the dynamics of particles under the influence of a deterministic force, stemming from an arbitrary potential, and thermal fluctuations. We will study a simple polynomial potential, a double Lennard-Jones (LJ) potential, as well as one containing multiple barriers/trapping wells, serving as a toy model of an energy landscape on which particles can diffuse. To solve (\ref{Wetterich eqn}) we employ a widely used approximation scheme, the gradient expansion, to second order. We find good agreement with simulations (and the Fokker-Planck equation where it proved amenable to a numerical solution) at moderate to high temperatures and that this correlates with the spectral representation of the relaxation process, i.e.~the overlap between the initial condition (a delta function initial probability distribution) and the spectrum of the Fokker Planck operator. In particular, the more the resulting spectrum is shifted towards lower-lying eigenvalues, the better the gradient expansion agrees with the exact evolution. This is in line with expectations about the validity of the gradient expansion, namely that it better captures slower evolution which should be associated with the lower eigenvalues. Furthermore, we expect that for a fixed initial condition lower temperatures correspond to a spectrum shifted towards higher eigenvalues and, correspondingly, worst performance of the gradient expansion.                

We start in Sec. \ref{sec:Brown} by reviewing the connection between Langevin dynamics and Supersymmetric Quantum Mechanics in imaginary time, first shown in \cite{Parisi:1982ud} - see e.g. \cite{zinn2002quantum} for a review of this connection. The path integral formulation then allows us to apply the fRG program directly. We include a brief summary of how the Langevin equation can be reformulated in terms of a probability distribution function whose evolution is described by the Fokker-Planck equation (\ref{eq: F-P}) and how the latter relates to a Euclidean Schr\"{o}dinger equation. We close the section by discussing the symmetries of the resulting theory and their implications through Ward-Takahashi type relations, paying attention to the fact that the initial state may not be that of equilibrium.     

In Sec. \ref{sec:fRG} we present the flow equations for the effective action utilising a slight modification of the results of \cite{Synatschke2009} for supersymmetric RG flows. As we explain, the flow equation derived from the supersymmetric formulation ensures compatibility with the equilibrium Boltzmann distribution, a feature not directly obvious from the application of the renormalisation group to the Onsager-Machlup form of the generating functional.  
To turn the functional integro-differential equation (\ref{Wetterich eqn}) into a mathematically more tractable form we employ two commonly used approximations for the effective action $\Gamma_k$: the Local Potential Approximation (LPA) as well as the LPA augmented by Wavefunction Renormalisation (WFR). In the LPA, the effect of fluctuations is progressively taken into account during the flow by the effective potential $V_k(x)$ experienced by the particle, which is altered compared to the bare, fundamental potential $V(x)$. Wavefunction Renormalisation (WFR) involves a second function $Z_k(x)$ which can be interpreted as a redefinition of position $x \rightarrow Z(x)$. These are also known as the first two (leading and subleading) orders in a gradient expansion of $\Gamma$.   
   
In Sec. \ref{sec:acc eom} we derive the effective equations of motion (EEOM) through variational derivatives of the effective action $\Gamma$. We do this first for the one point function, or average position $\chi$, whose equation of motion simply reduces to an over-damped equation in an effective potential with no noise i.e. purely classical. We then obtain an equation for the evolution of the two point function; this Green's function equation allows us to solve for the Variance and Covariance of the stochastic process. Solutions to these deterministic equations with appropriate initial conditions can then approximate the relaxation towards equilibrium.  

In Sec. \ref{sec:Sol Flow} we present numerical solutions to the flow equations for three types of potential: a simple polynomial one, an unequal LJ type potential and a ``rugged'' potential consisting of an underlying harmonic $x^2$ potential with the addition of six Gaussian bumps and dips. We also consider different temperatures, effectively controlling the strength of thermal fluctuations compared to the classical force. Naturally, we find that the end results of the flow equations differ, with higher temperatures resulting in potentials that carry less memory of the bare potentials' morphology. In section \ref{sec:Equilibrium Res} the numerical solution of the LPA flow equation accurately reproduces static equilibrium quantities, as it should.  We also examine the characteristic decay behaviour of the connected 2-point function $\left\langle x(0)x(t) \right\rangle$ or covariance at equilibrium by utilizing both the effective potential $V_{k\rightarrow 0}$ and the WFR function $Z_{k\rightarrow 0}$ and find good agreement with exact results down to relatively low temperatures when deviations start becoming pronounced. 

In Sec. \ref{sec:Accelerated} we examine how well the LPA + WFR approximation to the effective action can handle relaxation towards equilibrium in our potentials starting from a fixed initial condition ($P_{\rm ini}(x)=\delta(x-x_{\rm ini})$). Numerical solutions to the EEOM for the one- and two-point functions are compared to direct numerical simulation of equation (\ref{eq:langevin}) and/or numerical solutions to the Fokker Planck equation. We find that under the LPA + WFR approximations, the fRG solutions give a good description of the relaxational evolution and are able to capture overshoots of the variance in the potentials we examine. Similarly to equilibrium, the approximations fare worse as the temperature is decreased and the classical force determined by the potential's slope becomes dominant. We suggest that this behaviour is correlated to the spectral decomposition of the relaxation in terms of eigenvalues of the relevant Fokker-Planck operator: the more the spectral decomposition of $\langle x(t)\rangle$,  $\langle x(t_{\rm ini}) x(t)\rangle$ and  $\langle x^2(t)\rangle$ is shifted towards lower eigenvalues, the better the LPA+WFR approximation performs.   

We conclude in Sec. \ref{sec:Summary}. Appendix  \ref{app:equil-flow} derives the zero-dimensional analogue of the Wetterich equation (\ref{Wetterich eqn}) for the effective potential corresponding to the equilibrium Boltzmann distribution while in Appendix \ref{app:2ptfuncderiv} we include a derivation of the solution of (\ref{eq:2pointfuncdef}) that provides the two-point function. In Appendix \ref{app:det} we explicitly show that the determinant appearing in the path integral directly follows from the boundary conditions imposed in \ref{subsec:BMPI}.

\section{\label{sec:Brown} Brownian Motion as SuperSymmetric Quantum Mechanics}
This section reviews Brownian motion and its path integral formulation in terms of the action functional of Supersymmetric Quantum Mechanics. In the next section we will exploit this link to derive the RG flow equations. 

Brownian motion for a single particle of mass $m$ moving in a potential $\bar{V}(x)$, coupled to an external heat bath with temperature $T$, can be described by the Langevin equation:
\begin{eqnarray}
	m\ddot{x} + \gamma \dot{x} &=& -m\, \partial_{x} {V}(x) + f(t) \label{eq:langevinfull} \\
	\left\langle f(t)f(t') \right\rangle &=& 2D\gamma^2 \delta (t-t')\label{eq:f defn}
\end{eqnarray}
where $\gamma$ is a frictional coefficient, $f(t)$ is a gaussian ``noise" term and $V(x)$ is the potential in which the particle moves. $D = k_bT/\gamma$ is the diffusion constant, given so as to match the Boltzmann equilibrium distribution (should an equilibrium state exists). Hereafter, we will be concerned with the overdamped limit:
\begin{eqnarray}
	\dot{x} &=& -\varepsilon \partial_{x} V(x) + \eta(t) \label{eq:langevin} \\
	\left\langle \eta(t)\eta(t') \right\rangle &=& 2D \delta (t-t')\label{eq:eta defn}
\end{eqnarray}
to which the system settles over a timescale $\varepsilon \equiv m/\gamma$ which we assume to be short. Note that the overdamped equations are a consistent approximation to the full dynamics as long as $\varepsilon^2 V''\ll 1$.

We will be examining the impact of changing the temperature, and hence changing the strength of the fluctuating force $\eta$, on the coarse-grained effective theory. Let us therefore introduce a reference temperature $T_0$ and a dimensionless parameter $\Upsilon$ which allows us to dial the temperature around $T_0$.  Writing $D=D_0\Upsilon$, we further define dimensionless variables 
\begin{equation}
	x=\sqrt{2D_0\varepsilon} \,\hat{x}\,,\quad t=\varepsilon \, \hat{t} 
\end{equation}
\begin{equation}
	V(x)=\frac{2D_0}{\varepsilon} \, \hat{V}(\hat{x})\,,\quad \eta(t)=\sqrt{\frac{2D_0}{\varepsilon}} \, \hat{\eta}(\hat{t})
\end{equation}
in terms of which the dynamical equation becomes
\begin{eqnarray}\label{eq:Langevin-dimless}
	\frac{\mathrm{d}\hat{x}}{\mathrm{d}\hat{t}} &=& -\frac{\partial\hat{V}}{\partial\hat{x}} + \hat{\eta}(\hat{t}) \label{eq:langevindimless}\\
	\langle \hat{\eta}(\hat{t}) \hat{\eta}(\hat{t}')\rangle &=& \Upsilon \delta(\hat{t} - \hat{t}')
\end{eqnarray}   
From here onwards we will be dropping the hats for simplicity of notation but  generally refer to dimensionless quantities unless otherwise stated. 

\subsection{The Brownian Motion Path Integral}\label{subsec:BMPI}
In order to bring the tools of Quantum Field Theory \cite{Peskin:1995ev, Vasiliev1998} to bear, we will need to reformulate the stochastic differential equation (\ref{eq:langevindimless}) in terms of a path integral. In this subsection we will outline one known way to obtain this path integral, aiming to link this to Supersymmetric Quantum Mechanics. Our final expression, and the starting point of our subsequent analysis, is the Brownian Motion transition probability (\ref{eq:TransProb}), expressed in terms of an integral over possible histories weighted by the action (\ref{eq: PDF action}), to which the busy reader may progress if uninterested in the details of the derivation. We will be using a condensed functional notation of infinite dimensional functional integrals but all expressions can be considered as limits of finite, high dimensional ordinary integrals. This derivation is based on the path integral reformulation by De Dominicis, Peliti and Janssen \cite{DEDOMINICIS1976, Janssen1976, DeDominicis1978} of the well known Martin-Siggia-Rose approach for stochastic dynamics, first developed in \cite{Martin1973}.  More details on these path integrals, including the corresponding finite discretisation of the stochastic process can be found in \cite{Lau2007} - see also \cite{Hertz2017} for a pedagogical exposition. 

The dynamics of the (dimensionless) Langevin equation (\ref{eq:langevindimless}) can be captured in terms of the Probability Distribution Function (PDF) $\mathcal{P}(x_{f}\vert x_{i})$ of observing the particle at $x_f$ at time $t = t_f$ given that initially, at $t = t_i$, the particle was at $x_i$. By definition this can be expressed as:
\begin{equation}
	\mathcal{P}(x_{f}\vert x_{i}) = \left\langle \delta\left( x(t_f)-x_f \right) \right\rangle \label{eq:PDF defn}
\end{equation}
where the expectation value is taken over all possible realisations of the noise $\eta(t)$ and $\delta\left( x(t_f)-x_f \right) $ is the Dirac delta function. Put another way, $x(t_f)$ is the position at $t_f$ for a given noise history $\eta(t)$ and the brackets indicate averaging over all possible noise histories, or stochastic paths, which start at $x_i$ and end up at $x(t_f)=x_f$ at $t_f$. To express this in a path integral form, we can rewrite the PDF using a Gaussian measure for the noise (\ref{eq:eta defn}) and express the average as
\begin{eqnarray}
	\mathcal{P}(x_{f}\vert x_{i}) = \int\mathcal{D}\eta(t)\delta\left( x(t_f)-x_f \right)\text{exp}\left[-\int \text{d}t \, \dfrac{\eta^2(t)}{2\Upsilon}\right]\nonumber \\
	\label{eq: PDF first PI}
\end{eqnarray}
where each noise history is weighted by the exponential factor in the above expression.  We now consider the identity (see e.g. \cite{zinn2002quantum}):
\begin{eqnarray}
	1 &=& \int dx_f\int_{x_i}^{x_f}\mathcal{D}x(t)\,\delta\left( x(t)-x_\eta(t)\right)\\
	& = & \int dx_f\int_{x_i}^{x_f}\mathcal{D}x(t)\,\delta\left( \dot{x} + V_{,x} -\eta (t)\right) \text{det}\textbf{M}\nonumber \\
	&=& \int_{x_i}\mathcal{D}x(t)\,\delta\left( \dot{x} + V_{,x} -\eta (t)\right) \text{det}\textbf{M} \label{eq: useful ident}
\end{eqnarray}
where the matrix $\textbf{M}(t,t')$ is:
\begin{eqnarray}
	\textbf{M} &\equiv &\dfrac{\delta \left[ \dot{x} + V_{,x} -\eta (t)\right]}{\delta x(t')} = \left(\dfrac{d}{dt} +  V_{,xx}\right) \delta(t-t')\label{eq: M defn}\,.
\end{eqnarray}
This identity expresses the obvious fact that, if the particle starts at some $x_i$ and follows a particular history $x_\eta(t)$ dictated by the Langevin equation without disappearing, it will end up somewhere after time $t_f$. We have used the standard subscript notation to denote derivative with respect to that variable e.g. $V_{,xx} = \partial_{xx}V$.
Note that the path integral in (\ref{eq: useful ident}) is over all paths starting at $x_i$ at $t_i$ and ending at any $x$ at $t_f$. Inserting this `fat unity' factor (\ref{eq: useful ident}) into (\ref{eq: PDF first PI}) and noting that the delta function there restricts $x(t_f)$ to be $x_f$ we obtain:
\begin{eqnarray}
	\mathcal{P}(x_{f}\vert x_{i}) &=& \int\limits_{x(t_i)=x_i}^{x(t_f)=x_f}\mathcal{D}\eta\mathcal{D}x\,\delta\left[ \dot{x} + V_{,x} -\eta \right] \text{ det}\textbf{M}\nonumber \\
	&& \times \text{ exp}\left[-\int \text{d}t \,\dfrac{\eta^2(t)}{2\Upsilon}\right] \label{eq: PDF 2nd PI}
\end{eqnarray}
where the $\mathcal{D}x(t)$ integral is taken over all paths beginning at $x_i$ and ending at $x_f$. 
We can rewrite the delta function as a functional Fourier transform using a new variable $\tilde{x}$ which is usually called the response field:
\begin{eqnarray}
	\delta\left[ \dot{x} + V_{,x} -\eta \right] = \int \mathcal{D}\tilde{x}\text{ exp}\left[ i\int \text{d}t~\tilde{x}\left( \dot{x} + V_{,x} -\eta \right)\right] \nonumber \\
	\label{eq: tilde x PI}
\end{eqnarray}
There are a couple of standard ways we can incorporate $\rm{det} \textbf{M}$ into an exponential. We can formally write  
\begin{equation}
	\textbf{M}=\left(\frac{d}{dt}\right)\left(1+\left(\frac{d}{dt}\right)^{-1} V_{,xx}\right)\equiv\left(\frac{d}{dt}\right)\tilde{\textbf{M}}\,,
\end{equation}
where
\begin{equation}
\left(\frac{d}{dt}\right)^{-1}(t,t')=\lambda\Theta(t-t') - \left(1-\lambda\right)\Theta(t'-t)\,.
\end{equation}
Imposing retarded (causal) boundary conditions, which are appropriate for the problem at hand, requires that we set $\lambda=1$ and we find    
\begin{eqnarray}
	\text{det} \textbf{M} &=& \text{det}\left(\frac{d}{dt}\right) \times \text{det} \tilde{\textbf{M}} \propto\text{exp}\left[ \text{Tr}\,\text{log}\left(\tilde{\textbf{M}} \right)\right] \nonumber\\
	&\propto& \text{exp}\left[\dfrac{1}{2}\int \text{d}t~V_{,xx}\right]\label{eq:det-Alg}
\end{eqnarray}
where we used the Stratonovich prescription $(\theta (0) = 1/2)$.
Alternatively, and to make the link with SUSY, we can use anticommuting Grassmann variables $c$ and $\bar{c}$  such that:
\begin{eqnarray}
	\text{det} \textbf{M} = \int \mathcal{D}c\mathcal{D}\bar{c}\text{ exp}\left[\int\text{d}t \, \bar{c}\left( \partial_{t} +  V_{,xx} \right)c \right] \label{eq: ccbar PI}
\end{eqnarray}
The determination of $\text{det}\textbf{M}$ then requires appropriate boundary conditions for $\bar{c}$ and $c$, which are
\begin{equation}    \label{eq:ghost_bcs}
c(t_{\rm in})=0\,,\quad \bar{c}(t_{\rm f})=0\,.
\end{equation}
Other choices are possible but lead to determinant values that are different from (\ref{eq:det-Alg}), corresponding to non-causal boundary conditions for the stochastic problem. The boundary condition (\ref{eq:ghost_bcs}) is implied by the discretised form of the path integral, see \cite{Hertz2017}. Its appropriateness is verified by direct computation in Appendix \ref{app:det}. The discretised path integral can also be consulted to infer that we must further impose
\begin{eqnarray}
\tilde{x}(t_{\rm f}) = 0 . \label{eq:auxiliary_bc}
\end{eqnarray} 
Although introducing $c$ and $\bar{c}$ is not strictly necessary, it pays to keep the determinant expressed in this form since, as we see below, it allows us to conveniently express hidden symmetries of the resulting action. Inserting equations (\ref{eq: tilde x PI}) \& (\ref{eq: ccbar PI}) into (\ref{eq: PDF 2nd PI}) we obtain:
\begin{eqnarray}
	\mathcal{P}(x_{f}\vert x_{i}) &=& \int\mathcal{D}\eta\mathcal{D}x\mathcal{D}\tilde{x} \mathcal{D}c\mathcal{D}\bar{c} \nonumber \\
	&&\text{ exp}\Bigg[\int\text{d}t \Big\{-\dfrac{\eta^2}{2\Upsilon} + i\tilde{x}\left( \dot{x} + V_{,x} -\eta \right) \nonumber \\
	&&\quad \quad \quad + \bar{c}\left( \partial_{t} +  V_{,xx} \right)c \Big\} \Bigg]
\end{eqnarray}
We can now trivially perform the gaussian integral over $\eta$ to obtain the path integral in terms of the Brownian Motion (BM) action $\mathcal{S}_{BM}(x,\tilde{x},\bar{c},c)$:
\begin{eqnarray}
	\mathcal{P}(x_{f}\vert x_{i}) &=& \int\mathcal{D}x\mathcal{D}\tilde{x} \mathcal{D}c\mathcal{D}\bar{c} \text{ exp}\left[-\mathcal{S}_{BM}(x,\tilde{x},\bar{c},c)\right] \nonumber \label{eq:TransProb} \\
	\label{eq: PDF final}\\
	\mathcal{S}_{BM}(x,\tilde{x},\bar{c},c) &=& \int \text{d}t\bigg[ \frac{\Upsilon}{2}\tilde{x}^2 - i\tilde{x}(\dot{x}+ V_{,x}) \nonumber \\
	& & \quad \quad \quad - \bar{c}\left( \partial_{t} +  V_{,xx} \right)c \bigg] \label{eq: PDF action}
\end{eqnarray} 
Computing this path integral, which henceforth shall be called the Brownian Path Integral (BPI), without resorting to some approximation is in general impossible analytically. Instead, we will be using numerical solutions to the fRG flow equations to compute it in the LPA + WFR approximations. 

Redefining our fields as :
\begin{eqnarray}
	x(t) &\equiv & \sqrt{\Upsilon}\,\varphi(t) \nonumber \\
	V(x) &\equiv & {\Upsilon} \, W(\varphi) \nonumber \\
	\tilde{x} &\equiv & \dfrac{1}{\sqrt{\Upsilon}}\,(i\dot{\varphi} - \tilde{F})\nonumber \\
	\bar{c}c &\equiv & i\bar{\rho}\rho \nonumber \\ \label{eq:LPAidentifications}
\end{eqnarray}
we obtain 
\begin{equation}
	\mathcal{S}_{BM}[\varphi, \tilde{F}, \bar{\rho}, \rho] = \left[W(\varphi_f) -  W(\varphi_i)\right] + \mathcal{S}_{SUSY}
\end{equation}
where
\begin{eqnarray}
	\mathcal{S}_{SUSY}[\varphi, \tilde{F}, \bar{\rho}, \rho] = \int dt\bigg[&\dfrac{1}{2}&\dot{\varphi}^2 + \dfrac{1}{2}\tilde{F}^2 + i\tilde{F}W_{,\varphi}(\varphi) \nonumber \\
	&-&i \bar{\rho}(\partial_{t} +  W_{,\varphi\varphi}(\varphi))\rho\bigg]  \label{eq:SUSYClass}
\end{eqnarray} 
Action (\ref{eq:SUSYClass}) describes the dynamics of Euclidean, or imaginary time, Supersymmetric Quantum Mechanics where $\rho$ \& $\bar{\rho}$ are the fermionic fields and $\varphi$ \& $\tilde{F}$ are the bosonic fields \cite{Synatschke2009}. The same action also describes Brownian motion and the BM action is equivalent to the SUSY QM one up to a factor depending on the initial and final positions $x_i$ \& $x_f$; these terms can be simply taken outside the path integral as an exponential prefactor.

Variation of $\mathcal{S}_{SUSY}$ with respect to $\tilde{F}$ yields its ``equation of motion'' $\tilde{F}=-iW_{,\varphi}$ which when substituted back into $\mathcal{S}_{SUSY}$ yields the ``on mass-shell'' action
\begin{eqnarray}
	\mathcal{S}_{OM}[\varphi, \bar{\rho}, \rho] = \int dt\bigg[&\dfrac{1}{2}&\dot{\varphi}^2 + \dfrac{1}{2}W_{,\varphi}{}^2 \nonumber \\
	&-&i \bar{\rho}(\partial_{t} +  W_{,\varphi\varphi})\rho\bigg]
\end{eqnarray} 
We will keep working with the auxiliary field $\tilde{F}$ and (\ref{eq:SUSYClass}) as it allows for the symmetry transformations to take on a simpler form, linear in all fields.

It is illuminating to express the above action in terms of the original dimensional variables and perform the integration over $\bar{\rho}$ and $\rho$, leading to the alternative form of the term stemming from the determinant:
\begin{eqnarray}
	\mathcal{S}_{OM}[x] = \int \frac{dt}{2Dm}\bigg[&\dfrac{1}{2}&m\dot{x}^2 + \dfrac{1}{2}\varepsilon^2 mV_{,x}{}^2 - Dm\varepsilon V_{,xx}\bigg] \nonumber \\\label{eq:dim-action}
\end{eqnarray} 
Note that $2Dm$ has the dimensions of action and therefore plays in the thermal problem a role analogous to $\hbar$ in quantum mechanics - see also section \ref{sec:F-P} in this respect. Unlike $\hbar$ of course, it can be varied by changing the temperature, therefore controlling the strength of fluctuations.


\subsection{\label{sec:F-P}The Fokker-Planck equation and the spectral expansion}
Before moving on to the fRG we outline the more standard procedure as to how the on-mass shell action (\ref{eq:dim-action}) can be obtained from the Fokker-Planck equation which resembles a Euclidean Schr\"{o}dinger equation. We will see later -- in section \ref{sec:comparspec} -- that the spectral expansion method outlined here confirms the validity of the fRG approach at moderate to high temperatures.

Instead of working with the Langevin equation directly once can deal with the probability distribution of position:
\begin{equation}
	P(x,t) = \left\langle \delta (x-x_{\eta})\right\rangle
\end{equation} 
where $x_{\eta}$ is the solution to (\ref{eq:langevin}) for a given noise function $\eta$ (i.e. a specific particle trajectory). It can be shown that this evolves according to the following PDE:
\begin{equation}
	\dfrac{\partial P(x,t)}{\partial t} = \partial_{x}(P(x,t)\partial_x V) + \dfrac{\Upsilon}{2}\partial_{xx} P(x,t) \label{eq: F-P}
\end{equation}
which is known as the Fokker-Planck (F-P) equation. It is usually more useful however to rescale the PDF like so:
\begin{equation}
	P(x,t) = e^{-V/\Upsilon}\tilde{P}(x,t) \label{eq:Ptransform}
\end{equation}
This leads to the F-P equation taking the form:
\begin{eqnarray}\label{eq:FP1}
	\dfrac{\Upsilon}{2}\dfrac{\partial\tilde{P}(x,t)}{\partial t} &=& \left(\dfrac{\Upsilon}{2}\right)^2\partial_{xx}^{2}\tilde{P}(x,t)  + \bar{U}\tilde{P}(x,t) \label{eq: rescaled F-P}\\
	\bar{U} &\equiv & \dfrac{\Upsilon}{4}\partial_{xx}^{2} V - \dfrac{1}{4}(\partial_{x} V)^2 \label{eq: Ubar=}
\end{eqnarray}
which resembles a Euclidean Schr\"{o}dinger equation with $\Upsilon /2$ playing the role of $\hbar$ in controlling the fluctuation amplitude, as one might expect. 

Equation (\ref{eq:FP1}) can be solved in terms of a spectral expansion (see e.g. \cite{gardiner2009stochastic,Markkanen:2019kpv}). Writing 
\begin{equation}
	\tilde{P}(x,t) = \sum\limits_{n=0}^\infty c_n p_n(x)e^{-E_n t}
\end{equation}  
we find that $p_n$ satisfy the corresponding, time independent  Euclidean Schr\"{o}dinger equation
\begin{equation}
	-\frac{\Upsilon}{2}\dfrac{d^2 p_n}{dx^2}  +\frac{1}{2}\left(\frac{\left(V_{,x}\right)^2}{\Upsilon}- V_{,xx}\right)p_n=E_np_n
\end{equation} 
The lowest eigenfunction with $E_0=0$ is 
\begin{equation}
	p_0(x) =\mathcal{N} e^{-V(x)/\Upsilon}
\end{equation} 
corresponding to the equilibrium distribution $P_{\rm eq}(x)=p_0(x)^2$. The $p_n(x)$ eigenfunctions are complete and orthonormal  
\begin{eqnarray}
\int\limits_{-\infty}^{\infty} dx \, p_n(x)p_m(x)=\delta_{mn} \\
\sum\limits_{n=1}^{\infty} p_n(x)p_n(x_0)=\delta(x-x_0)
\end{eqnarray}
The conditional probability, a quantity akin to the evolution operator or propagator in quantum mechanics, can be expressed in terms of the spectral expansion as 
\begin{eqnarray}\label{eq:specral-prop1}
\tilde{P}(x,t|x_0,0) &=& \sum\limits_{n=0}^\infty p_n(x)p_n(x_0)e^{-E_n t}\\
P(x,t|x_0,0) &=& e^{-{V(x)}/{\Upsilon}}\tilde{P}(x,t|x_0,0)e^{+{V(x_0)}/{\Upsilon}} \label{eq:spectral-prop2}
\end{eqnarray}
Any correlation function can then be expressed by using (\ref{eq:spectral-prop2}). An economic notation can be achieved by using Dirac bra-ket notation in terms of which e.g.
\begin{equation}
	\tilde{P}(t,0)=\sum\limits_{n=0}^{\infty}\left|n\rangle e^{-E_n t}\langle n \right|
\end{equation} 
Correlation functions can then be expressed in the spectral expansion as:  
 \begin{eqnarray}
 \left\langle f(x(t)) g(x(0))\right\rangle = \sum\limits_{n=0}^{\infty}\left\langle 0 \right| f \left| n \right\rangle e^{-E_n t} \left\langle n \right|g \left| {\rm in} \right\rangle
 \end{eqnarray}
where, explicitly 
\begin{eqnarray}
\left\langle 0 \right| f \left| n \right\rangle &=& \int\limits_{-\infty}^{\infty}dx\, p_0(x)f(x)p_n(x) \\
\left\langle n \right| g \left| {\rm in} \right\rangle &=& \int\limits_{-\infty}^{\infty}dx\, p_n(x)\,g(x) \tilde{P}(x,0) 
\end{eqnarray}  
Note that the ``out state'' in the stochastic problem is always $\langle 0 |$ and the ``in state'' is defined in terms of $\tilde{P}(x,t=0)$. 

We can also write the conditional probability
\begin{equation}
P(x,t|x_0,0)= \left \langle x\right| e^{-{V(x)}/{\Upsilon}} \tilde{P}(t,0)e^{+{V(x_0)}/{\Upsilon}}\left|x_0\right\rangle 
\end{equation}  
governed by the above Euclidean Schr\"{o}dinger equation, as a path integral
\begin{equation}
	\begin{split}
		&P(x,t|x_0,0) = \mathcal{N}\text{ exp}\left(\dfrac{\varepsilon}{2D}\left[V(x)-V(x_0)\right]\right) \\
		&\times\!\!\!\!\int\limits_{x(0)=x_0}^{x(t)=x} \!\!\!\!\!\!\mathcal{D}x(\tau) \text{ exp}\left( -\int \dfrac{d\tau}{2Dm} \left\{\dfrac{1}{2}m(\partial_{\tau}{x})^2 - \bar{U}(x)\right\}\right) \label{eq:PI F-P}
	\end{split}
\end{equation}
where we have reinstated the dimensionful variables. We therefore recover the ``on mass-shell" path integral (\ref{eq:dim-action}) obtained earlier. Note the importance of including the determinant (\ref{eq:det-Alg}) in order to obtain the $\partial_{xx}^2V$ term in the Schr\"{o}dinger potential $\bar{U}$. 

Before proceeding to the next sections we should add a comment regarding the above path integrals. Beyond being expressions that allow formal manipulations, they can also be understood as limits of large multi-variate integrals arising from the discretization of time evolution into small discrete time intervals. In general, this discretization results into apparent ambiguities \cite{Leschke1977}\footnote{We would like to thank the anonymous referee for highlighting this point.}. In our case this can also be linked to the discretization of the Langevin equation (\ref{eq:Langevin-dimless}) (the index $i$ refers to the $i^{\rm th}$ time interval)
\begin{equation}
x_i=x_{i-1} + V'(x_i^\star)\Delta t + \int\limits_{t_{i-1}}^{t_i}ds \, \xi(s)
\end{equation}
where $x_i^\star = \alpha x_i + \left(1-\alpha\right)x_{i-1}$. In the above discussions and manipulations we have tacitly assumed $\alpha=1/2$ (Stratonovich) which, in the continuous limit is equivalent to setting $\Theta(0)=1/2$ \cite{Lau2007}. It is well known however that, as long as the noise amplitude does not depend on $x$ (known as additive noise), the solution to the Langevin equation is unique, as is the corresponding Fokker-Planck equation. Our path integrals respect this and all results derived from them are independent of the choice of $\alpha$. Calculations with the discrete version, show that contributions from terms involving $\alpha$ cancel. For example, one can start from the discretized path integral with an arbitrary choice of $\alpha$ and obtain a unique Fokker-Planck equation - see e.g. \cite{Lau2007}. In the continuum formulation, the supersymmetry discussed in subsection \ref{WT} below imposes the cancellation of terms where the ambiguous quantity $\Theta(t=0)$ appears. This is also reflected in the path integral (\ref{eq:PI F-P}) through the special form of the potential $U(x)$ in terms of the ``superpotential" $V(x)$. Overall, this non-dependence on $\alpha$ ultimately stems from the inclusion of the determinant  $\text{det}\textbf{M}$ in (\ref{eq: useful ident}). These considerations support the view that the formal path integrals considered here are well defined and free from any ambiguity.

\subsection{Correlation functions from the generating functional}
One way to compute correlation functions is through the use of objects called \textit{generating functionals}. In this subsection we will outline how these generating functionals yield correlators in practice. We will then outline in section \ref{sec:fRG} how the fRG can be used to compute these generating functionals in the first place and therefore how to obtain correlation functions in section \ref{sec:acc eom}.

The first generating functional we examine is the partition functional $\mathcal{Z}(\rm J)$ which depends on source terms ${\rm J}(t)$ (in analogy with a magnetic field source term $M(x)$ in spin systems) 
\begin{equation}\label{eq:partition functional}
	\mathcal{Z}({\rm J}) = \int\mathcal{D}\Phi \text{ exp}\left[-\mathcal{S}_{BM}[\Phi]  + \int dt \,{\rm J} \Phi\right]\,,
\end{equation}
which, under variation w.r.t. ${\rm J}$ will give any required correlator. In the above functional integral, $\Phi$ stands collectively for $(\varphi(t),\tilde{F}(t),\rho(t),\bar{\rho}(t))$ and ${\rm J}(t)$ for all the corresponding currents
\begin{equation}
 \int dt \,{\rm J} \Phi \equiv \int dt \left(  J_\varphi \varphi + J_{\tilde{F}} \tilde{F} + \bar{\rho}\zeta + \bar{\zeta}\rho    \right) \label{eq:currents}
\end{equation}
The only constraint we will require of the currents is that they satisfy ${\rm J}(t_{\rm in}) ={\rm J}(t_{\rm f}) = 0$ at the initial and final times $t_{\rm in}$ and $t_{\rm f}$. 

The averages of the fields are defined by 
\begin{eqnarray}
	\left\langle \Phi(t)\right\rangle &\equiv&  {\int\mathcal{D}\Phi\,\, \Phi(t)\,\,\text{ exp}\left[-\mathcal{S}_{BM}[\Phi] \right]}\\
	&=&	\dfrac{\delta \mathcal{Z}[{\rm J}]}{\delta {\rm J}(t)}\Bigg|_{{\rm J}=0}\,,
\end{eqnarray}
the two point correlation function is:
\begin{eqnarray}
	\left\langle \Phi(t_{1})\Phi(t_{2})\right\rangle &\equiv&  \frac{\int\mathcal{D}\Phi\,\, \Phi(t_1)\Phi(t_2)\,\,\text{ exp}\left[-\mathcal{S}[\Phi] \right]}{\int\mathcal{D}\Phi \text{ exp}\left[-\mathcal{S}[\Phi] \right]}\\
	&=& \dfrac{\delta^2\mathcal{Z}(J)}{\delta J(t_{2})\delta J(t_{1})}\bigg\rvert_{J = 0}\,,
\end{eqnarray}	
and similarly for higher correlation functions. Note that the usual normalization by a factor $\mathcal{Z}(0)^{-1}$ that is included in general, is omitted here since for this theory $\mathcal{Z}(0)=1$ by construction.    

Defining 
\begin{equation}
	\mathcal{W}[{\rm J}] \equiv \text{ln}\left(\mathcal{Z}({\rm J})\right)
\end{equation}
allows us to compute connected correlation functions (or Ursell functions) as:
\begin{equation}
	\left\langle \Phi(t_1)...\Phi(t_n)\right\rangle_{C} = \dfrac{\delta^n \mathcal{W}[{\rm J}]}{\delta {\rm J}(t_1)...\delta {\rm J}(t_n)}\Bigg|_{{\rm J}=0}
\end{equation}
For instance the connected 2-point function (more commonly known as covariance) $G(t_1,t_2)$ is:
\begin{eqnarray}
	G(t_1,t_2) \equiv \left\langle \Phi(t_1)\Phi(t_2) \right\rangle_{C} &=& \left\langle \Phi(t_1)\Phi(t_2) \right\rangle - \left\langle \Phi(t_1) \right\rangle\left\langle \Phi(t_2) \right\rangle \nonumber \\
	&=& \dfrac{\delta^2 \mathcal{W}[{\rm J}]}{\delta {\rm J}(t_1)\delta {\rm J}(t_2)}\Bigg|_{{\rm J}=0}
\end{eqnarray}

\subsection{Symmetry transformations for $\mathcal{S}_{BM}$ and Ward-Takahashi identities}\label{WT}
In this subsection we recall the transformations that leave the action  $\mathcal{S}_{BM}$ invariant, up to boundary terms. We comment on the implications of such symmetries, also paying attention to the boundary terms that are usually dropped under the assumption of equilibrium, or, equivalently, a corresponding infinite amount of elapsed time between initial and final states \cite{Mallick2011}. For us to later use SUSY fRG flow equations in section \ref{sec:fRG} it is crucial to verify the presence of this symmetry in an out-of-equilibrium context.

In general, invariances of the action imply relations between various correlation functions in field theory, generally known as Ward-Takahashi (W-T) identities. Their derivation can be summarized as follows: A general infinitesimal transformation of the fields $\Phi\rightarrow \Phi'= \Phi+\Delta\Phi$ will generically change the action $S\rightarrow \mathcal{S}'=\mathcal{S}+\Delta \mathcal{S}$. Also shifting ${\rm J}\rightarrow {\rm J}+\Delta {\rm J}$ leads to 
\begin{eqnarray}
\Delta \mathcal{Z}=\int\!\! dt  \dfrac{\delta \mathcal{Z}}{\delta {\rm J}(t)}\Delta J(t) &=&\! \int\!\!\mathcal{D}\Phi e^{-\mathcal{S}[\Phi]  + \int \!\!dt \,{\rm J} \Phi}\nonumber\\
&\times&\left(-\Delta S+{\rm J}\Delta\Phi + \Delta{\rm J}\,\Phi\right) 
\end{eqnarray}
where we used that a) $\Phi$ is simply an integration variable in (\ref{eq:partition functional}) and $\mathcal{Z}$ is not altered by a change in $\Phi$ but only via J and b) $\mathcal{D}\Phi = \mathcal{D}\Phi'$ i.e.~the transformation involves no non-trivial Jacobian determinant. Symmetries of the dynamical system comprise of transformations for which $\Delta S$ is, at most, a total derivative (or a total divergence for higher dimensions): $\Delta \mathcal{S}=\int dt \dfrac{d}{dt}\mathcal{A} = \mathcal{A}(t_f)-\mathcal{A}(t_i) = \left[\mathcal{A}\right]^{t_f}_{t_i}$. Further choosing $\Delta {\rm J}$ such that, for a given $\Delta\Phi$, ${\rm J}\Delta\Phi + \Delta{\rm J}\,\Phi=0$, leads to
\begin{equation}
	\int dt  \dfrac{\delta \mathcal{Z}}{\delta {\rm J}(t)}\Delta {\rm J}(t) = \int\mathcal{D}\Phi \left[\mathcal{A}\right]^{t_f}_{t_i}e^{-\mathcal{S}[\Phi]  + \int dt \,{\rm J} \Phi} 
\end{equation}
Differentiating this master equation w.r.t.~${\rm J}$ and setting ${\rm J}=0$, gives relations between correlations functions that are necessitated by the symmetry under $\Phi \rightarrow \Phi+\Delta\Phi$.

For our case, given two independent, infinitesimal Grassmann variables $\epsilon$ and $\bar{\epsilon}$, the following transformations of the fields \cite{Synatschke2009}
\begin{eqnarray}
	\varphi &\rightarrow & \varphi + i \bar{\epsilon}\rho - i \bar{\rho}\epsilon \label{SUSY1}\\
	\tilde{F} &\rightarrow & \tilde{F} - \bar{\epsilon}\dot{\rho} - \dot{\bar{\rho}}\epsilon\\
	\rho &\rightarrow & \rho +\left(\dot{\varphi}-i \tilde{F}\right)\epsilon \\
	\bar{\rho} &\rightarrow & \bar{\rho} + \bar{\epsilon}\left(\dot{\varphi} + i \tilde{F}\right) \label{SUSY4}
\end{eqnarray}
leave $\mathcal{S}_{BM}$ invariant up to a boundary term at the initial time $t_{\rm in}$:
\begin{equation}\label{eq:action change}
\mathcal{S}_{BM} \rightarrow \mathcal{S}_{BM} + \bar{\rho}_{\rm in}\left(i\dot{\varphi}+ \tilde{F} +2i W_{,\varphi}\right)_{\rm in}\epsilon
\end{equation}
where a subscript $`{\rm in}$ 'denotes the initial time $t_{\rm in}$. The boundary term at $t_{\rm f}$ has been eliminated using the boundary condition  (\ref{eq:ghost_bcs}).
Note that the $\bar{\epsilon}$ transformation leaves $\mathcal{S}_{BM}$ invariant identically, irrespective of the boundary conditions.

Adding source currents $\left(J_\varphi,\, J_{\tilde{F}},\, \zeta, \bar{\zeta}\right)$ to the action \cite{Damgaard1987}
\begin{equation}
	\mathcal{S}_{BM} \rightarrow \mathcal{S}_{BM} - \int dt \left(  J_\varphi \varphi + J_{\tilde{F}} \tilde{F} + \bar{\rho}\zeta + \bar{\zeta}\rho   \right)
\end{equation}
and requiring appropriate transformations of those currents, 
\begin{eqnarray}
	J_\varphi &\rightarrow& J_\varphi + \dot{\bar{\zeta}} \epsilon + \bar{\epsilon} \dot{\zeta}  \\
	J_{\tilde{F}} &\rightarrow& J_{\tilde{F}} + i\bar{\zeta} \epsilon - i \bar{\epsilon} \zeta \\
	\zeta &\rightarrow& \zeta + \epsilon \left(i J_\varphi -\dot{J}_{\tilde{F}}\right)\\
	\bar{\zeta} &\rightarrow& \bar{\zeta} - \bar{\epsilon} \left(i J_\varphi +\dot{J}_{\tilde{F}}\right)
\end{eqnarray} 
we have
\begin{equation}
	{\rm J}\Phi \rightarrow {\rm J}\Phi- \bar{\epsilon}\frac{d}{dt}\left(\rho J_{\tilde{F}}-\varphi\zeta\right)- \frac{d}{dt}\left(\bar{\rho} J_{\tilde{F}}-\varphi\bar{\zeta}\right){\epsilon}
\end{equation}
We therefore see that the transformations result in     
\begin{equation}
		\mathcal{S}_{BM} - 	{\rm J}\Phi \rightarrow  \mathcal{S}_{BM} - 	{\rm J}\Phi	+ \bar{\rho}_{\rm in}\left(i\dot{\varphi}+ \tilde{F} +2i W_{,\varphi}\right)_{\rm in} \!\! \epsilon
\end{equation}
and the exponent in the integrand of (\ref{eq:partition functional}) only changes by a lower boundary term that is also independent of $\bar{\epsilon}$. 

The field transformation (\ref{SUSY1}) - (\ref{SUSY4}) are linear shifts that leave the integration measure in the path integral invariant. Coupled with the shift in the currents we find, setting $\epsilon = 0$
\begin{equation}
	\int dt \left[\frac{\delta \mathcal{Z}}{\delta J_{\varphi}(t)}\dot{\zeta}-i\frac{\delta \mathcal{Z}}{\delta J_{\tilde{F}}(t)}\zeta  -  \frac{\delta \mathcal{Z}}{\delta \bar{\zeta}(t)}\left(iJ_\varphi+\dot{J}_{\tilde{F}}\right)\right] = 0 \label{BRST1}
\end{equation}     
while for $\bar{\epsilon}=0$ we obtain
\begin{eqnarray}
	\int dt \left[\frac{\delta \mathcal{Z}}{\delta J_{\varphi}(t)}\dot{\bar{\zeta}} + i\frac{\delta \mathcal{Z}}{\delta J_{\tilde{F}}(t)}\bar{\zeta}  -  \frac{\delta \mathcal{Z}}{\delta {\zeta}(t)}\left(iJ_\varphi-\dot{J}_{\tilde{F}}\right)\right] \nonumber \\
	= \! \int  \!\!\!  \mathcal{D}\Phi  \left[-\bar{\rho}_{\rm in}\left(i\dot{\varphi}_{\rm in}+\tilde{F}_{\rm in} +2iW'_{\rm in}\right)\right]e^{-\mathcal{S}_{BM}+\int \!\! dt {\rm J}\Phi} \label{BRST2}
\end{eqnarray}	
These are the master equations from which so-called Ward-Takahashi identities between various correlators can be obtained. For example, differentiating (\ref{BRST1}) w.r.t.  $J_{\varphi}(t')$, $\zeta(\tau)$ and setting ${\rm J}=0$ gives
\begin{equation}\label{eq:WT1}
	\frac{d}{d\tau}\left\langle \varphi(t')\varphi(\tau)\right\rangle + \left\langle\varphi(t')W'(\varphi(\tau))\right\rangle - i \langle \rho(t')\bar{\rho}(\tau)\rangle = 0
\end{equation}    
which, along with the original Langevin equation, allows us to infer that 
\begin{equation}
i \langle \rho(t')\bar{\rho}(\tau)\rangle = \frac{1}{\sqrt{\Upsilon}}\left\langle  \varphi(t)\eta(\tau)\right\rangle\,,
\end{equation}  
meaning that $\langle \rho(t')\bar{\rho}(\tau)\rangle$ is proportional to the response of $\varphi(t')$ to noise $\eta(\tau)$ (clearly a retarded quantity $\propto \Theta(t'-\tau)$). Furthermore, equation (\ref{eq:WT1}) can be rewritten as 
\begin{equation}\label{eq:ghost=response}
	\sqrt{\Upsilon}\left\langle \tilde{x}(\tau)\varphi(t')\right\rangle =- \langle \bar{\rho}(\tau)\rho(t')\rangle
\end{equation} 
which confirms that $\left\langle \varphi(t')\tilde{x}(\tau)\right\rangle$ is the retarded response function or propagator. Importantly, equation (\ref{eq:ghost=response}) also establishes that in a diagrammatic expansion closed ghost loops act to cancel closed loops involving the retarded propagator. This ensures $\mathcal{Z}[{\rm J}=0]=1$, which simply reflects conservation of probability, and furthermore that correlators do not depend on the ill-defined quantity $\Theta(0)$, reflecting the well-known fact that, for additive noise, the discretization of the stochastic differential equation (Ito, Stratonovic etc) does not matter.         

Differentiating (\ref{BRST2}) w.r.t. $J_\varphi(t')$, $\bar{\zeta}(\tau)$ and setting ${\rm J}=0$ gives, with the use of (\ref{eq:ghost=response}) and recalling that integration over $\tilde{F}$ gives $\tilde{F} \rightarrow -iW_{,\varphi}$, 
\begin{eqnarray}
	2\frac{d}{d\tau} \left\langle \varphi(t')\varphi(\tau)\right\rangle &-&i\sqrt{\Upsilon}\left\langle \tilde{x}(\tau)\varphi(t')\right\rangle + i\sqrt{\Upsilon}\left\langle \tilde{x}(t')\varphi(\tau)\right\rangle \nonumber \\
 &=& -i\Upsilon \left\langle \tilde{x}_{\rm in} \varphi(\tau)\right\rangle\left\langle \tilde{x}_{\rm in} \varphi(t')\right\rangle
\end{eqnarray}  
This is a modified Fluctuation-Dissipation relation with the term on the rhs accounting for the initial condition. Sending $t_{\rm in} \rightarrow -\infty$ makes the rhs vanish and we recover the Fluctuation-Dissipation relation at equilibrium \cite{Mallick2011}:
\begin{equation}
		\frac{d}{d\tau} \left\langle \varphi(t')\varphi(\tau)\right\rangle =  i\frac{\sqrt{\Upsilon}}{2}\left(\left\langle \tilde{x}(\tau)\varphi(t')\right\rangle - \left\langle \tilde{x}(t')\varphi(\tau)\right\rangle\right)
\end{equation}

\section{\label{sec:fRG}Applying the functional Renormalisation Group}

The fRG has already been applied to study non-equilibrium physics, see e.g. \cite{Canet2004, Gezzi2007, Jakobs2007, Gasenzer2008, Berges2009, Gasenzer2010, Kloss2011, Canet2011, Sieberer2014, Mukherjee2015, Pawlowski2015, Corell2019} for an incomplete selection of references. Recently, the fRG has further been used for averaging fluctuations in the \emph{temporal domain} of Langevin dynamics in \cite{Duclut2017} but without direct use of the supersymmetry. As discussed in \cite{Moreau2020}, the physically inspired conditions the authors of \cite{Duclut2017} require of their flow equations are straightforwardly imposed by the Supersymmetric flow. The Supersymmetric flow equation itself was first derived in \cite{Synatschke2009} but without making any connection to stochastic dynamics. This connection was made independently in \cite{Canet2011} -- see also \cite{Damgaard1987} -- which however considered a field theory in extended spatial dimensions and smoothing the corresponding \emph{spatial fluctuations}, not temporal fluctuations as we do here. In fact, the authors of \cite{Synatschke2009} obtain a slightly different flow equation when wavefunction renormalisation is included since they do not connect the action functional they study to Brownian motion and the corresponding equilibrium Boltzamnn distribution. This Supersymmetric fRG flow has only been very recently utilized in the context of stochastic dynamics in early universe inflation \cite{Prokopec2018, Moreau2020, Moreau2020a}. 

The formulation of the fRG involves the Wetterich equation \cite{Wetterich1993} which is a functional (infinite dimensional) integro-differential equation  describing the flow of the effective action between the microscopic and macroscopic scale. This flow is controlled by a parameter $k$ that ranges from the UV cutoff $\Lambda$ down to the IR regime as $k$ $\rightarrow$ 0. In our Brownian motion scenario, microscopic regime refers to a small timestep and macroscopic to a long timestep. The definition of $\Lambda \sim 1/\Delta t$ is analogous to the Condensed Matter interpretation of the cutoff being inversely proportional to the lattice size, the only difference here being that the Condensed Matter lattice is in space and ours is in time. We will use the fRG ultimately to calculate correlation functions of the particle position. As this derivation uses known techniques and results we refer the busy reader to our basic equations and main results of this section: equation (\ref{eq:dV/dk}) for the Local Potential Approximation to the RG flow and when we also include Wavefunction Renormalisation they are (\ref{eq:dV/dktilde2}) and (\ref{eq:dzetax/dktilde}).

The fRG formulation adds a regulating term to the action in our definition of the generating functional:
\begin{equation}
	\mathcal{Z}_k(J) = \int\mathcal{D}\Phi \text{ exp}\left[-\mathcal{S}[\Phi] - \Delta\mathcal{S}_{k}[\Phi] + \int_t {\rm J} \Phi \right] 
\end{equation}
where the regulating term $\Delta\mathcal{S}_{k}[\Phi]$ is quadratic in $\Phi$:
\begin{equation}
	\Delta\mathcal{S}_{k}[\Phi] = \dfrac{1}{2}\int_{t,t'}\Phi(t)R_{k}(t,t')\Phi(t') 
\end{equation}
Crucially $R_{k}$ is an IR regulator that depends on a Renormalisation scale $k$ and the momentum/frequency $p$ of the modes. The precise form of $R_{k}$ is not crucially important and it is chosen in order to optimize calculations but it should suppress  IR modes and vanish as $k \rightarrow 0$, $\lim\limits_{k\rightarrow 0} R_k=0$, ensuring that the full effective action is recovered in this limit. By defining the mean field as $X(t) \equiv \left< \Phi(t)\right>$ we can construct the Regulated Effective Action:
\begin{equation}
	\Gamma_{k}[X] = \int_t \rm{J} X - \mathcal{W}_{k}[\rm{J}]-\Delta\mathcal{S}_{k}[X] 
\end{equation}
where $\mathcal{W}_{k}[{\rm J}]= \text{ln}\left(\mathcal{Z}_{k}\right)$ and $X$ refers to all the relevant mean fields. 

From the Regulated Effective Action one can obtain the Wetterich equation \cite{Wetterich1993, Morris1994}:
\begin{equation}\label{Wetterich functional}
	\partial_{k}\Gamma_{k}[X] = \dfrac{1}{2}\,{\rm STr}\left\{\int_{t,t'}\partial_{k}R_{k}(t,t')\left[R_{k} + \Gamma_{k}^{(2)}\right]^{-1}\right\} 
\end{equation}
which is a functional equation determining how $\Gamma_k$ changes as $k \rightarrow 0$. $\Gamma^{(2)}$ is the second functional derivative w.r.t. the relevant fields and {\rm STr} refers to the supertrace - see \cite{Synatschke2009} for details. The equation evolves $\Gamma_{k}$ from the microscopic scale ($k = \Lambda$), where $\Gamma_{\Lambda} =\mathcal{S}$, down to the IR regime ($k  = 0$) where the full effective action $\Gamma[\chi] = \Gamma_{k=0}[\chi]$, encoding the effect of all fluctuations, is obtained. A simplified derivation for one degree of freedom at equilibrium, which however captures all the relevant manipulations, can be found in Appendix \ref{app:equil-flow}.

As demonstrated in the previous section, our Brownian motion problem is actually SUSY QM. We can therefore apply the fRG technology and incorporate the effect of thermal fluctuations by following the flow of the effective action $\Gamma_k$ via the Wetterich equation. Synatschke et. al have analysed a system with action $\mathcal{S}_{SUSY}$ in light of its underlying symmetries in \cite{Synatschke2009} - see also \cite{Canet2011}. We adopt their results here. They find that from a supersymmetric perspective, the appropriate regulating term takes the form    
\begin{eqnarray}
	\Delta\mathcal{S}_k\! &=&\! \int_{\tau\tau'} \hspace{-0.2cm} r_2(k,\Delta\tau)  \left[-\dot{\phi}(\tau)\dot{\phi}(\tau')+F(\tau)F(\tau')-i\bar{\psi}(\tau)\dot{\psi}(\tau')\right]\nonumber\\
	&+&2ir_1(k,\Delta\tau) \left[\phi(\tau)F(\tau')-\bar{\psi}(\tau)\psi(\tau'))\right]
\end{eqnarray}
where $\Delta\tau \equiv \tau-\tau'$. Such a form was also suggested in \cite{Duclut2017}, however we will see that compatibility with the Boltzmann distribution suggests setting $r_{2} \rightarrow 0$. The flow equations of \cite{Synatschke2009} are discussed below. 

\subsection{Local Potential Approximation}

In practice, calculating $\Gamma_k$ exactly is usually impossible and we must consider a truncation to make the functional equation (\ref{Wetterich functional}) tractable. The most common approximation is the so-called derivative expansion. The Local Potential Approximation (LPA), the leading order in the derivative expansion, is the assumption that the only part of the effective action that depends on our momentum scale $k$ is the superpotential $W$.  The effective action then takes the form:
\begin{eqnarray}
	\Gamma_{k}[\phi,F,\bar{\psi},\psi] = \int d\tau\bigg[&\dfrac{1}{2}&\dot{\phi}^2 + \dfrac{1}{2}F^2 + iFW_{k, \phi}(\phi) \nonumber \\
	&-& i\bar{\psi}\left(\partial_t + W_{k, \phi\phi} \right)\psi \bigg] \label{eq:SUSYEffective}
\end{eqnarray}
such that $\Gamma_{k = \Lambda} = \mathcal{S}_{SUSY}$ under the condition $W_{k = \Lambda}(\phi) = W(\phi)$ with 
\begin{equation}
\phi \equiv \left\langle \varphi\right\rangle\,,\quad F \equiv \langle \tilde{F} \rangle\ \,, \quad \psi \equiv \left\langle \rho\right\rangle \,, \quad \bar{\psi} \equiv \left\langle \bar{\rho}\right\rangle
\end{equation}
are the mean fields, and we also denote $\chi \equiv \left\langle x \right\rangle = \sqrt{\Upsilon}\phi$. In this approximation the only thing changing with $k$ directly, progressively incorporating the effect of fluctuations on different timescales, is $W_k$. This means we only have one flow equation to solve which turns out to be \cite{Synatschke2009}:
\begin{equation}
	\partial_{k}W_{k}(\phi) = \int_{-\infty}^{\infty}\dfrac{dp}{4\pi}\dfrac{(1+r_2)\partial_{k}r_{1}- \partial_{k}r_{2}~(r_{1} + \partial^2_{\phi}W_{k}(\phi))}{p^2+(r_{1} + \partial^2_{\phi}W_{k}(\phi))^2} \nonumber \\
\end{equation}
We notice that if we set $r_2 = 0$ and choose a local-in-time $r_1(k, \tau-\tau')=k\delta(\tau-\tau')$ the so-called Callan-Symanzik regulator then this choice\footnote{Physically speaking the final results should be independent of the regulator chosen. This is a subtlety we will not address in this work as it was shown in \cite{Synatschke2009} that even for other choice of regulators the difference in the final results was negligible, at least for the LPA.} effectively adds a quadratic term to the potential $W \rightarrow W + k\phi^2$ and leads to a relatively simple flow equation: 
\begin{equation}
	\partial_{k}W_{k}(\phi) = \dfrac{1}{4}\cdot\dfrac{1}{k+ \partial^2_{\phi}W_{k}(\phi)} \,.
	\label{eq:dW/dk}
\end{equation}
In terms of the physical variables we have 
\begin{equation}\label{eq:dV/dk}
	\partial_{k}V_{k}(\chi) = \dfrac{\Upsilon}{4}\cdot\dfrac{1}{k+ \partial^2_{\chi}V_{k}(\chi)} \,,
\end{equation}
which shows explicitly the effect of dialling the temperature $\Upsilon$: the higher the temperature the faster the flow as a result of stronger thermal fluctuations. Equation (\ref{eq:dV/dk}) can be discretised in the $\chi$ direction and become a set of coupled ODEs that can be solved in the $k$ direction in order to obtain a numerical solution.  

It is important to note that equation (\ref{eq:dV/dk}) is identical to the flow of the effective potential that corresponds to the equilibrium Boltzmann distribution, see \cite{Guilleux2015} and Appendix C with $R \rightarrow k$. We  therefore see that the form of $\mathcal{S}_{SUSY}$ and deriving flow equations that respect its symmetries establishes automatic consistency with the equilibrium Boltzmann distribution. If one started  directly from the Onsager-Machlup functional (\ref{eq:dim-action}) and naively treated it as an $N=1$ Euclidean scalar theory in one-dimension with the combination $U = \frac{1}{2}\left(V_{,x}\right)^2 - \frac{\Upsilon}{2}V_{,xx}$ as the scalar potential to be evolved along the RG flow, one would have obtained a different flow equation  
\begin{eqnarray}
	\partial_{k}U_{k}(\phi) = \dfrac{1}{2}\int_{-\infty}^{\infty}\dfrac{dp}{2\pi}\dfrac{\partial_{k}R_{k}}{p^2+ R_{k} + \partial^2_{\phi}U_{k}(\phi)}\,.
\end{eqnarray}
The corresponding Callan-Symanzik regulator would be  $R_k=k^2$, giving
\begin{equation}
	\partial_{k}U_{k}(\phi) = \frac{1}{2} \frac{k}{\sqrt{k^2+\partial^2_{\phi}U_{k}(\phi)}}\,.
\end{equation}
It is unclear how or if the end-of-the-flow potential $U_{k=0}$ from this equation would relate to the physical potential $V_{k=0}$.

\subsection{Wave Function Renormalisation}

In the previous subsection we assumed that the effective action $\Gamma_k$ only depends on the renormalisation scale through the form of the potential. We now allow for the field $\varphi$ itself to be renormalised which results in a scaling of the kinetic term. The new effective action in the SUSY formalism is \cite{Synatschke2009}:
\begin{eqnarray}
	\Gamma_{k}[\phi,\bar{\psi},\psi] = \int dt~\dfrac{1}{2}Z_{,\phi}^{2}\dot{\phi}^{2} + \dfrac{1}{2}\left(\dfrac{W_{,\phi}}{Z_{,\phi}}\right)^2 \nonumber \\
	- i\bar{\psi}\left(Z_{,\phi}^2 \partial_{t} + Z_{,\phi} Z_{,\phi\phi}\dot{\phi} - Z_{,\phi\phi} \dfrac{W_{,\phi}}{Z_{,\phi}} + W_{,\phi\phi}\right)\psi
\end{eqnarray}
where we have suppressed the explicit dependence on $k$ of $W$ \& $Z$ to avoid overly cluttered notation. From now on we will in general drop this explicit dependence on $k$ for $W$, $V$, $Z$ \& $\zeta$, defined below, only restoring it when we are directly comparing it to the original cutoff value. We introduce another identification\footnote{This $\zeta$ is unrelated to the one appearing in (\ref{eq:currents})} in addition to (\ref{eq:LPAidentifications}):
\begin{eqnarray}
	\zeta(x) = \sqrt{\Upsilon} Z(\phi) \Rightarrow \zeta_{,x} = Z_{,\phi} \\
	\bar{c}c = -i\zeta_{,x}\bar{\psi}\psi
\end{eqnarray}
such that the (on-shell) effective action for Brownian motion is now written as:
\begin{eqnarray}
	\Gamma_{k}[\chi,\bar{c},c] = \int dt~\dfrac{1}{2\Upsilon}\zeta_{,\chi}^{2}\dot{\chi}^{2} + \dfrac{1}{2\Upsilon}\left(\dfrac{V_{,\chi}}{\zeta_{,\chi}}\right)^2 \nonumber \\
	- \bar{c}\left(\zeta_{,\chi}^2 \partial_{t} + \zeta_{,\chi} \zeta_{,\chi\chi}\dot{\chi} - \zeta_{,\chi\chi} \dfrac{V_{,\chi}}{\zeta_{,\chi}} + \cdot V_{,\chi\chi}\right)c
\end{eqnarray}
The regulator term becomes more complicated for this action and we do not reproduce it here, see \cite{Synatschke2009} for details of this. Following their approach one arrives at the LPA + WFR flow equations: 
\begin{eqnarray}\label{eq:dV/dktilde2}
	\partial_{{k}}V_{{k}}(\chi) &=& \dfrac{\Upsilon}{4}\cdot\dfrac{1}{{k}+ \partial^2_{\chi \chi}V_{{k}}(\chi)} \\
	\partial_{{k}}\zeta_{,\chi} &=& \dfrac{\Upsilon}{4}\cdot\dfrac{\mathcal{P}}{\zeta_{,\chi}\cdot\mathcal{D}^2 } \label{eq:dzetax/dktilde}\\
	\mathcal{D} &\equiv & V_{,\chi\chi} + k\,\zeta_{,\chi}^{2} \\
	\mathcal{P} &\equiv & \dfrac{4\zeta_{,\chi\chi} V_{,\chi\chi\chi}}{\mathcal{D}} - \left( \zeta_{,\chi\chi}\zeta_{\chi}\right)_{,\chi} - \dfrac{3\zeta_{,\chi}^{2}V_{,\chi\chi\chi}^{2}}{4\mathcal{D}^2}
\end{eqnarray}
which now consist of the previous LPA equation for the effective potential (\ref{eq:dV/dk}) as expected, augmented by one more flow equation for the wavefunction renormalisation $\zeta_{,\chi}$. 

As before we will integrate the LPA equation (\ref{eq:dV/dk}) by discretising along the $\chi$ direction and solving the resulting set of coupled ODEs in $k$. Once the effective potential $V_k(\chi)$ has been obtained the second PDE can be solved for $\zeta_{,\chi}$ in a similar way. It is worth pointing out here that our approach differs slightly from \cite{Synatschke2009} in that the effective potential obeys the same equation as in the LPA approximation even with the inclusion of WFR.\footnote{For the WFR approximation the authors of \cite{Synatschke2009} use a spectrally adjusted regulator which is evaluated on a background field $\bar{\phi}$. They make the simple choice of identifying this background field with the fluctuation field (i.e. $\bar{\phi} = \phi$). This approach however modifies the flow of $V_k$: equation (\ref{eq:dV/dktilde2}) differs from the LPA version (\ref{eq:dV/dk}) and the flow no longer correctly approaches the Boltzmann equilibrium distribution's effective potential. In the approach of \cite{Guilleux2017} which uses a simplified version of the WFR, this can be corrected by a further rescaling of $k$.} This is because the equilibrium state is described exactly by the LPA equation \cite{Guilleux2015,Guilleux2017, Moreau2020a}, as we mentioned above and explicitly recall in Appendix \ref{app:equil-flow}. The LPA flow equation was first solved in \cite{Synatschke2009, Guilleux2015, Guilleux2017}, while more recently WFR was included for a double well potential in \cite{Moreau2020a}.

\section{\label{sec:acc eom}The effective equations of motion}

A standard formulation of classical mechanics involves the principle of least action. If one considers the classical action $\mathcal{S}$:
\begin{equation}
	\mathcal{S} = \int dt~ \left( L(x,\dot{x}) -jx\right)
\end{equation}
where $L(x,\dot{x})$ is the Lagrangian and a source term has been added, one can obtain the equations of motion by requiring the variational derivative of $\mathcal{S}$ to be zero:
\begin{equation}
	\dfrac{\delta \mathcal{S}}{\delta x} = j \label{eq:delta S = 0}
\end{equation}
The Effective Action (EA) $\Gamma$ is so named because its definition makes it look like a classical action but includes the effect of fluctuations that have been integrated out. Defining 
\begin{equation}
	e^{\mathcal{W}[{\rm J}]} = \int\mathcal{D}\Phi ~ e^{-\mathcal{S}[\Phi]+\int dt  {\rm J}\Phi}
\end{equation}
the effective action $\Gamma[X]$ is then defined as
\begin{equation}
	\Gamma[X] = \int_t {\rm J} X - \mathcal{W}[{\rm J}]
\end{equation}
where
\begin{equation}
	X = \left\langle \Phi  \right\rangle \,.
\end{equation}
We then have
\begin{eqnarray}
	\dfrac{\delta \Gamma}{\delta X}= {\rm J}
\end{eqnarray}
Therefore $\Gamma$, the central object of the fRG, leads to effective equations of motion that incorporate the aggregate effects of the thermal fluctuations.\footnote{These equations are equivalent to those obtained by the 1-PI generating functional \cite{Vasiliev2004}.} 
  
\subsection{\label{sec:1-point function}The EEOM for the one point function}
In a similar way to how the classical action $\mathcal{S}(x)$ can yield the classical equations of motion through variational derivatives, so too does $\Gamma [\chi]$ yield the \textit{effective equation of motion} for the one point function (or average position) $\chi$:
\begin{equation}
	\dfrac{\delta\Gamma}{\delta \chi(t)} = 0 \label{eq:gen QEOM}
\end{equation}
Here we have assumed there are no external sources\footnote{Note that this is not the same as assuming that the noise term (\ref{eq:eta defn}) is zero as this is true for $\Gamma$ by definition} (J = 0).  
Under the LPA equation (\ref{eq:gen QEOM}) is:
\begin{equation}
	\dfrac{\delta\Gamma_{k = 0}}{\delta \chi(t)} = \ddot{\chi} -  \partial_\chi V_{k=0}(\chi)\,\partial_{\chi}^2V_{k=0}(\chi) = 0
\end{equation} 
where the final equality comes by assuming that source terms have been set to zero (i.e $J(t) = 0$). The WFR version of (\ref{eq:gen QEOM}) reads:
\begin{equation}
	\left(\zeta_{,\chi}\dot{\chi}\right)\dot{} - \frac{\partial_\chi V_{k=0}}{\zeta_{,\chi}^2}\left(\partial_{\chi\chi}^2V_{k=0}-\frac{\zeta_{,\chi\chi}}{\zeta_{,\chi}}\partial_\chi V_{k=0}\right) = 0
\end{equation} 
where $\partial_\chi\zeta$ and $\partial^2_\chi\zeta$ are also evaluated at $k=0$.
Both of these second order differential equations can actually be reduced to a first order differential equation like so: 
\begin{eqnarray}
	\dot{\chi} = -\tilde{V}_{,\chi}(\chi) \label{eq:QEOM}
\end{eqnarray} 
where we have introduced the \textit{effective dynamical potential} $\tilde{V}$ defined by 
\begin{eqnarray}
	\tilde{V}_{,\chi}(\chi) \equiv 
	\begin{cases}
		V_{,\chi}(k = 0, \chi), & \text{ for LPA}  \\[10pt]
		\dfrac{V_{,\chi}(k = 0, \chi)}{\zeta_{,\chi}^{2}(k = 0, \chi)}, & \text{ for WFR} 
	\end{cases}\label{eq: Vtilde}
\end{eqnarray}
Here we can clearly see that for LPA the \textit{effective} and \textit{effective dynamical} potentials are equivalent whereas WFR provides an additional factor for the latter.  

Equation (\ref{eq:QEOM}) tells us that the equation of motion for the average position $\chi$ is an extremely simple first order differential equation that appears like a Langevin equation with no noise. This means that once you have obtained the \textit{effective dynamical} potential you can compute the evolution of the average position $\chi$ trivially from any starting position.

At equilibrium the average position of the particle should not change, this means that $\dot{\chi} = 0$. It naturally follows from this condition and the EEOM for $\chi$ (\ref{eq:QEOM}) that equilibrium is defined for both LPA \& WFR by the condition 
\begin{eqnarray}
	\partial_\chi V_{k = 0}(\chi_{eq}) = 0 \label{eq:equilbirum position}
\end{eqnarray}  
As the potential $V_{k=0}(\chi)$ should be convex (by definition of $\Gamma$) equation (\ref{eq:equilbirum position}) tells us that $\chi_{eq}$ corresponds to the minimum of  $V_{k=0}(\chi)$. Or more concretely: 
\begin{equation}
	\lim\limits_{t \to \infty} \left\langle x(t) \right\rangle =  x \text{ that minimises } V_{k=0}(x)
\end{equation}
The equilibrium position is obviously the same for both LPA and WFR as they both lead to the same effective potential. As the equilibrium position is straightforwardly computed from the Boltzmann distribution verifying that the minimum of the effective potential matches the predicted equilibrium position is a good first test for the numerical solution of $V_{k=0}(x)$, at least close to its minimum.

\subsection{The EEOM for the two point function}
The connected 2-point  function $G(t,t')=\left\langle x(t)x(t')\right\rangle_{C}=\delta^2\mathcal{W}/\delta J(t)\delta J(t')$ and the second functional derivative of the effective action $\Gamma_{k=0}$ are inverse to each other
\begin{equation}
	\int d\tau~\dfrac{\delta^2\Gamma_{k=0}}{\delta \chi(t)\delta \chi(\tau)}\dfrac{\delta^2\mathcal{W}_{k=0}}{\delta J(\tau)\delta J(t')} = \delta(t-t')\label{eq:2ptW and gamma}
\end{equation} 
Concretely, this means that the connected 2-point function $G(t,t')$ satisfies the following equation:
\begin{eqnarray}
	\left(\dfrac{d^2}{dt^2} -  \mathcal{U}(\chi(t)) \right)G(t,t') &=& -2\Delta\delta(t-t')  \label{eq:2pointfuncdef} 
\end{eqnarray}
where $\mathcal{U}(\chi)$ is:
\begin{eqnarray}
	\mathcal{U}(\chi)= 
	\begin{cases}
		V_{,\chi\chi}^{2} + V_{,\chi}V_{,\chi\chi\chi}, & \text{for LPA}\\[10pt]
		\dfrac{V_{,\chi\chi}^{2}}{\zeta_{,\chi}^{4}}+ \dfrac{V_{,\chi}V_{,\chi\chi\chi}}{\zeta_{,\chi}^{4}}- \dfrac{V_{,\chi}^{2}\zeta_{,\chi\chi\chi}}{\zeta_{,\chi}^{5}}  \\
		-\dfrac{5V_{,\chi}V_{,\chi\chi}\zeta_{,\chi\chi}}{\zeta_{,\chi}^{5}}    + \dfrac{5V_{,\chi}^{2}\zeta_{,\chi\chi}^{2}}{\zeta_{,\chi}^{6}} , & \text{for WFR}
	\end{cases} \label{eq:U = }
\end{eqnarray}
and
\begin{eqnarray}
	\Delta\equiv
	\begin{cases}
		\dfrac{\Upsilon}{2}, & \text{for LPA}\\[10pt]
		\dfrac{\Upsilon}{2\zeta_{,\chi}^{2}}, & \text{for WFR}
	\end{cases}
	\label{eq:Deltadef}
\end{eqnarray}
The derivation of the full solution to (\ref{eq:2pointfuncdef}) can be found in Appendix~\ref{app:2ptfuncderiv} but here we just highlight the two main results:\\
The EEOM for the \textit{Variance} $t'\rightarrow t$:
\begin{eqnarray}
	\textbf{Var}(x) \equiv G(t,t) &=& \dfrac{\Upsilon}{2\lambda P(t)} \tilde{Y}_1(t)\tilde{Y}_2(t)  \nonumber \\
	&& +~ \dfrac{P(0)}{P(t)}\left[G_{00} - \dfrac{\Upsilon}{2\lambda P(0)}\right]\tilde{Y}_{2}^{2}(t)\nonumber \\
	\label{eq:QEOM Variance}
\end{eqnarray}
and the EEOM for the \textit{Covariance} $t'\rightarrow 0$, $t > 0$:
\begin{eqnarray}
	\textbf{Cov}(x(0)x(t)) \equiv G(t,0) = G_{00}\tilde{Y}_2(t) \label{eq:QEOM Covariance}
\end{eqnarray}
where $\tilde{Y}_i(t) \equiv Y_i(t)/Y_i(0)$ are the `normalised' solutions to the homogeneous equation (\ref{eq:appendhomo2ptfunc}) which can be obtained numerically. $P(t) = 1$ or $\zeta_{\chi}(\chi(t))$ for LPA and WFR respectively and $\lambda$ is defined below by (\ref{eq:lambdadef}). $G_{00} = G(0,0)$ is the initial variance at $t = 0$

If we take the equilibrium limit $\chi \rightarrow \chi_{eq}$ of the full EEOM for the 2-point function (\ref{eq:2pointfuncdef}) we find that it simplifies to:
\begin{eqnarray}
	\left(\dfrac{d^2}{dt^2} -  \lambda^2\right)G_{eq}(t_1,t_2) =  -2\Delta \vert\delta(t_2-t_1)  \label{eq:2pointfuncequi} 
\end{eqnarray}
where
\begin{eqnarray}\label{eq:lambdadef}
	\lambda^2 \equiv  
	\begin{cases}
		V_{,\chi\chi}^{2}\vert, & \text{for LPA}\\[10pt]
		\dfrac{V_{,\chi\chi}^{2}\vert}{\zeta_{,\chi}^{4}\vert}, & \text{for WFR}
	\end{cases} 
\end{eqnarray}
and $\Delta$ is defined as in (\ref{eq:Deltadef}). The notation $\vert$ means we have evaluated the function at $k = 0$ and at equilibrium $\chi = \chi_{eq}$.

The appropriate solution to (\ref{eq:2pointfuncequi}) providing the connected correlation function at equilibrium is  
\begin{eqnarray}
	G_{eq}(t_1,t_2) = \textbf{Cov}_{eq}(x(t_1)x(t_2))  &=& \dfrac{\Upsilon}{2V_{,\chi\chi}\vert} e^{-\lambda|t_1-t_2|}\nonumber \\ \label{eq:2-pointfunc} \\
	\Rightarrow G_{eq}(t,t) = \textbf{Var}_{eq}(x) &=& \dfrac{\Upsilon}{2V_{,\chi\chi}\vert}
	\label{eq:equal2pt1}
\end{eqnarray}
As the equilibrium variance is also easily computed from the Boltzmann distribution, equation (\ref{eq:equal2pt1}) gives us a second test to verify that the effective potential has been computed correctly, at least around the minimum. 

In the LPA approximation the variance and the decay rate of the autocorrelation function are both directly given by the curvature of the effective potential at its minimum. The inclusion of WFR however alters the decay rate without changing the equilibrium variance. This is as it should be since the latter is fixed by the equilibrium Boltzmann distribution. As we will see, WFR improves the decay rate which is indeed not exactly determined by the effective potential's curvature alone.

\section{\label{sec:Sol Flow}Solutions to the flow equations}
 
In this section we obtain the resulting effective potential and wave function renormalisation for 3 types of potential. The first we consider is a simple polynomial in line with \cite{Synatschke2009}:
\begin{eqnarray}
V(x) = 1 + x + \dfrac{x^2}{2} + \dfrac{2x^3}{3} + \dfrac{x^4}{4} \label{eq:poly defn}
\end{eqnarray}
We will also consider a doublewell made by two LJ potentials back to back:
\begin{eqnarray}
V(x) &=& 4\epsilon_1\left(\dfrac{\sigma^{12}}{(x+3)^{12}}-\dfrac{\sigma^6}{(x+3)^6} \right) \nonumber \\
&+& 4\epsilon_2\left(\dfrac{\sigma^{12}}{(x-3)^{12}}-\dfrac{\sigma^6}{(x-3)^6}\right)
\end{eqnarray} 
where $\sigma$ will be taken to be 1 from here on in and $\epsilon_1$ \& $\epsilon_2$ represents the depth of each well. We will choose $\epsilon_1 = 1,$ $\epsilon_2 = 10$ meaning the left well is 1 unit deep and the right well is 10 units deep (in $2D_0/\varepsilon$ units) and the potential is asymmetric. Clearly here the domain of interest is $x \in (-3,3)$ as the potential diverges at $x = \pm 3$. \\
Finally we also consider a ``bumpy'' bare potential consisting of a simple $x^2$ underlying potential with additional gaussian bumps (or dips):
\begin{eqnarray}
V(x) &=& x^2 + \sum_{i=1}^{n} \alpha_i \text{exp} \left[ -\dfrac{(x-\beta_i)^2}{\mu}\right]
\end{eqnarray}
where there are $n$ bumps or dips with the prefactor $\alpha_i$ being positive or negative respectively. We will focus on an $x^2$ plus 3 bumps and 3 dips in an asymmetrical setup. This potential could represent a rudimentary toy model for motion over a ``potential energy landscape'' with a series of local energy minima. This serves to clearly demonstrate the effect of local extrema on the final shape of the effective potential since the underlying $x^2$ potential does not alter its shape under the RG flow. We chose the parameters to be for $x^2$ + 6 bumps/dips:  
\begin{eqnarray}
	&& \alpha_1 = \alpha_4 = \alpha_5 = -1.5  \nonumber \\
	&& \alpha_2 = \alpha_3 = \alpha_6 =1.5, \mu = 0.06 \nonumber \\
	&& \beta_1 = -\beta_2 = 0.7, \beta_3 = -\beta_4 = 1.4, \beta_5 = -\beta_6 = 2.1  \nonumber \\
\end{eqnarray}
expressed in dimensionless units (we have set $ \hat{x}\rightarrow x$ to avoid notational clutter). 

\begin{figure}[t!]
	\centering
		\includegraphics[width=0.4\textwidth]{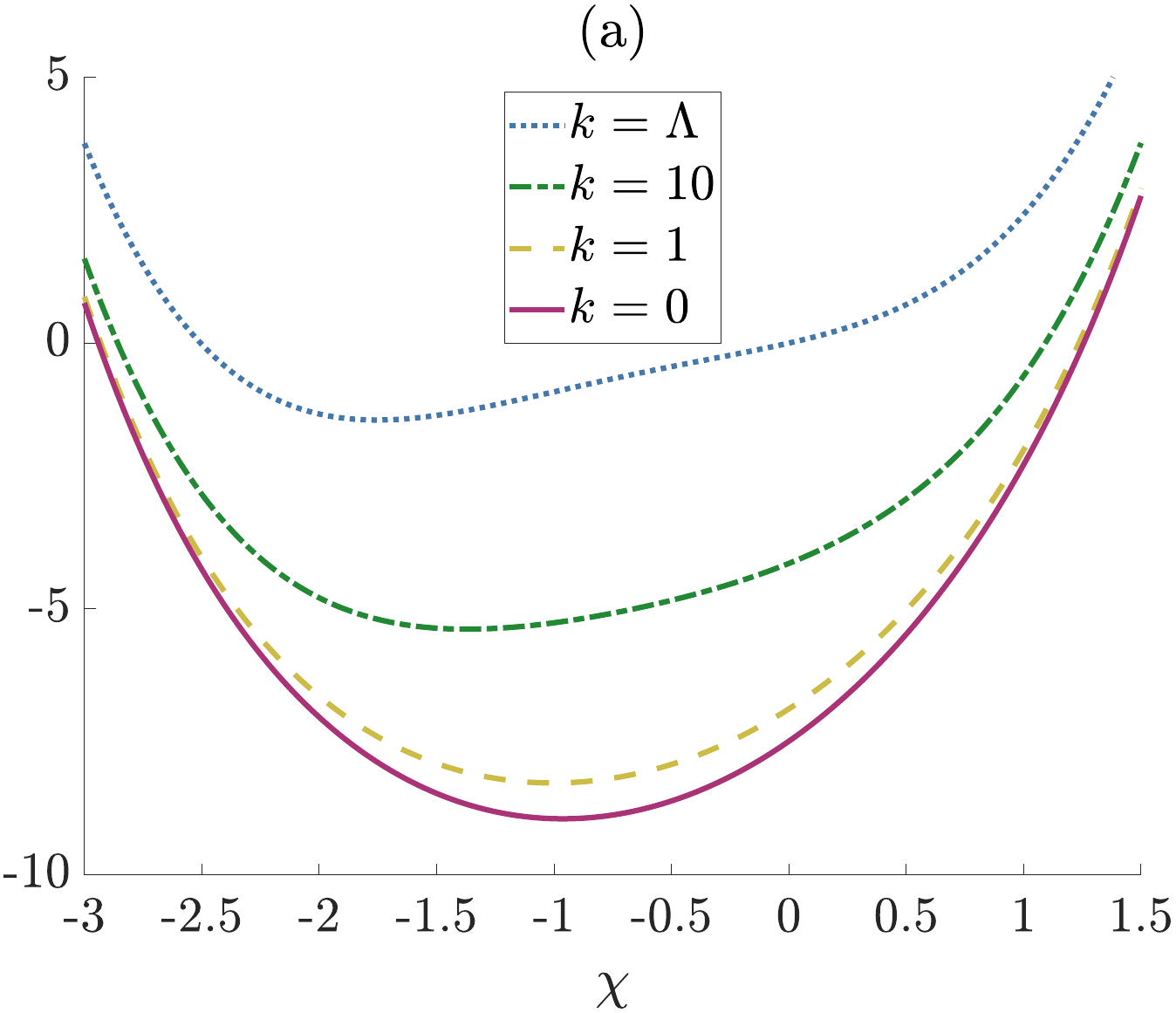}
		\includegraphics[width=0.4\textwidth]{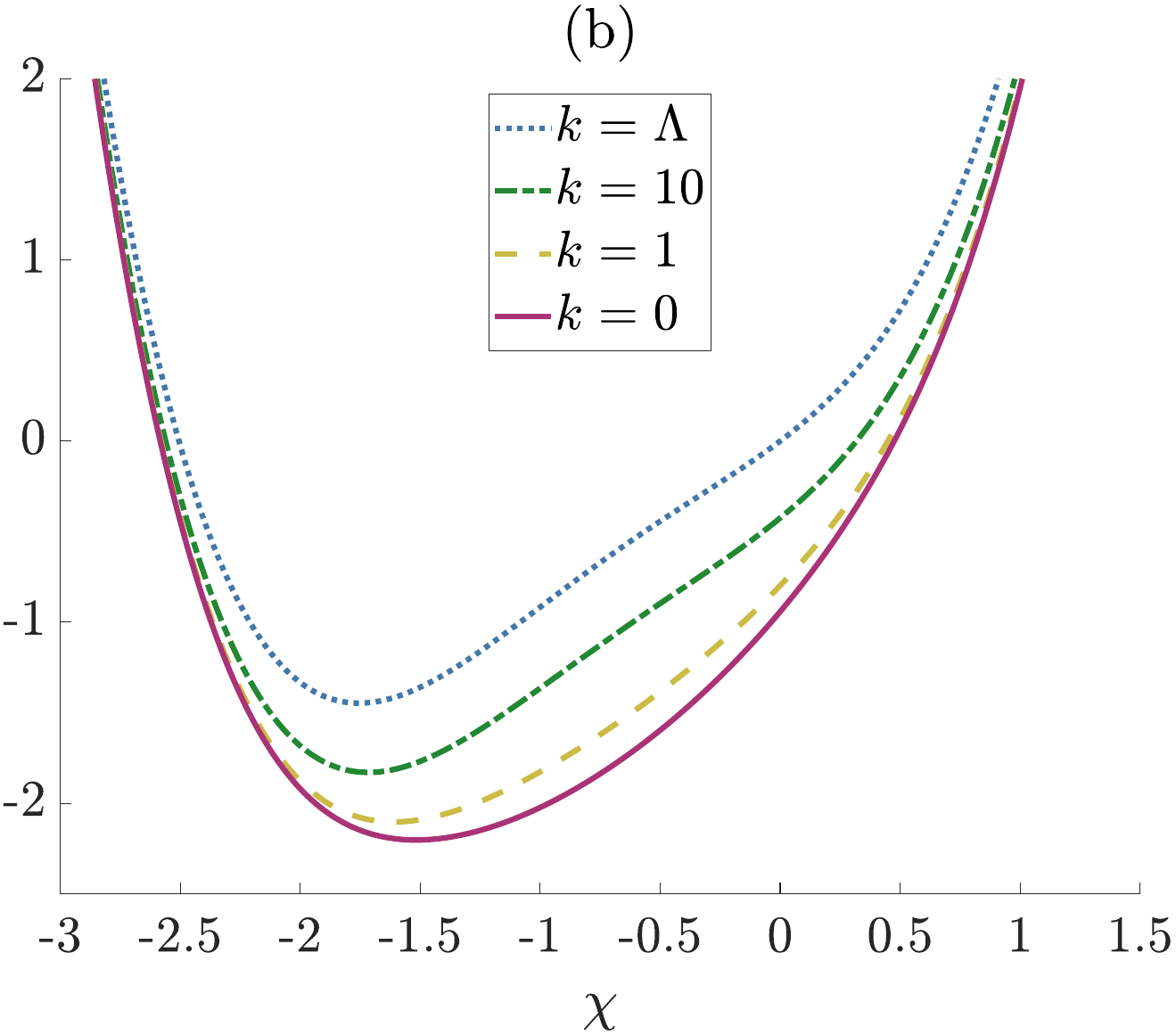}
		\caption{The flow of the polynomial Langevin potential V in the LPA for (a) $\Upsilon = 10$ (High temperature/strong fluctuations) and (b) $\Upsilon = 1$ (Low temperature/Weak fluctuations). The dotted blue curve indicates the bare potential which is progressively changed, through dot-dashed green and dashed yellow, into the solid red, effective potential, as fluctuations are integrated out.}\label{fig: Gies_LPA_V_5T.pdf}
\end{figure}
%
\begin{figure}[t!]
	\centering
		\includegraphics[width=0.4\textwidth]{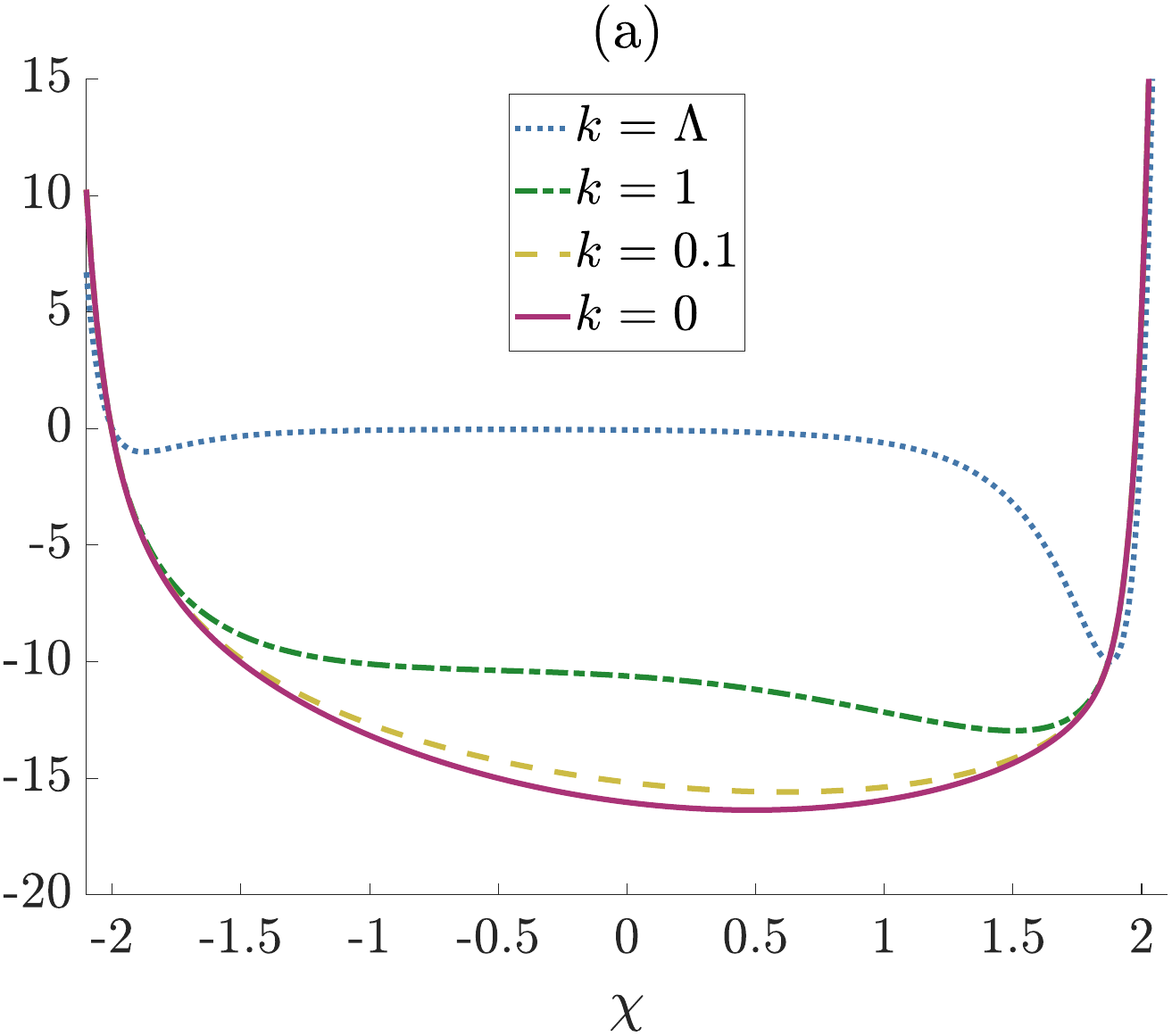}
		\includegraphics[width=0.4\textwidth]{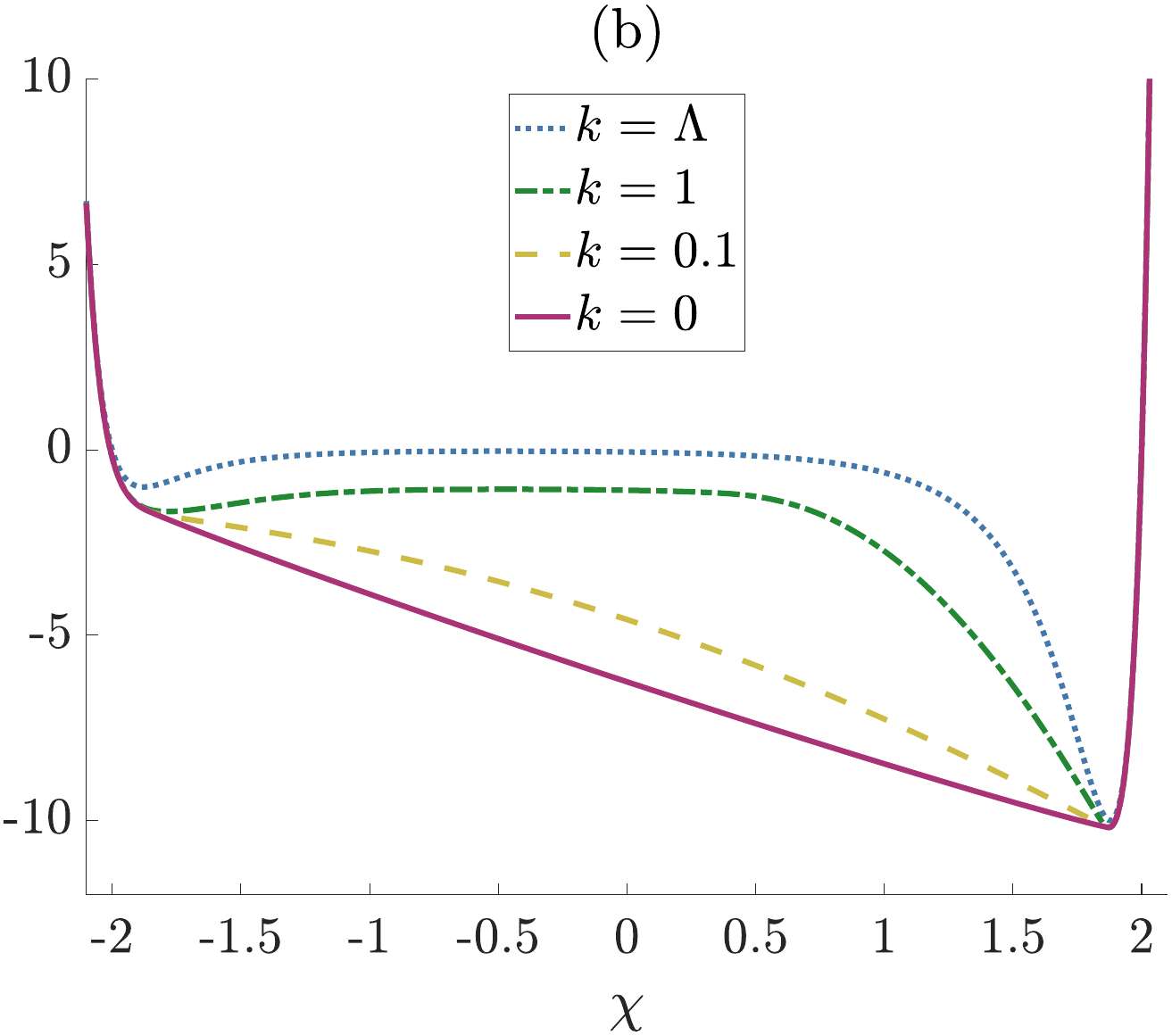}
		\caption{The flow of the unequal L-J Langevin potential V in the LPA for (a) $\Upsilon$ = 10 and (b) $\Upsilon$ = 1. Again, the bare potential is denoted by the dotted blue curve and the $k=0$ effective potential by the solid red one.}
		\label{fig: ULJ_LPA_V_5T.pdf}
\end{figure}
\begin{figure}[t!]
 	\centering
 	\includegraphics[width=0.4\textwidth]{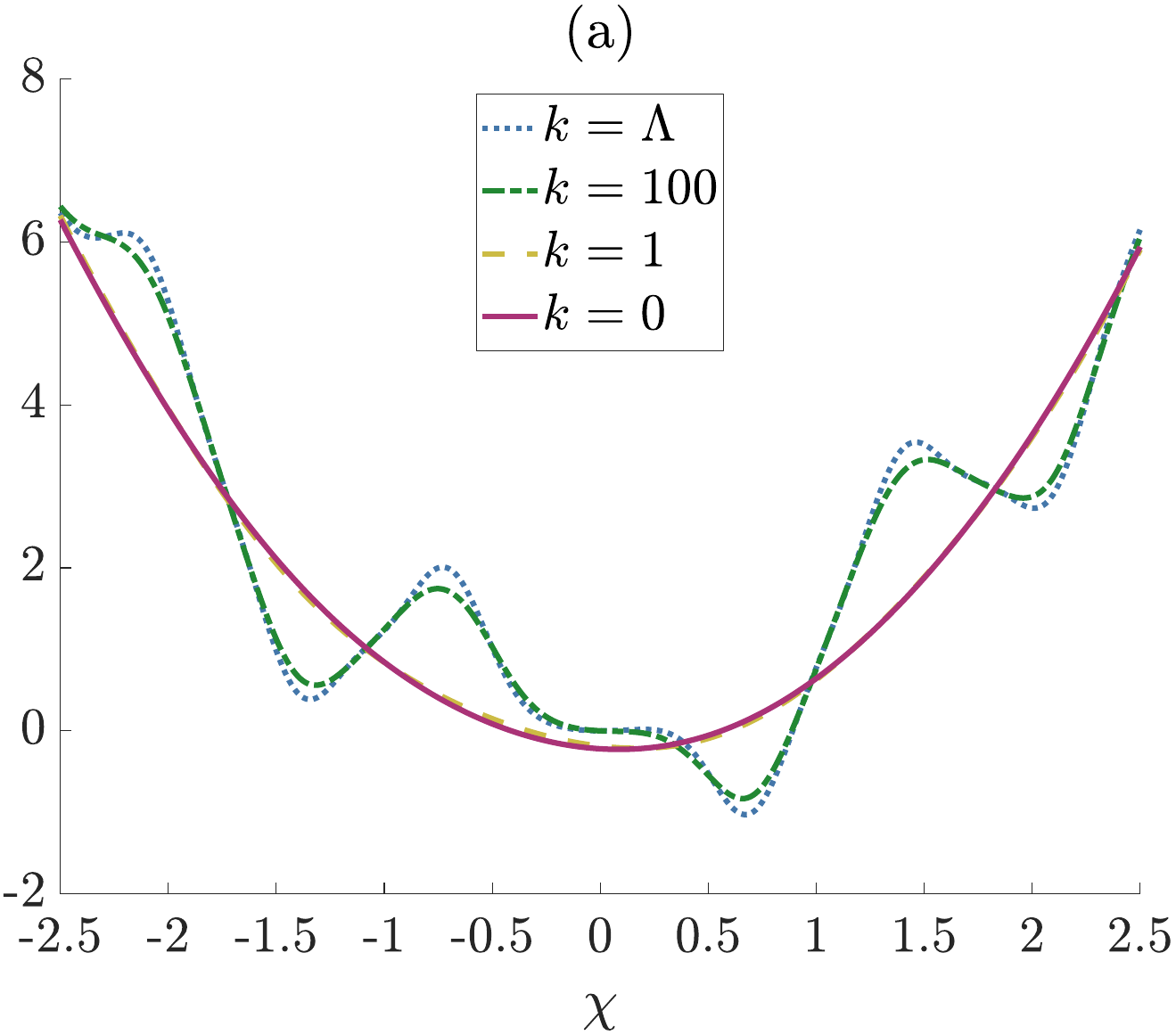}
 	\includegraphics[width=0.4\textwidth]{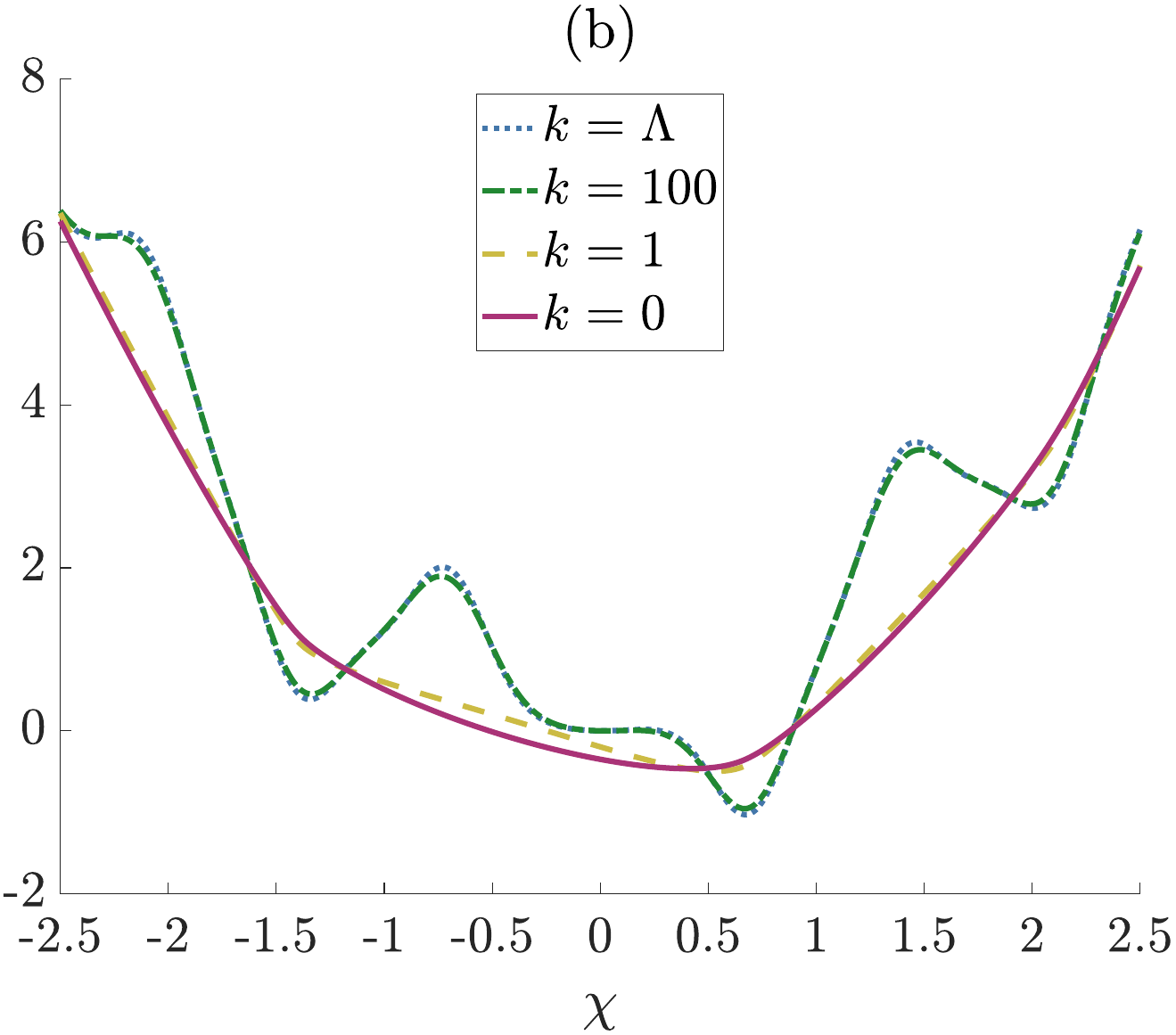}
 	\caption{The flow of the $x^2$ potential with 3 additional gaussian bumps and 3 dips in the LPA for (a) $\Upsilon$ = 3 and (b) $\Upsilon$ = 1. As before, the bare potential is denoted by the dotted blue curve and the $k=0$ effective potential by the solid red one.}
 	\label{fig: X2GaussMany_LPA_V_15T.pdf}
 \end{figure}
 \begin{figure}[t!]
 	\centering
 	\includegraphics[width=0.4\textwidth]{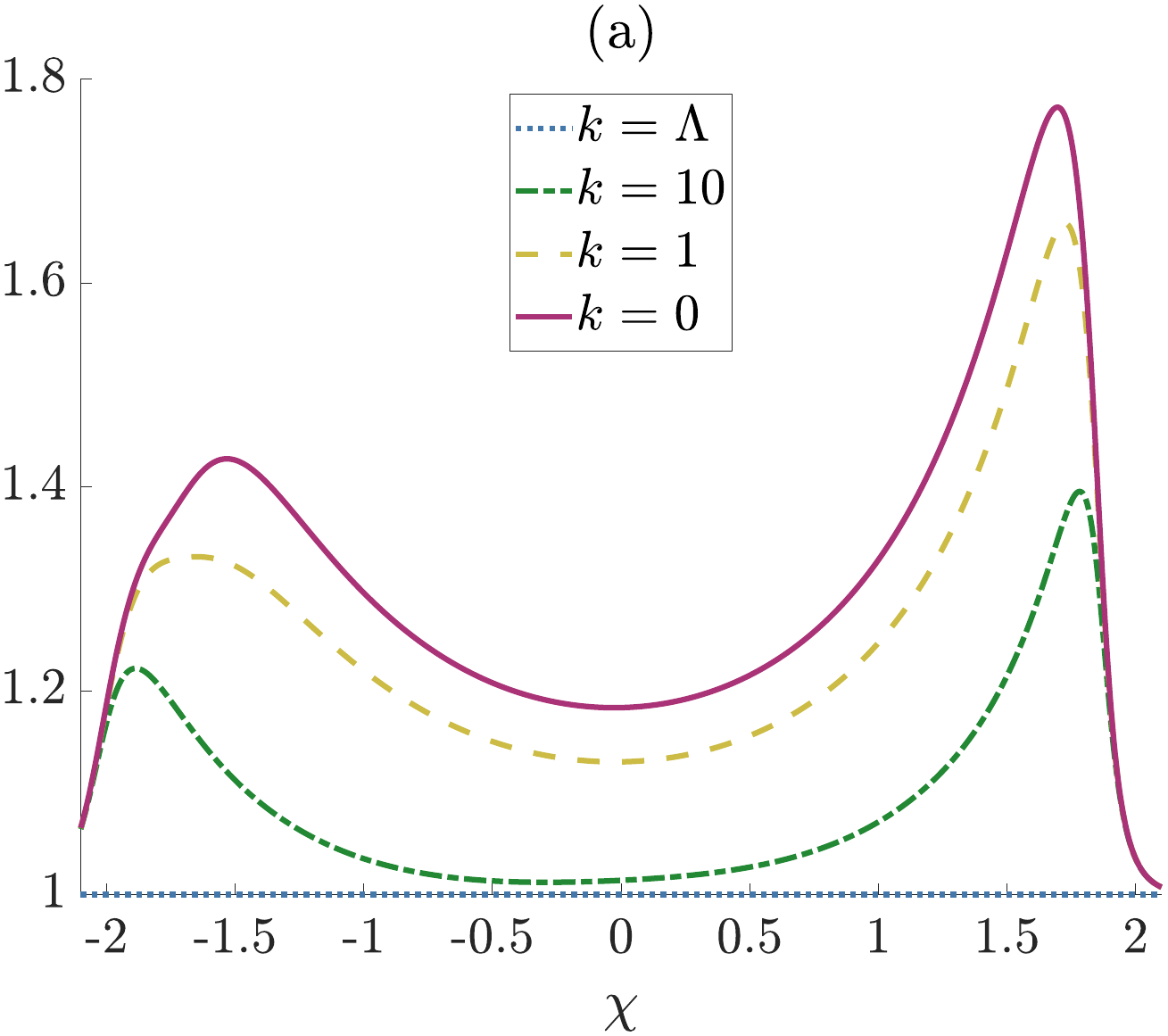}
 	\includegraphics[width=0.4\textwidth]{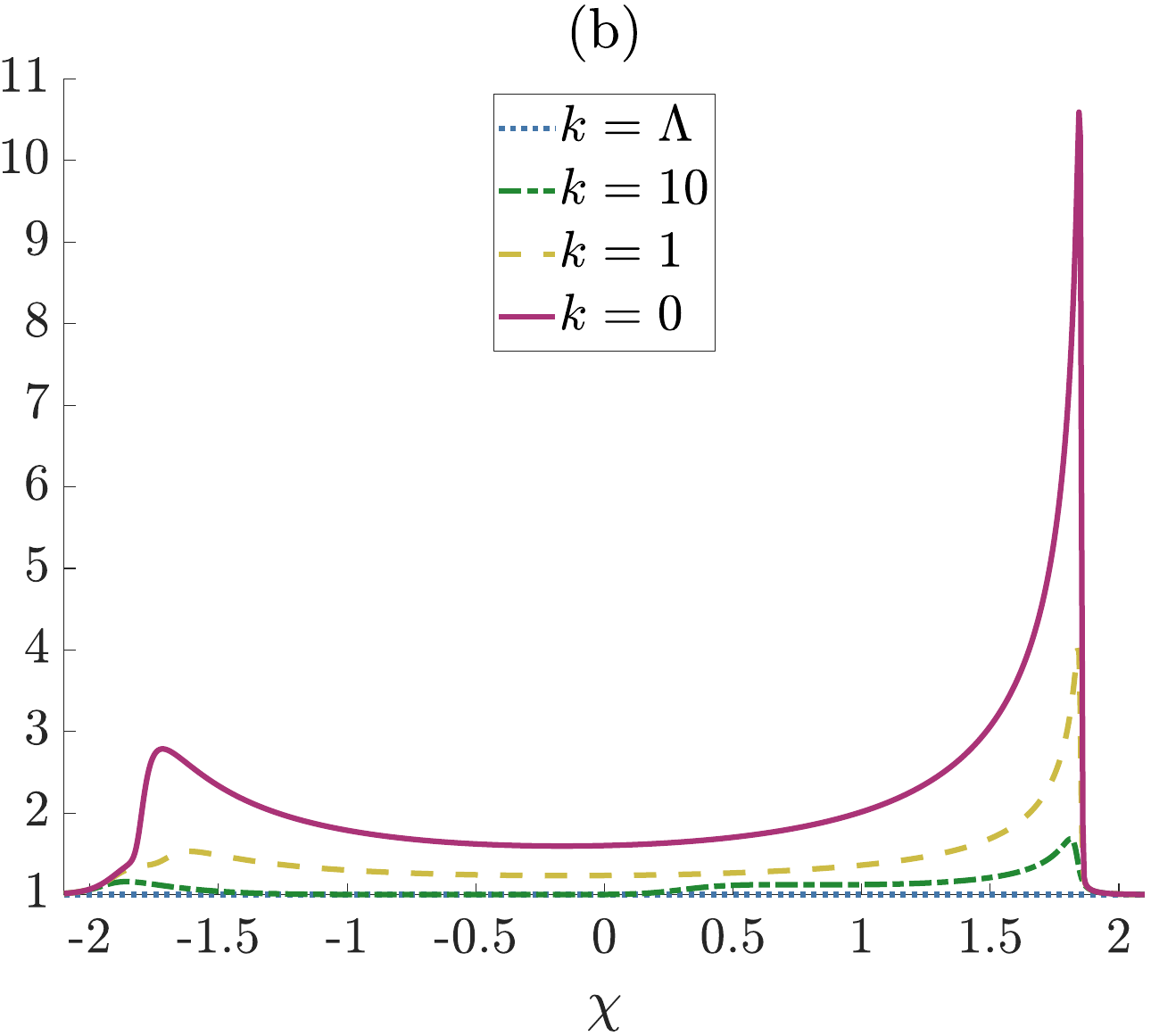}
 	\caption{The flow of $\zeta_{x}$ for of the unequal L-J Langevin potential for (a) $\Upsilon$ = 10 and (b) $\Upsilon$ = 2.}
 	\label{fig: ULJ_WFR_Zetax_5T.pdf}
 \end{figure}
 \begin{figure}[t!]
 	\centering
 	\includegraphics[width=0.4\textwidth]{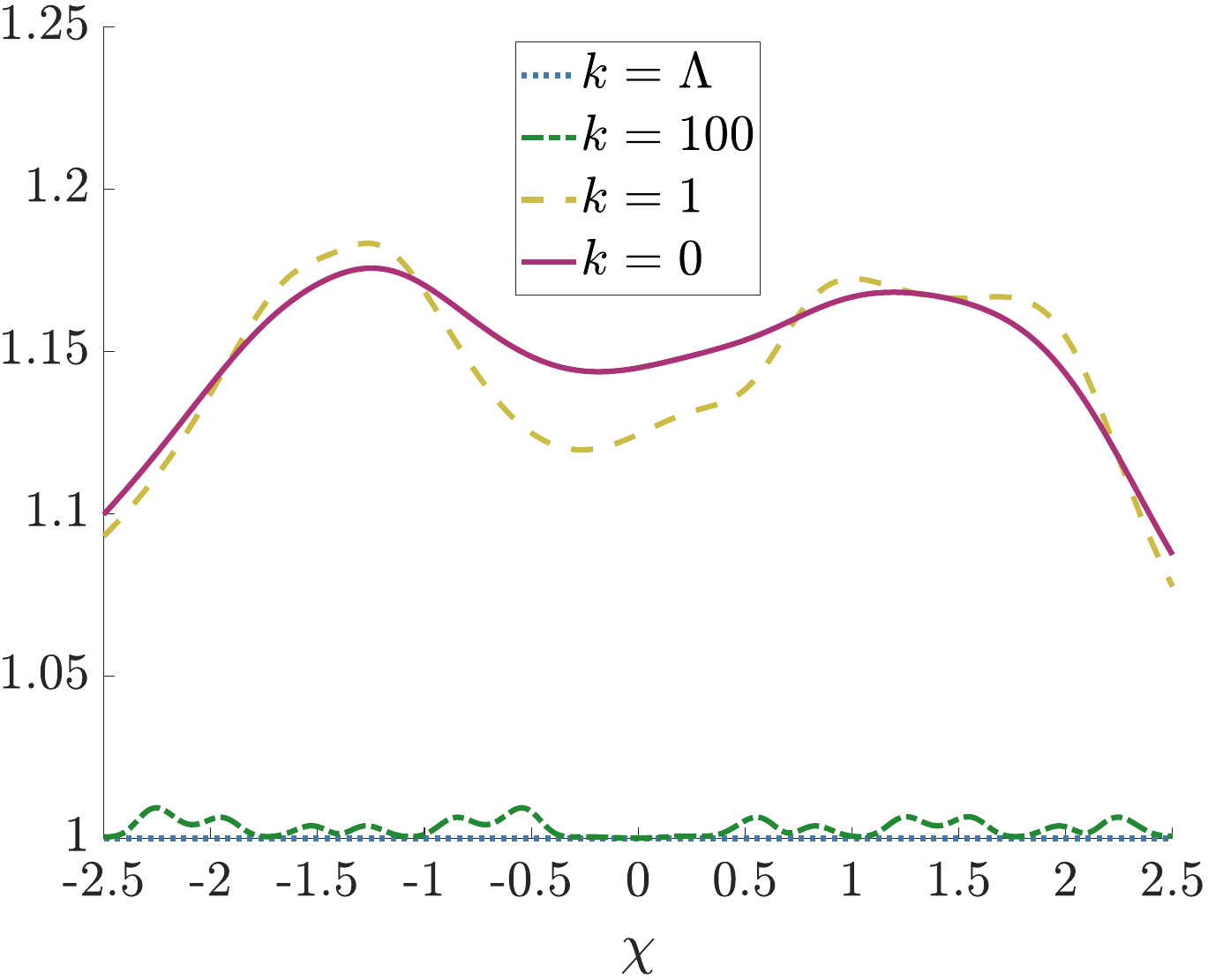}

 	\caption{The flow of $\zeta_{x}$ for an $x^2$ potential with 6 gaussian bumps/dips for $\Upsilon = 3$. }
 	\label{fig: X2GaussMany_WFR_Zetax_15T.pdf}
 \end{figure}
 
 We solve the LPA flow equation (\ref{eq:dV/dk}) on a grid in the $\chi$ direction 
 and using an adaptive step size Runge-Kutta ODE solver to evolve in the $k$ direction. A similar approach was used for including (\ref{eq:dzetax/dktilde}). The numerical derivatives in the $\chi$ direction were based on a finite difference scheme using the Fornberg method with a stencil size of 5 for the potentials under study. While increasing the grid size improves the accuracy of the numerical derivative it also increases the number of coupled ODEs to be solved, making the integration much more computationally expensive. A balance must be drawn depending on the potential in question. For our cases we considered 1001 points and $x \in (-3,3)$ for the unequal L-J potential and $x \in (-5,5)$ for the other two. Figs. \ref{fig: Gies_LPA_V_5T.pdf},~\ref{fig: ULJ_LPA_V_5T.pdf},~\ref{fig: ULJ_WFR_Zetax_5T.pdf},~\ref{fig:  X2GaussMany_LPA_V_15T.pdf} and \ref{fig: X2GaussMany_WFR_Zetax_15T.pdf}, display the results.

Figs.~\ref{fig: Gies_LPA_V_5T.pdf} and \ref{fig: ULJ_LPA_V_5T.pdf} show the flow from the bare to the effective potential for a high, $\Upsilon = 10$, and a low, $\Upsilon = 1$, temperature for the polynomial and unequal L-J potentials respectively. As $k\rightarrow 0$ is approached, a distinct single minimum develops, indicating the average position of the particle. As expected, the lower the temperature, the closest the effective minimum is to the bare potential's global minimum, reflecting the relative weakness of fluctuations to force the particle to spend time away from it. As one might expect, it takes `longer' in $k$ evolution for local features -- e.g. barriers -- to disappear in the $\Upsilon = 1$ case as fluctuations at each $k$ scale have less energy than their equivalent for the $\Upsilon = 10$ case and the particle's stochastic motion between barriers is less frequent. Physically this means that the fluctuations we have integrated out up to scale $k$ do not contribute significantly to the particle moving between minima. 

We see a similar phenomenon in Fig.~\ref{fig: X2GaussMany_LPA_V_15T.pdf} for the flow from the bare to the effective potential for $\Upsilon = 3$, and $\Upsilon = 1$ for an $x^2$ potential with 6 gaussian bumps/dips. Here the original Langevin potential is much more complicated than in the previous two cases but the fRG is still able to smoothen out these features in a non-trivial way. This example further demonstrates how the flow of the effective potential is driven by the local curvature, the gaussian features imposed here, since for an $x^2$ potential the fRG flow equation (\ref{eq:dV/dk}) yields no change beyond an unphysical shift by an overall additive constant. 

It is clear in all three figures that at the high temperature there is not much change in the shape of the potential when $k$ has reached the value given by the green, dot-dashed line. Physically this means that the fluctuations integrated out in this range do not contribute significantly to the particle evolution and transition between minima. However, by the time $k$ has been lowered to the value of the yellow, dashed line we have started to integrate over fluctuations over timescales relevant for inter-minima transitions. Naturally, when $k = 0$ is reached the potential is fully convex (as it must be by definition of $\Gamma$) with no local features to overcome. Similar behaviour is obtained where again we consider the corresponding lower temperature. As one might expect it takes `longer' in $k$ evolution for the barrier to disappear as fluctuations at each $k$ scale have less energy than their equivalent for the high temperature case. Of note is that not only is the evolution different but the final shape of $V_{k=0}(x)$ is different for the two different temperature regimes. Both the position and gradient near the global minimum are changed. This is suggestive of longer time scales required at lower temperatures to reach equilibrium. It also indicates longer times for the equilibrium covariance to decay, as we discuss below. 

Regarding WFR, the full numerical solutions to (\ref{eq:dzetax/dktilde}) for the unequal L-J potential at $\Upsilon = 10,$ $2$ and for an $x^2$ potential with 6 gaussian bumps/dips at $\Upsilon = 3$, are shown in Figs.~\ref{fig: ULJ_WFR_Zetax_5T.pdf} and \ref{fig: X2GaussMany_WFR_Zetax_15T.pdf} respectively. We see that from the initial condition, $\zeta_x = 1$ everywhere, features appear as $k \rightarrow 0$ in direct contrast with the evolution of the effective potential. At higher temperatures it is clear that at $k = 0$ a local minima appears at the same place as the global minimum for the effective potential where the height of the local minima is linked to the equilibrium covariance -- see equation (\ref{eq:lambdadef}). Looking at Fig.~\ref{fig: ULJ_WFR_Zetax_5T.pdf} (b) however we can see that this is no longer the case and the features generated by the fRG flow are much greater than in the high temperature case.  We will see later that the neighbouring features in $\zeta_x$ will help to better describe dynamical evolution than the bare potential alone. 

\section{\label{sec:Equilibrium Res} Equilibrium}
As mentioned above and recalled in Appendix \ref{app:equil-flow}, the LPA flow equation (\ref{eq:dV/dk}) exactly corresponds to the effective potential of the equilibrium Boltzmann distribution  
\begin{equation}
	P(x) = N\text{exp}\left(-\dfrac{2 V(x)}{\Upsilon}\right)
\end{equation}
We have verified that both the equilibrium position, given by the minimum of the effective potential, 
\begin{eqnarray}
	\partial_\chi V_{k = 0}(\chi_{eq}) &=& 0 \label{eq:equilbirum position}
\end{eqnarray}
and the equilibrium variance, defined from the effective potential's curvature through
\begin{eqnarray}
	\textbf{Var}_{eq}(x) &=& \dfrac{\Upsilon}{2V_{,\chi\chi}\vert}
	\label{eq:equal2pt}
\end{eqnarray}
are reproduced to sub-percent accuracy, indicating the accuracy of the numerical solution to the LPA flow equation, at least around the minimum of the effective potential. 

 In addition to the static variance at equilibrium, the curvature of the effective potential around the minimum also determines the time dependence of correlations in equilibrium, quantified by the time dependent covariance or connected 2-point function
\begin{equation}
\textbf{Cov}_{eq}(x(t_1)x(t_2))  = \dfrac{\Upsilon}{2V_{,\chi\chi}\vert} e^{-\lambda|t_1-t_2|} \,. \label{eq:2-pointfunc} 
\end{equation}
Here, $\lambda$  corresponds to $V_{,\chi\chi}\vert$ within the LPA but the solution to the WFR flow equation (\ref{eq:dzetax/dktilde}) for $\zeta_{,\chi}$ also contributes, providing a correction to $\lambda$ according to (\ref{eq:lambdadef}).

In Table.~\ref{tabel:effective mass for different potentials and methods} we collect the values of $\lambda$ obtained using the fRG under LPA \& WFR for different  $\Upsilon $ values, and compare this directly to high accuracy numerical simulations of the Langevin equation (\ref{eq:langevin}). We can clearly see from Table. \ref{tabel:effective mass for different potentials and methods} that the LPA can have good agreement with the simulation value for simple potentials at high temperature but can deviate drastically as temperature is lowered. Inclusion of the WFR factor $\zeta_x$ reduces the deviation error from the value obtained in the simulations substantially to $\sim 1 \%$ for the simplest cases and order of magnitude agreement for the most complicated, low temperature systems.
\begin{figure}[t!]
	\centering
		\includegraphics[width=0.4\textwidth]{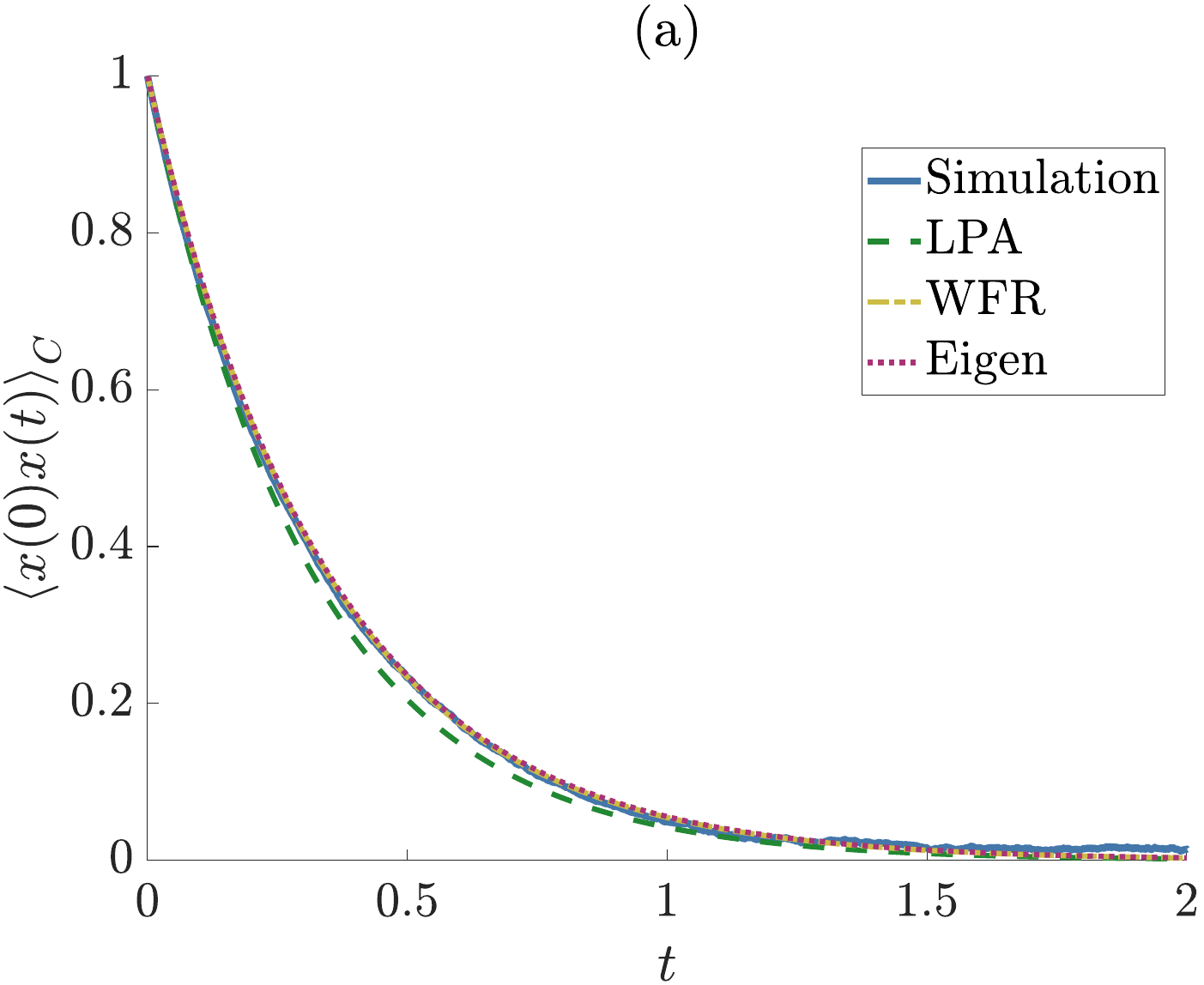}
		
		\includegraphics[width=0.4\textwidth]{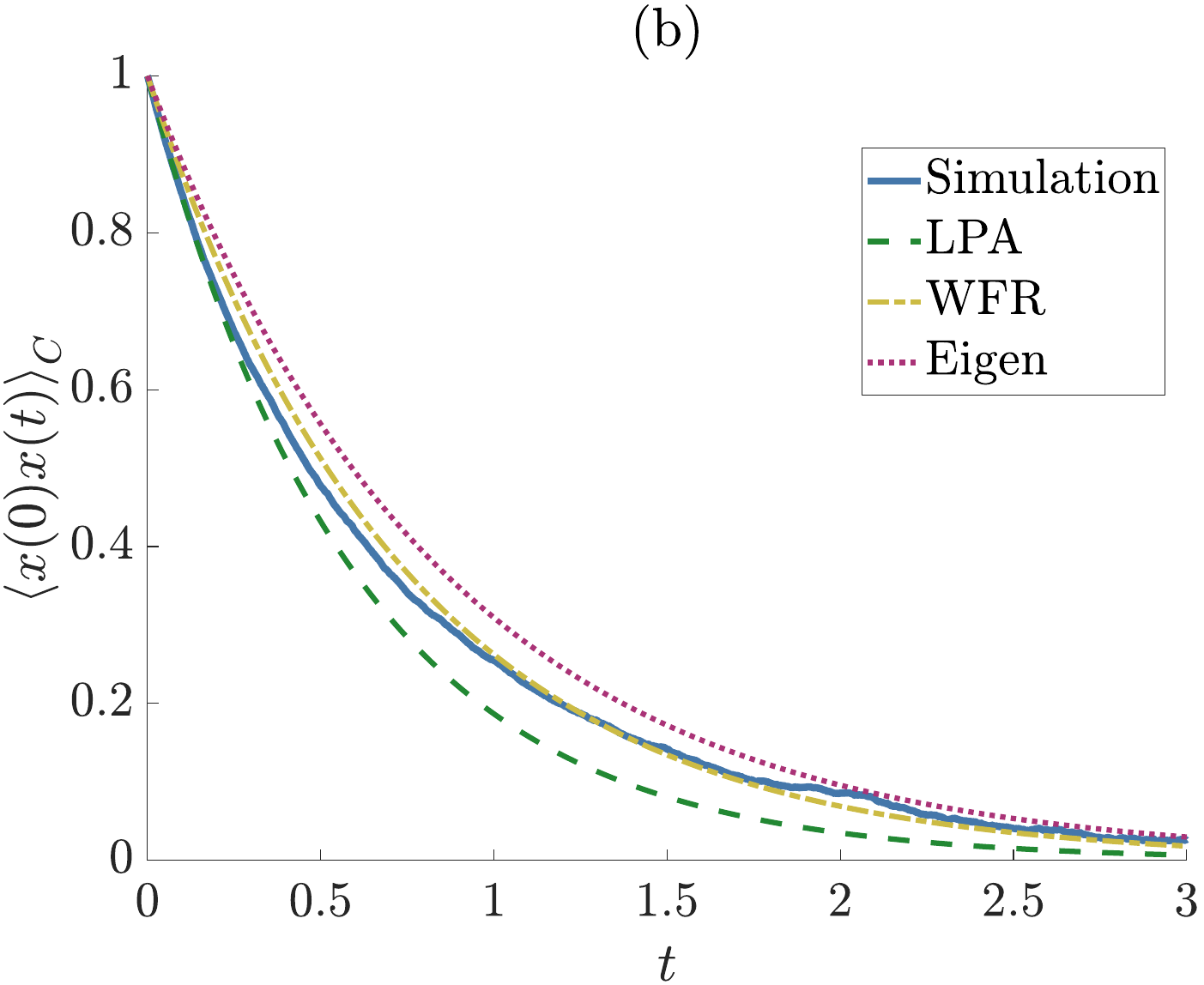}
		
		\caption{The decay of the (normalised) covariance $\left\langle x(0)x(t)\right \rangle_C$ at equilibrium in a polynomial potential for (a) $\Upsilon = 10$ and (b) $\Upsilon = 1$.}
		\label{fig:2pt_eq_Poly_5T.pdf}
\end{figure}

\begin{figure}[t!]
	\centering
		\includegraphics[width=0.4\textwidth]{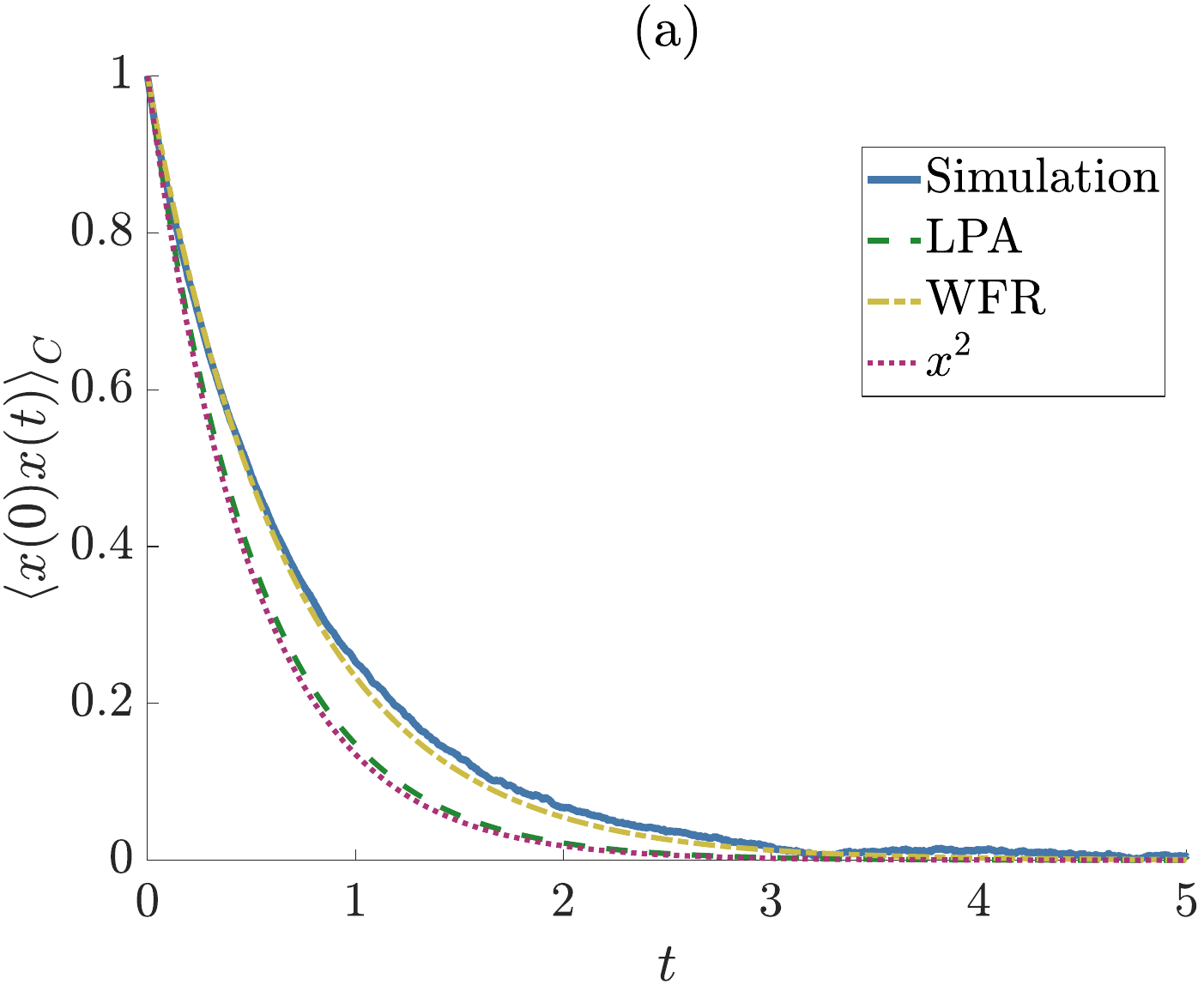}
		
		\includegraphics[width=0.4\textwidth]{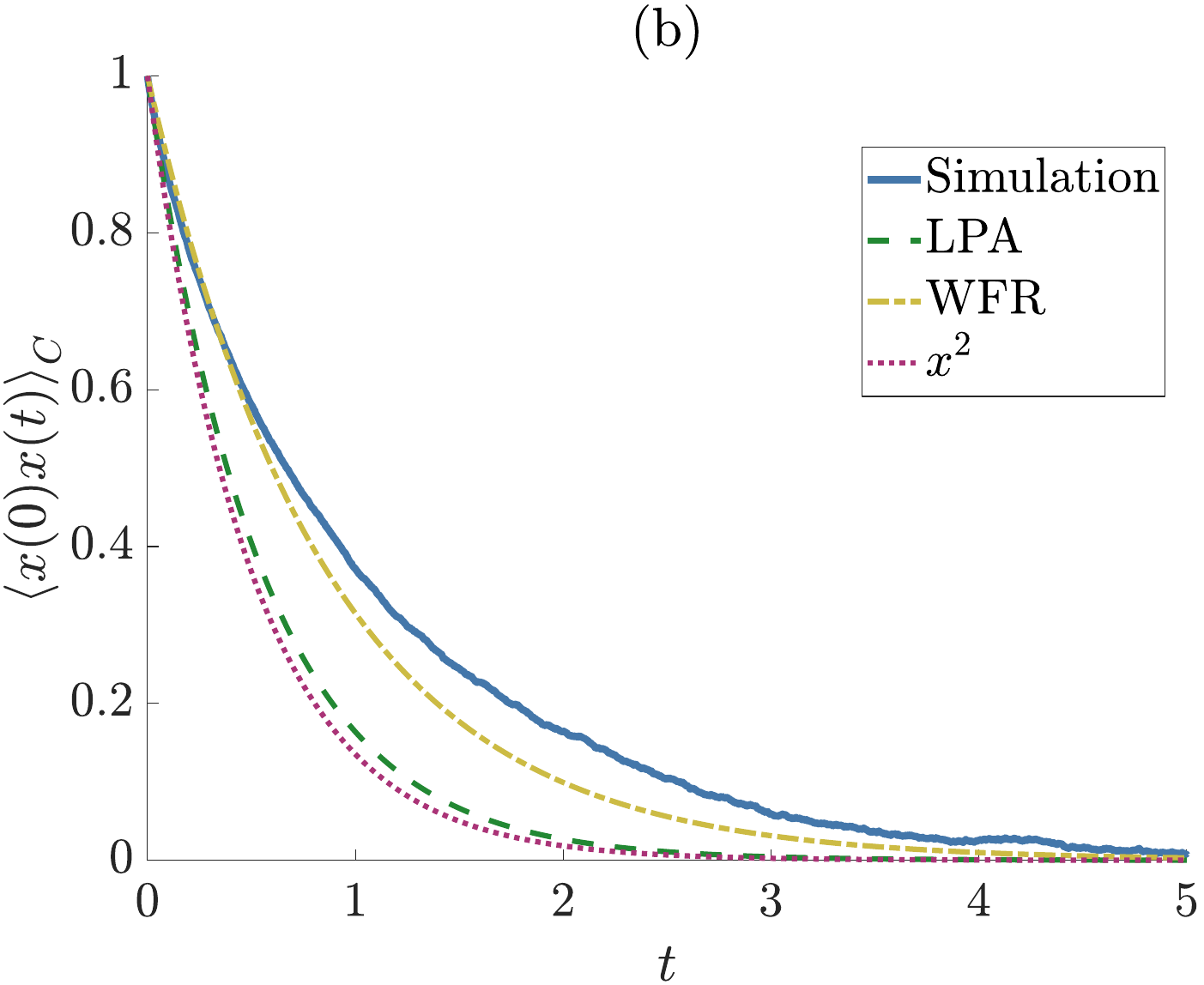}
		
		\caption{The decay of the (normalised) covariance $\left\langle x(0)x(t)\right \rangle_C$ at equilibrium in an $x^2$ plus 6 gaussian bumps/dips potential for (a) $\Upsilon = 3$ and (b) $\Upsilon = 2$.}
		\label{fig:2pt_eq_X2GaussMany_15T.pdf}
\end{figure}
\begin{table}[t!]
	\centering
		\begin{tabular}{l l l l l}
			\toprule
			Potentials & $\Upsilon$ & LPA & WFR & Sim \\
			\midrule
			Poly & 10 & 3.1857 & 2.9101 & 2.9191 \\
			& 4 & 2.0842 & 1.8664 & 1.8767   \\
			& 1 & 1.7112 & 1.3381 & 1.3585 \\
			Unequal L-J & 10 & 2.9655 & 2.0146 & 1.8783 \\
			& 2 & 50.93 & 0.4806 & 0.3691 \\
			$x^2$ + 6 b/d & 3 & 1.9199 & 1.4529 & 1.3977\\
			& 2& 1.8185 & 1.1552 & 0.9710 \\
			
			\bottomrule 
		\end{tabular}
	 \captionof{table}{Value of the autocorrelation decay rate at different temperatures. The LPA \& WFR columns display $\lambda $ as calculated from the fRG flow. The simulation values were generated by averaging over 50,000 runs.
	 }
	  	 \label{tabel:effective mass for different potentials and methods}
\end{table}

The decay of the equilibrium covariance is shown in Figs.~\ref{fig:2pt_eq_Poly_5T.pdf} \& ~\ref{fig:2pt_eq_X2GaussMany_15T.pdf} for the polynomial and $x^2$ plus 6 bumps/dips potentials respectively. We can see -- for the polynomial potential -- in Fig.~\ref{fig:2pt_eq_Poly_5T.pdf} (a), $\Upsilon = 10$, that the LPA and WFR are both in good agreement with simulations and in Fig.~\ref{fig:2pt_eq_Poly_5T.pdf} (b), $\Upsilon = 1$, that the WFR offers better agreement than the LPA. In Fig.~\ref{fig:2pt_eq_X2GaussMany_15T.pdf} (a) we can see that for $\Upsilon = 3$ the WFR prediction closely matches the simulations offering significant improvement over the LPA which closely matches the bare $x^2$ potential.  This indicates that even for highly non-trivial systems -- where the computation of eigenvalues for these potentials is a non-trivial exercise -- that the simulated decay is vastly different from the bare $x^2$ potential -- see Table. \ref{tabel:effective mass for different potentials and methods} and compare to the $x^2$ prediction for $\lambda /2$ which is 1 -- the fRG can appropriately capture these effects.
 
It is also worth pointing out that the simulated decay rate does not follow a pure exponential at all times in all cases. This can be best seen in the top plot of Fig.~\ref{fig:2pt_eq_Poly_5T.pdf} (b) where the decay is initially close to the LPA, then the WFR decay before moving towards the eigen decay rate at late times. This sort of behaviour has been identified in similar systems in the early universe \cite{Markkanen2020} where it was noticed that the smallest non-zero eigenvalue's spectral coefficient was sufficiently small that higher order eigenvalues would dominate the decay at earlier times. As LPA matches the decay rate predicted by the Boltzmann distribution and WFR is closer to the decay predicted by $E_1$. We discuss this further below.

\section{\label{sec:Accelerated} Relaxation towards Equilibrium}

In order to solve the equations of motion for the one point function $\chi (t)$ and two point function $G(t,t')$ we must first solve the PDEs for the LPA \& WFR to obtain the \textit{dynamical effective potential} $\tilde{V}$ and the function $\mathcal{U}$. We will use the solutions obtained in Section \ref{sec:Sol Flow} in order to compute these parameters and then solve the appropriate Effective Equation of Motion (EEOM).  
\subsection{The dynamical effective potentials}
\begin{figure}[t!]
\centering
\includegraphics[width=0.45\textwidth]{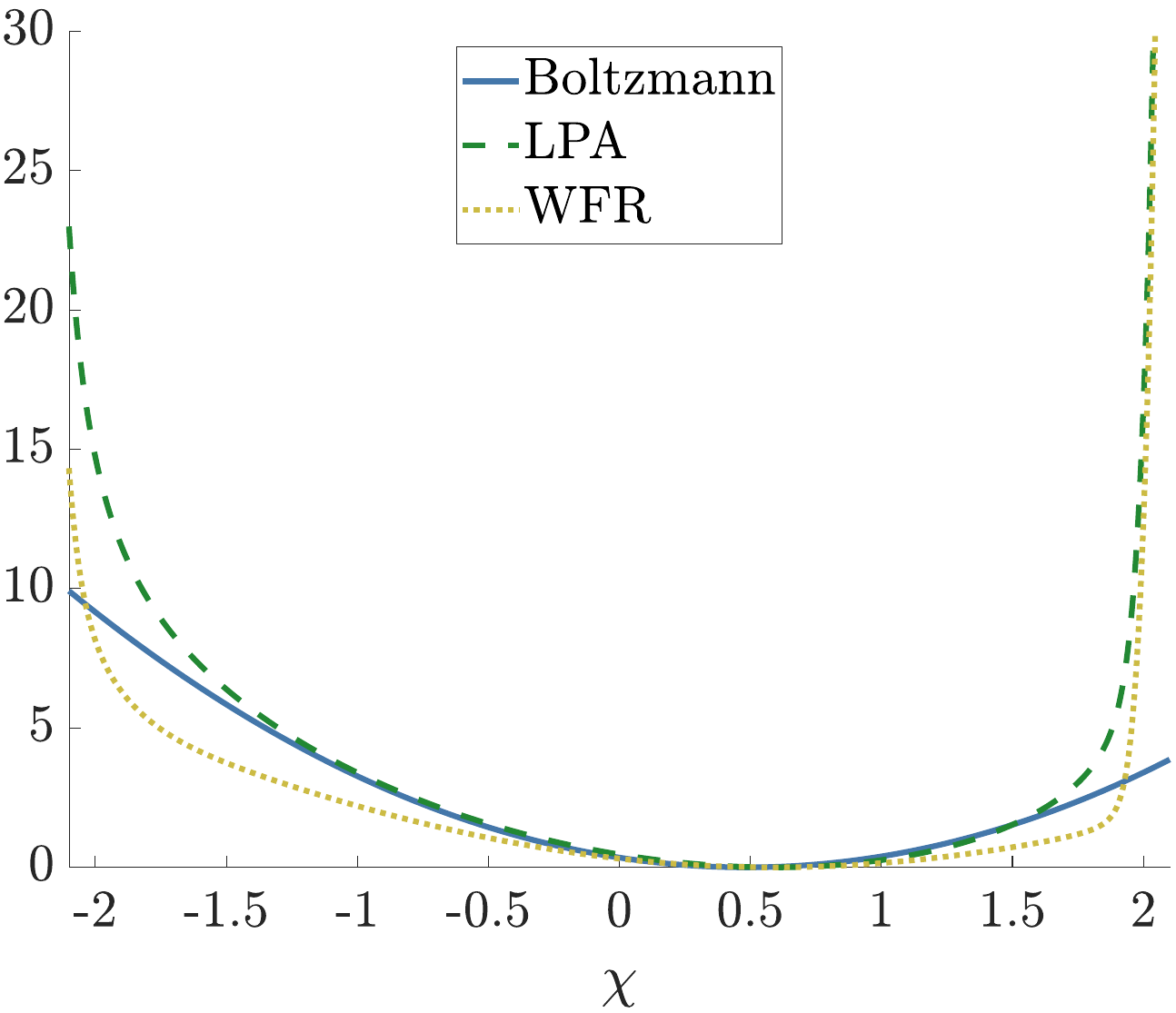}
\caption{Comparison of $\tilde{V}$ for unequal LJ potential at $\Upsilon = 10$ as calculated using fRG methods LPA and WFR compared to the Boltzmann ``near-equilibrium" approximation given by equation (\ref{eq: BoltztildeV}). All potentials have been vertically shifted so that their minima (corresponding to the equilibrium position) coincide. }\label{fig: ULJ_LPAvsBoltz_V_5T}
\end{figure}
In section \ref{sec:acc eom} we introduced the notion of the \textit{dynamical effective potential} $\tilde{V}$ given by equation (\ref{eq: Vtilde}) which together with (\ref{eq:QEOM}) describes the evolution of the average position, $\chi$. As the fRG guarantees that the fully effective potential V will be convex this implies that the dynamical effective potential $\tilde{V}$ will also be either fully or extremely close to fully convex for LPA and WFR respectively thus greatly simplifying dynamical calculations. In the previous section we emphasised how the fRG LPA effective potential gives us the Boltzmann equilibrium quantities such as equilibrium position and variance. What we would like to emphasise now however is that away from the minimum of the effective potential the fRG gives us information that the near equilibrium Boltzmann assumption does not. To be concrete, an approximate Gaussian Boltzmann distribution would assume that the potential is of the form:
\begin{eqnarray}
\tilde{V}_{Boltz}(x) = \dfrac{\Upsilon}{4\cdot \text{Var}_{eq}}\left( \chi - \chi_{eq}\right)^2 \label{eq: BoltztildeV}
\end{eqnarray}
where $\chi_{eq}$ and $\text{Var}_{eq}$ are the equilibrium position and variance respectively. We show in Fig.~\ref{fig: ULJ_LPAvsBoltz_V_5T} how this approximation can break down dramatically as one moves away from the equilibrium position suggesting that the fRG captures the far away from equilibrium dynamics well. In principle one could attempt to include higher order cumulants of the Boltzmann distribution such as skewness and kurtosis into an approximate effective potential, however the relationship between these cumulants and higher derivatives of the effective potential is non-trivial and cumbersome to include. In any case it is not expected including these corrections would lead to significant improvement away from equilibrium.


\subsection{Accelerated trajectories}
\begin{figure}[t!]
\centering
\includegraphics[width=0.45\textwidth]{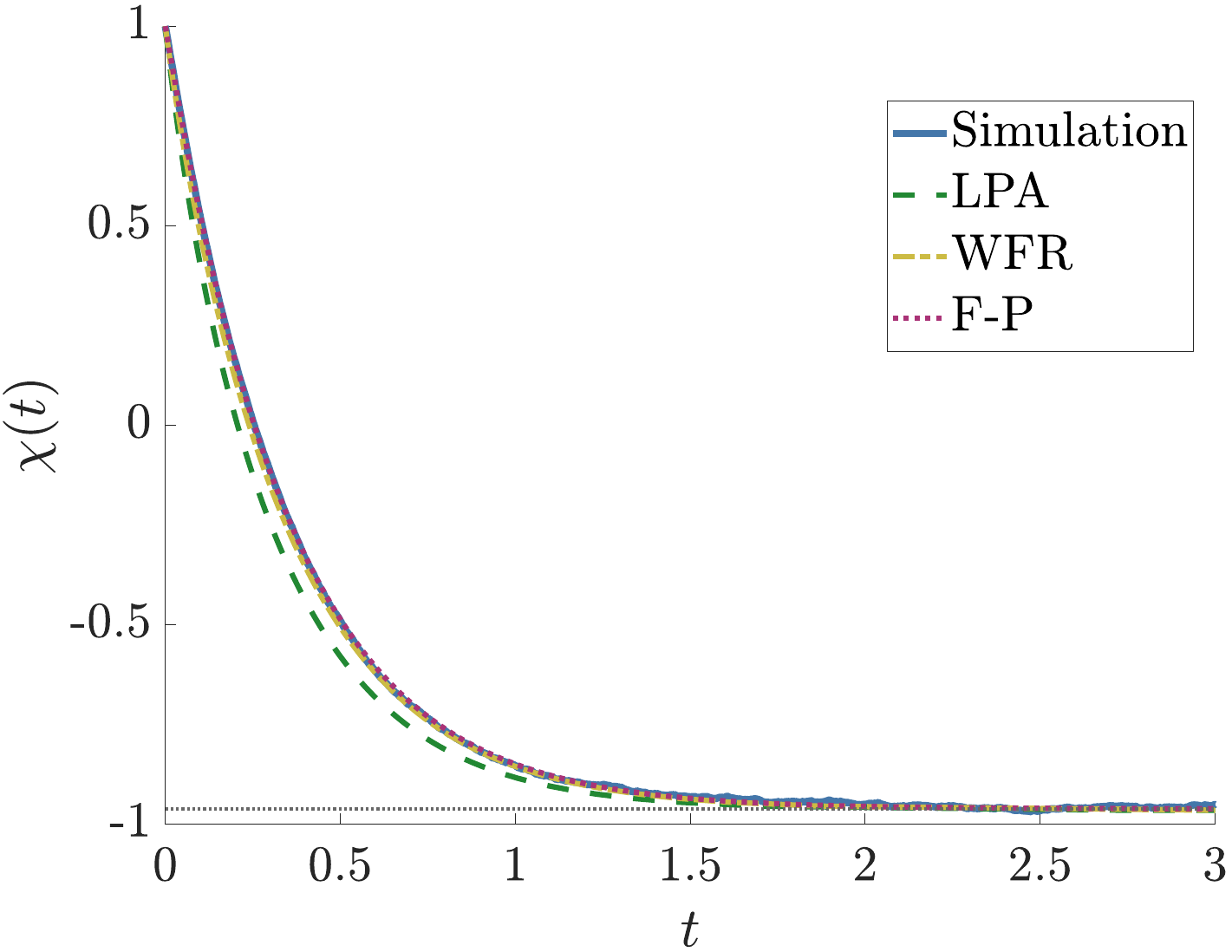}
\caption{The trajectory of the average position $\chi$ in a polynomial potential $\tilde{V}$ by direct simulation \& solving the EEOM (\ref{eq:QEOM}) using LPA and WFR for $\Upsilon = 10$. The horizontal dotted line is the equilibrium value. }\label{fig: Chi_Gies}
\end{figure}

\begin{figure}[t!]
\centering
\includegraphics[width=0.45\textwidth]{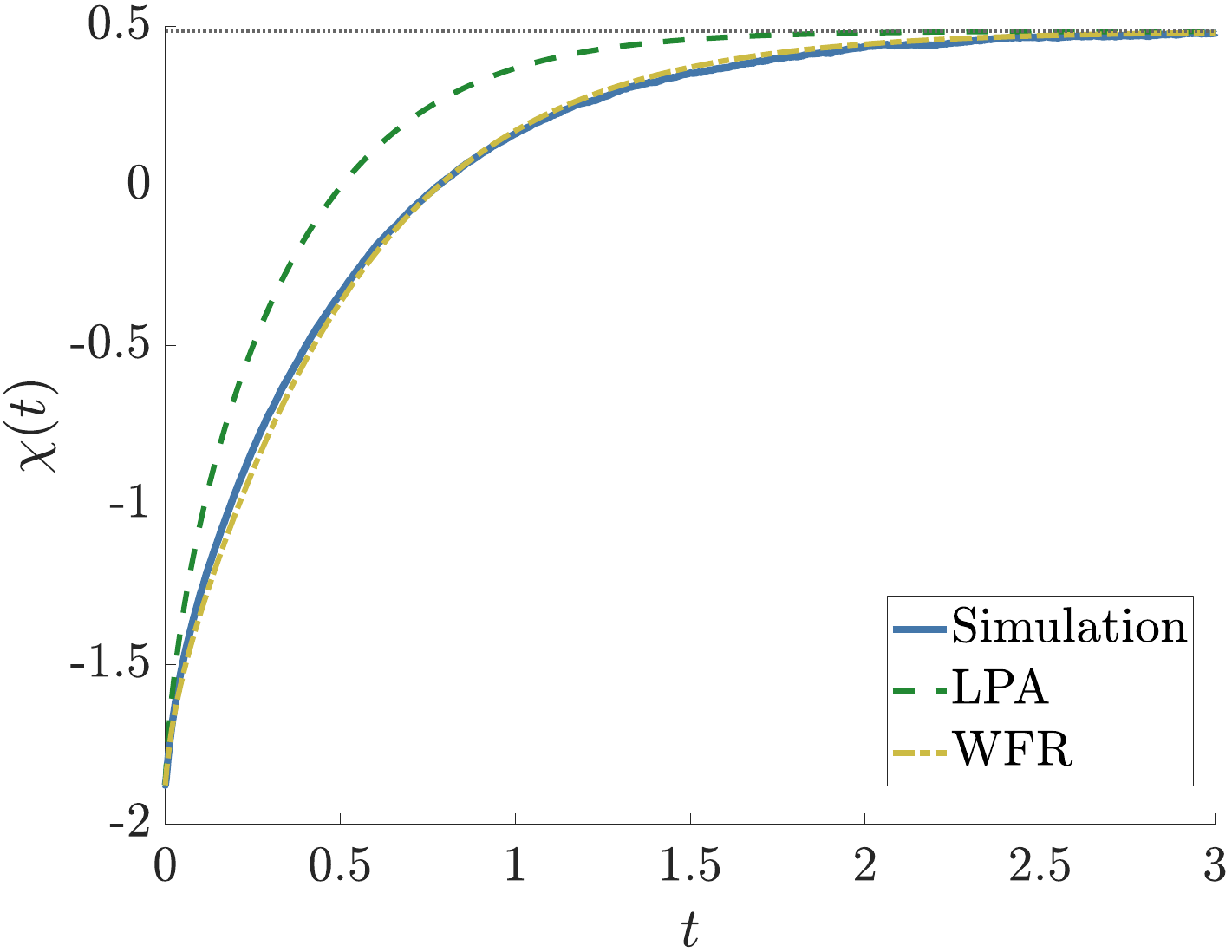}
\caption{The trajectory of the average position $\chi$ in an unequal LJ potential $\tilde{V}$ by direct simulation \& solving the EEOM (\ref{eq:QEOM}) using LPA and WFR for $\Upsilon = 10$. The horizontal dotted line is the equilibrium value.}\label{fig: Chi5T_ULJ}
\end{figure}

\begin{figure}[t!]
	\centering
	\includegraphics[width=0.4\textwidth]{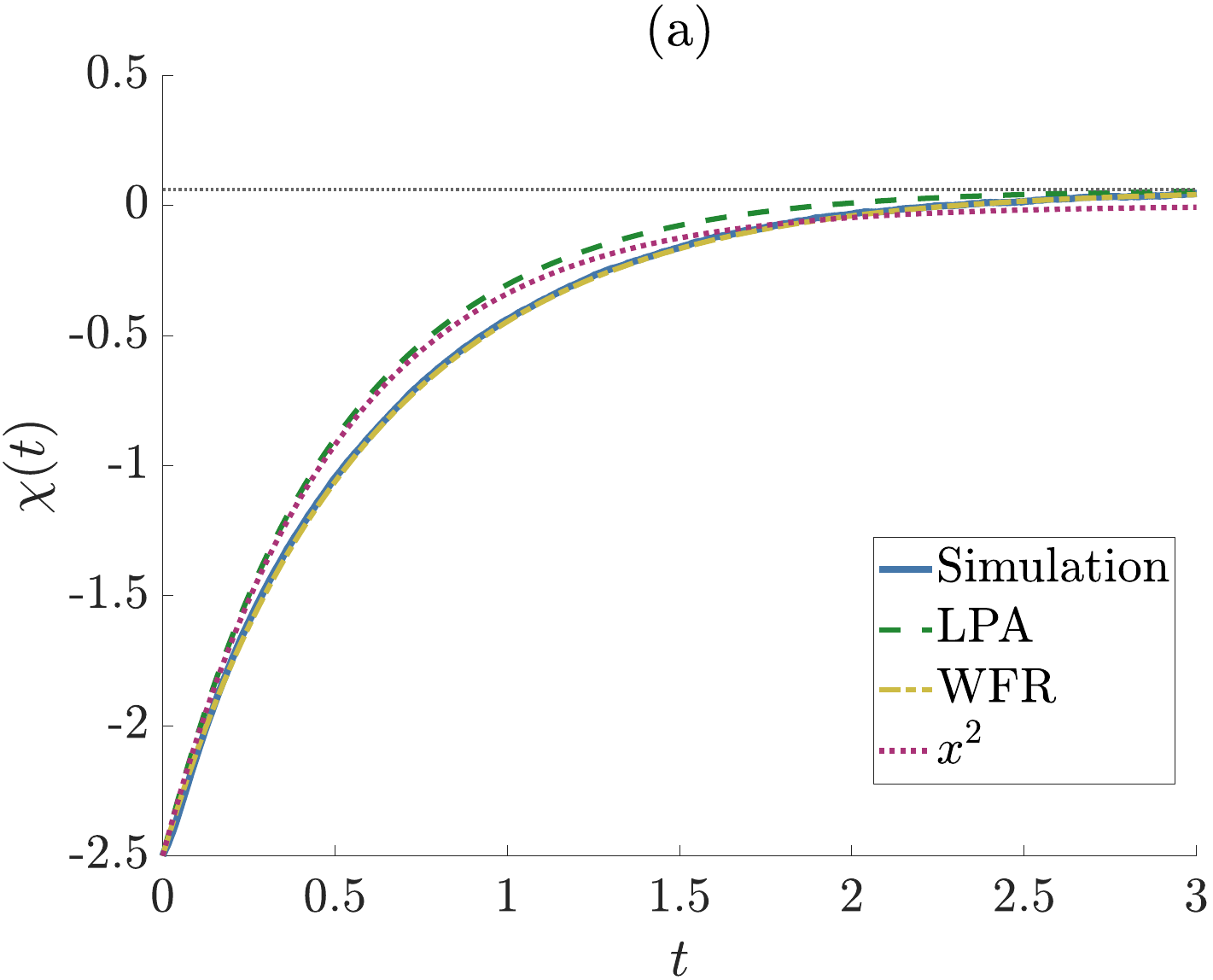}
	
	\includegraphics[width=0.4\textwidth]{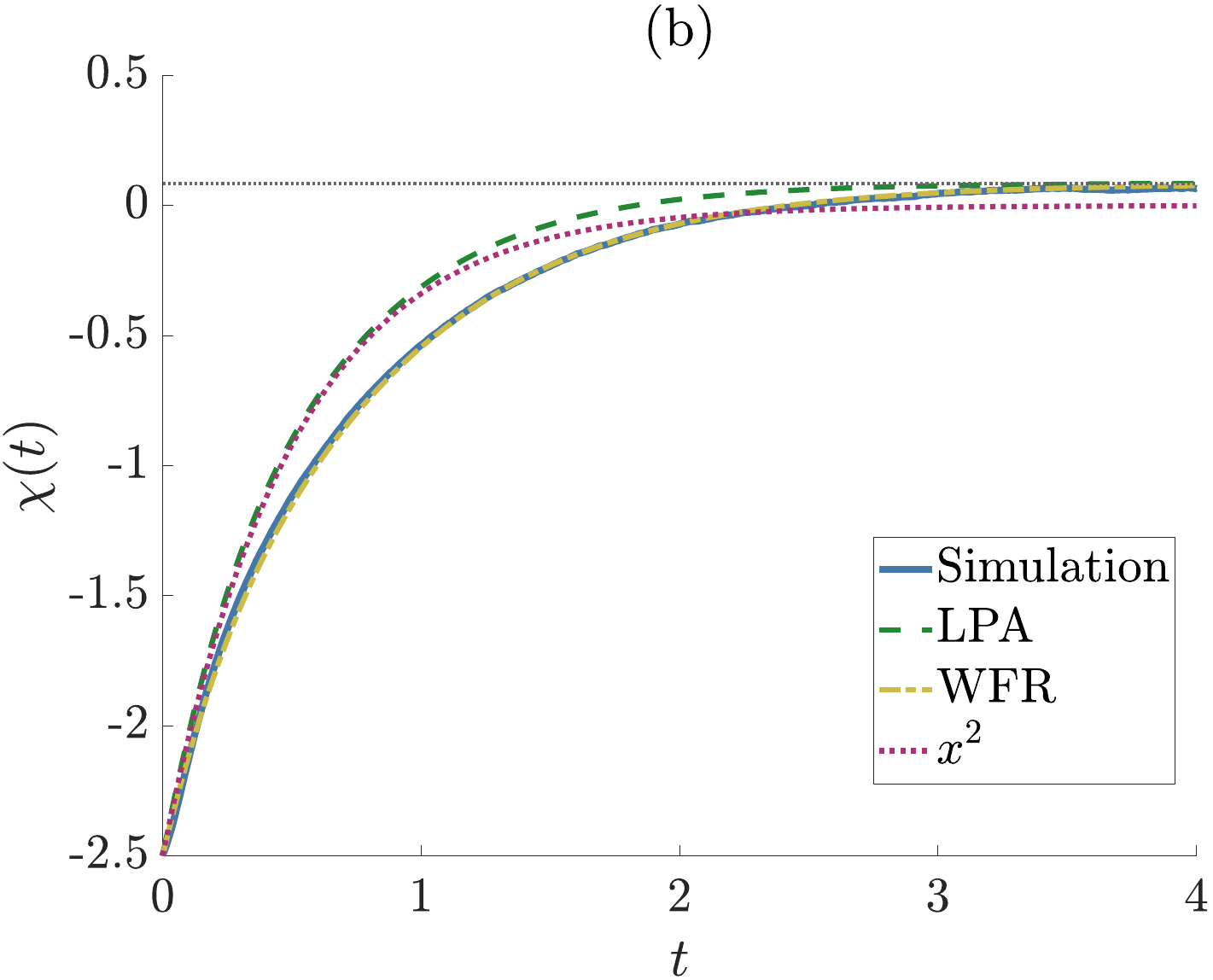}
	
	\includegraphics[width=0.4\textwidth]{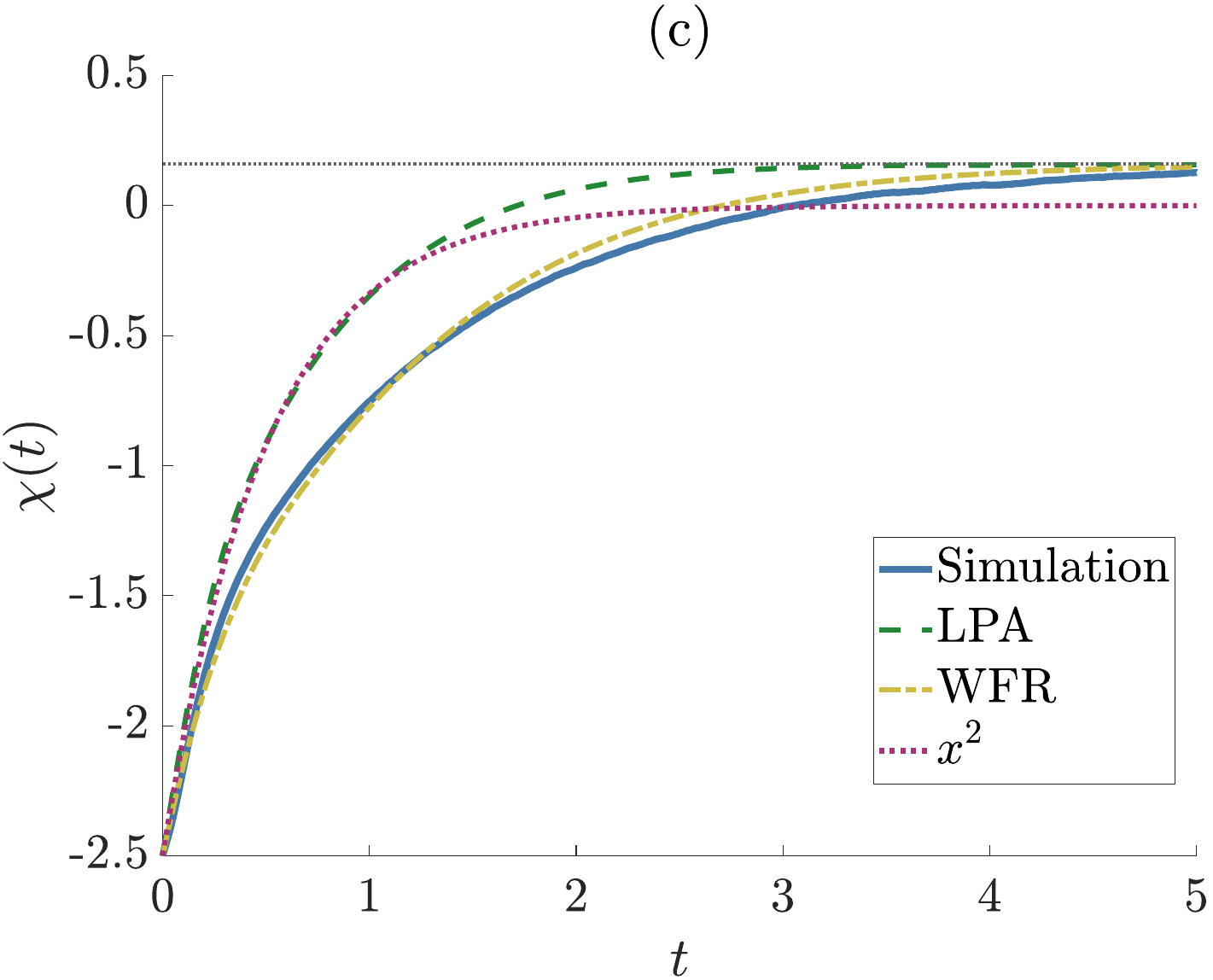}
	
	\caption{The trajectory of the average position $\chi$ for $x^2$ potential plus 6 gaussian bumps/dips $\tilde{V}$ by direct simulation \& solving the EEOM (\ref{eq:QEOM}) using LPA and WFR for (a) $\Upsilon = 4$, (b) $\Upsilon = 3$ and (c) $\Upsilon = 2$. For comparison the predicted evolution for the ``bare" $x^2$ potential is shown by the dotted red curve. The horizontal dotted line is the equilibrium value.}
	\label{fig: ChiT_X2GaussMany}
\end{figure}
Having solved the appropriate flow equations to obtain the dynamical effective potentials we can now solve (\ref{eq:QEOM}). Given the dynamical effective potential $\tilde{V}$ it only takes a couple of seconds to obtain the full trajectory of $\chi$ from some initial position $x_i = \chi_i$ to the equilibrium position. For the polynomial potential we initialised the particle far away from the equilibrium position at $x =  4$. In Fig.~\ref{fig: Chi_Gies} we show how the average position of the particle changes with time using direct simulation of the Langevin equation (\ref{eq:langevin}) over 50,000 runs, by numerically solving the F-P equation (\ref{eq: rescaled F-P}) and as calculated by the evolution in the dynamical effective potentials $\tilde{V}$ given using the LPA and WFR methods at $\Upsilon = 10$. All four trajectories agree to a very high precision. 

In Fig.~\ref{fig: Chi5T_ULJ} we plot the evolution of $\chi (t)$ for the unequal LJ potential where the particle begins in the smaller well at $x = - 1.878$ and moves towards its equilibrium position. We see as before that the WFR trajectory closely matches the simulated trajectory offering significant improvement over the LPA computation. Note that for this system it was impossible to get convergent numerics for the evolution of the F-P equation (\ref{eq: rescaled F-P}) 
This ability of the fRG to capture the non-trivial evolution of average position is also shown in Fig.~\ref{fig: ChiT_X2GaussMany} for the $x^2$ potential plus 6 bumps/dips which is a much more complex potential landscape at three different temperatures. Here the LPA trajectory offers improvement over the $x^2$ ``prediction" by converging to the correct equilibrium position and including WFR more closely matches the true simulated trajectory. Lowering the temperature generically decreases the accuracy of the fRG results. It is noteworthy that the fRG is able to reasonably capture these difficult dynamics well in systems where the F-P solution is difficult to obtain.

It is important to note the time advantage offered by the fRG. Solving the fRG flow equations is comparable in computation time to direct simulation while solving the F-P equation (\ref{eq: rescaled F-P}) takes longer than both. However the latter two methods obtain solutions that are only valid for a single initial condition. A huge advantage of the fRG is that once the dynamical effective potential $\tilde{V}$ is obtained it is trivial to solve the EEOM (\ref{eq:QEOM}) in a couple of seconds for any initial position whereas for both direct numerical simulation of (\ref{eq:langevin}) and solving the F-P equation (\ref{eq: rescaled F-P}) one has to start again from scratch. 

\subsection{Evolution of \textbf{Var}(x)}
For our accelerated trajectories we initialised the particles at the exact same point every time. This means that at t = 0 the probability distribution of the particles had zero variance \textbf{Var}(x) = 0. Using this as our initial condition we solved numerically the EEOM for the variance (\ref{eq:QEOM Variance})
, derived in appendix \ref{app:2ptfuncderiv}.
In Fig.~\ref{fig: VarT_Gies} we show how the variance evolves with time for the polynomial potential for $\Upsilon = 10$. We can see that the LPA closely matches the numerical and F-P evolution until $t = 0.5$ before departing slightly although it still tends towards the correct equilibrium distribution. 

\begin{figure}[t!]
\centering
\includegraphics[width=0.45\textwidth]{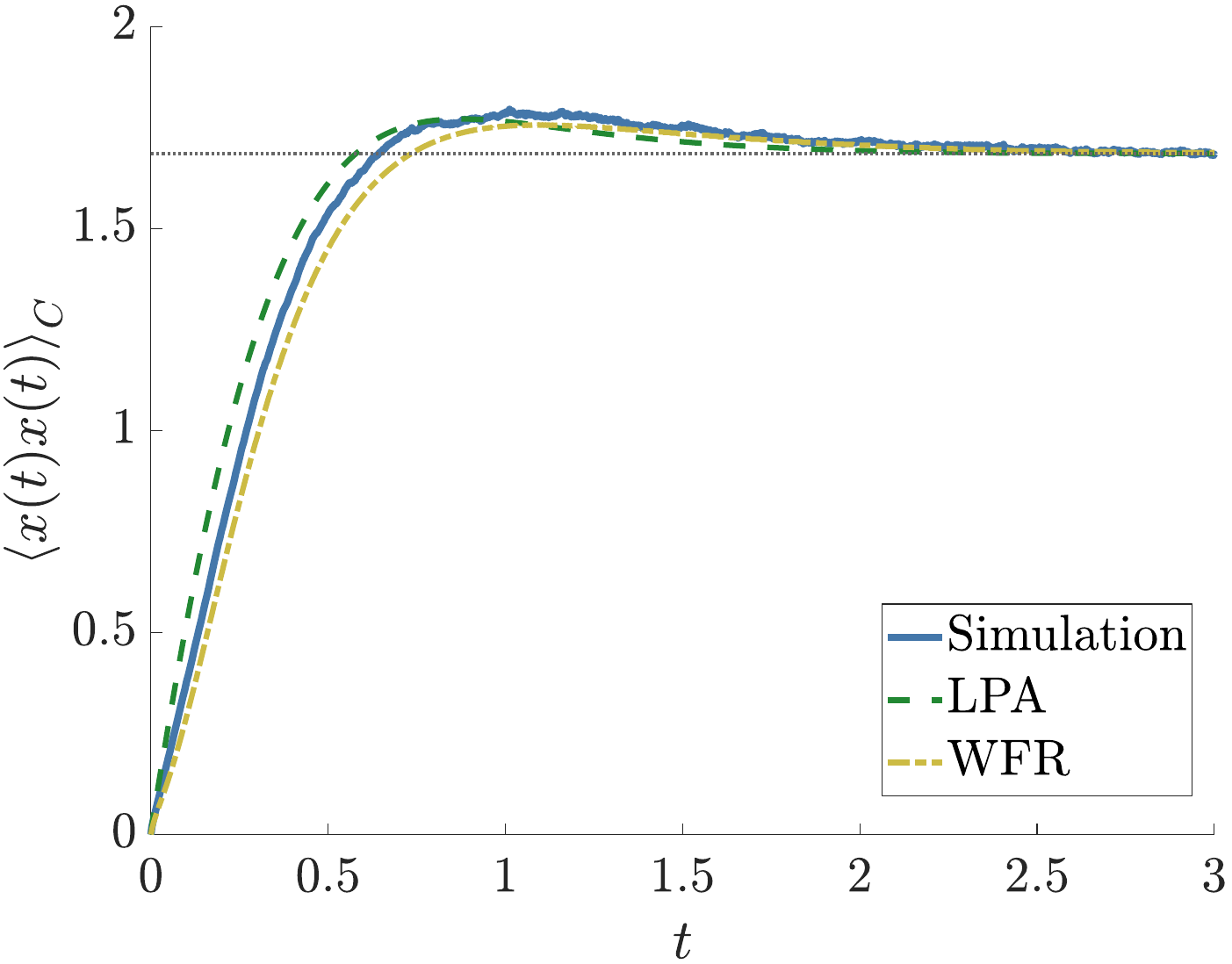}
\caption{The evolution of the variance \textbf{Var}(x) in an unequal LJ potential by direct simulation \& solving the EEOM (\ref{eq:QEOM Variance}) for $\Upsilon = 10$. The horizontal dotted line is the equilibrium value.}\label{fig: Var5T_ULJ}
\end{figure}

In Fig.~\ref{fig: Var5T_ULJ} 
we show how the variance evolves with time for an unequal LJ. As with the one-point function the F-P was unable to give sensible statistics however the LPA is able to very well match the early simulated trajectory even capturing the overshooting of the variance. The WFR on the other hand is better at capturing the late-time decay to equilibrium.

\begin{figure}[t!]
	\centering
	\includegraphics[width=0.4\textwidth]{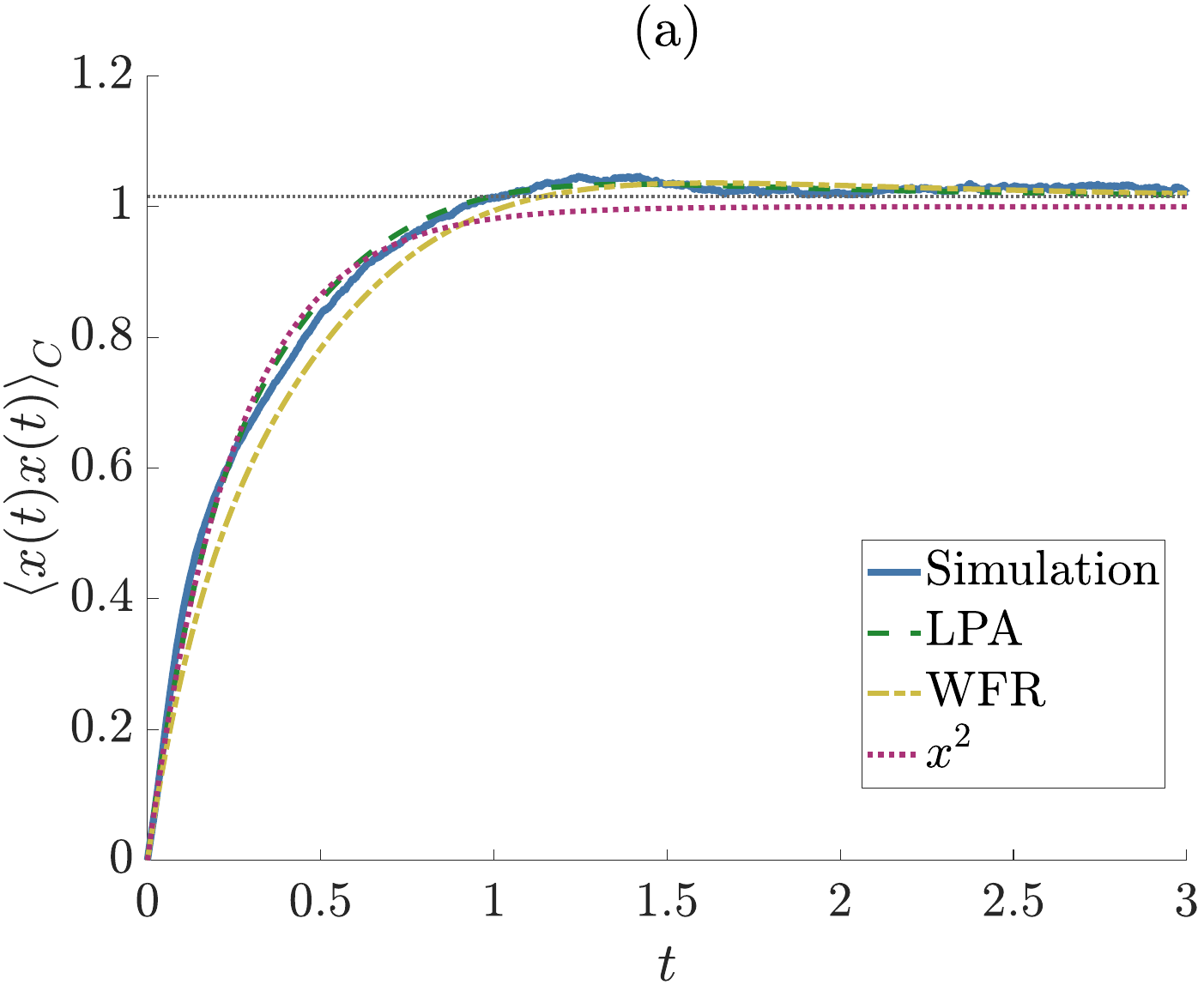}
	
	\includegraphics[width=0.4\textwidth]{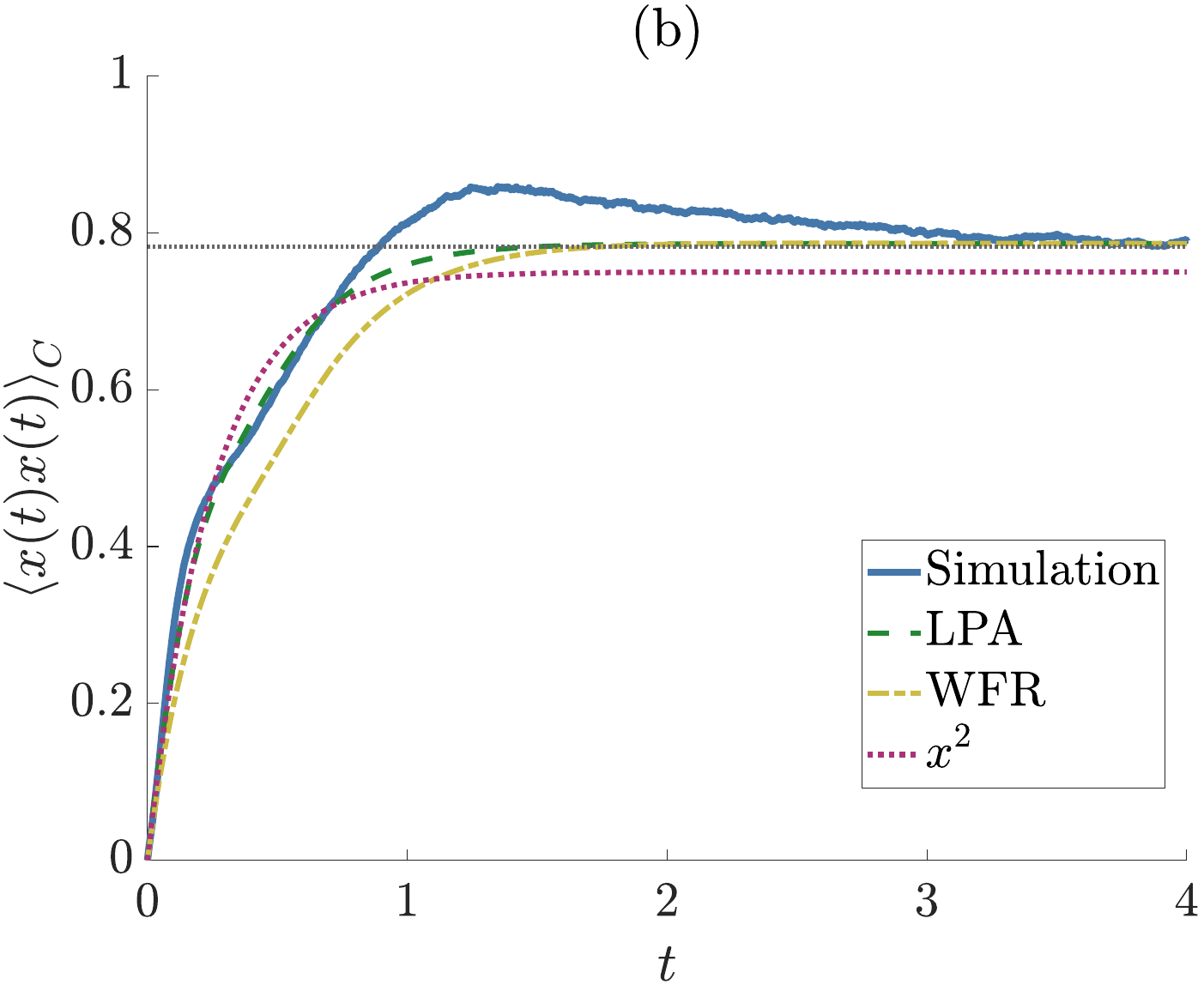}
	
	\includegraphics[width=0.4\textwidth]{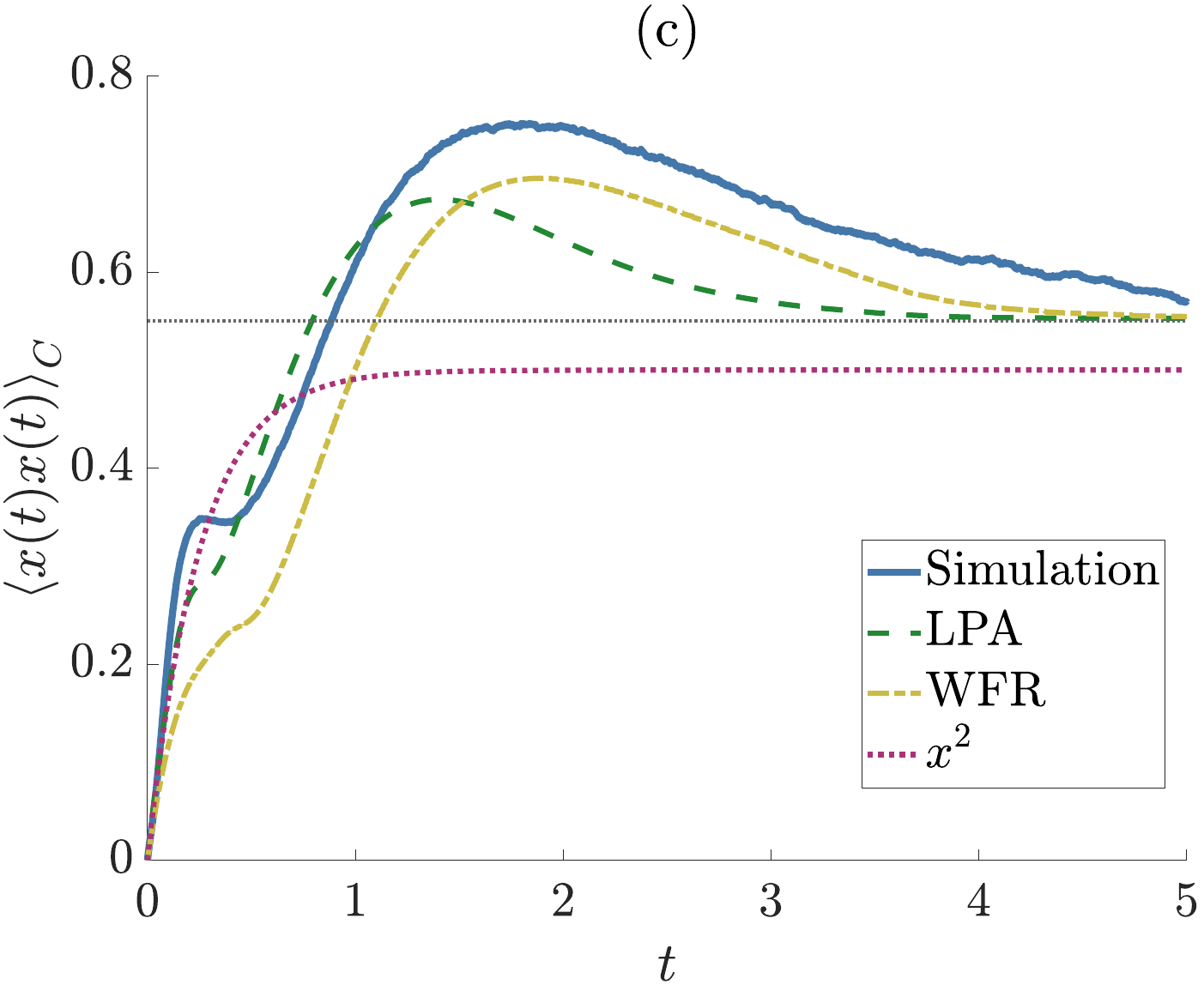}
	
	\caption{The evolution of the (normalised) variance $\left\langle x(t)x(t)\right \rangle_C$ in an $x^2$ plus six Gaussian bumps/dips potential for (a) $\Upsilon = 4$, (b) $\Upsilon = 3$ and (c) $\Upsilon = 2$. The horizontal dotted line is the equilibrium value.}
	\label{fig: VarT_X2Gaussmany.png}
\end{figure}

Finally in Fig.~\ref{fig: VarT_X2Gaussmany.png} we show how the variance evolves for the $x^2$ plus 6 gaussian bumps/dips potential at three different temperatures. As before, lowering the temperature decreases accuracy. In Fig.~\ref{fig: VarT_X2Gaussmany.png} (c) the fRG once again clearly captures the overshooting which is a feature of the gaussian bumps' existence; the bare $x^2$ evolution does not capture this behaviour and overall describes the evolution poorly, converging to the wrong equilibrium variance. Again as before the LPA much better describes the early evolution while WFR more accurately describes late time evolution. 
\begin{figure}[t!]
\centering
\includegraphics[width=0.45\textwidth]{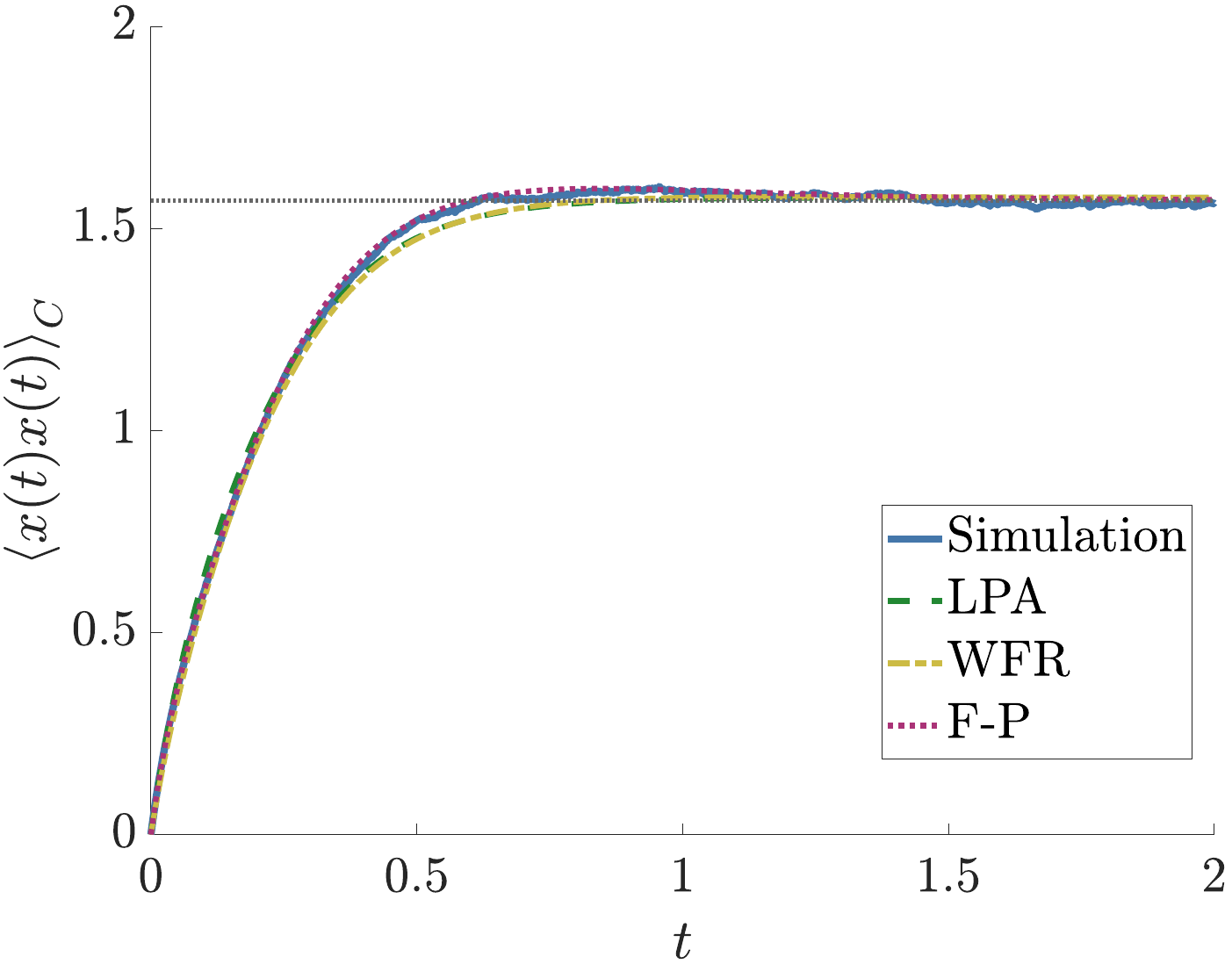}
\caption{The evolution of the variance \textbf{Var}(x) in a polynomial potential by direct simulation, solving the Fokker-Plank equation \& solving the EEOM (\ref{eq:QEOM Variance}) for $\Upsilon = 10$. The horizontal dotted line is the equilibrium value.}\label{fig: VarT_Gies}
\end{figure}

\subsection{\label{sec:comparspec}Comparison with the spectral expansion}
\begin{figure}[t!]
	\centering
	\includegraphics[width=0.4\textwidth]{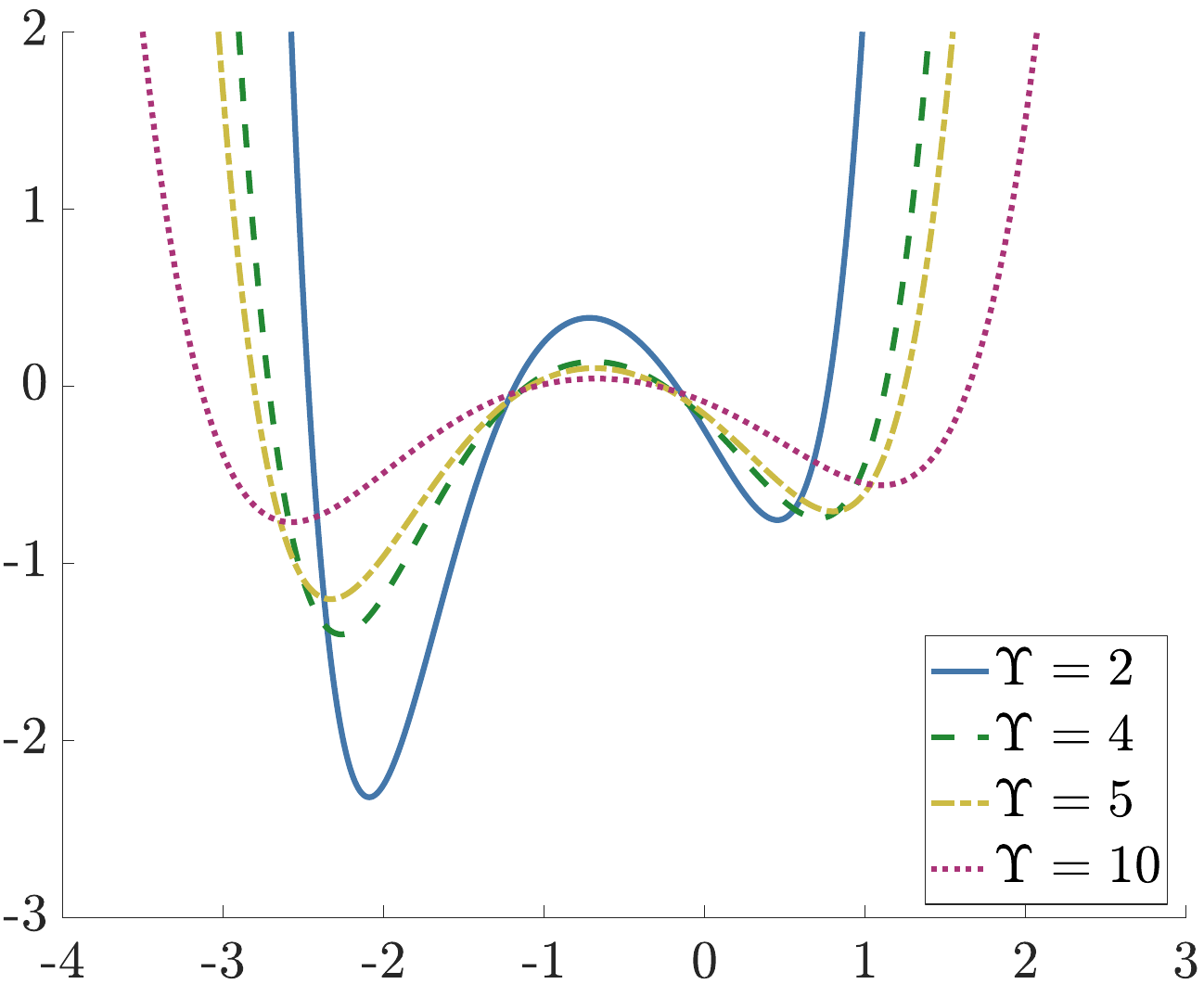}
	
	\caption{The dependence of the Schr\"{o}dinger potential $-\bar{U}$ (\ref{eq: Ubar=}) on the temperature $\Upsilon $ for the polynomial potential.}
	\label{fig: multiT_Gies.png}
\end{figure}

	

\begin{figure}[t!]
	\centering
	\includegraphics[width=0.4\textwidth]{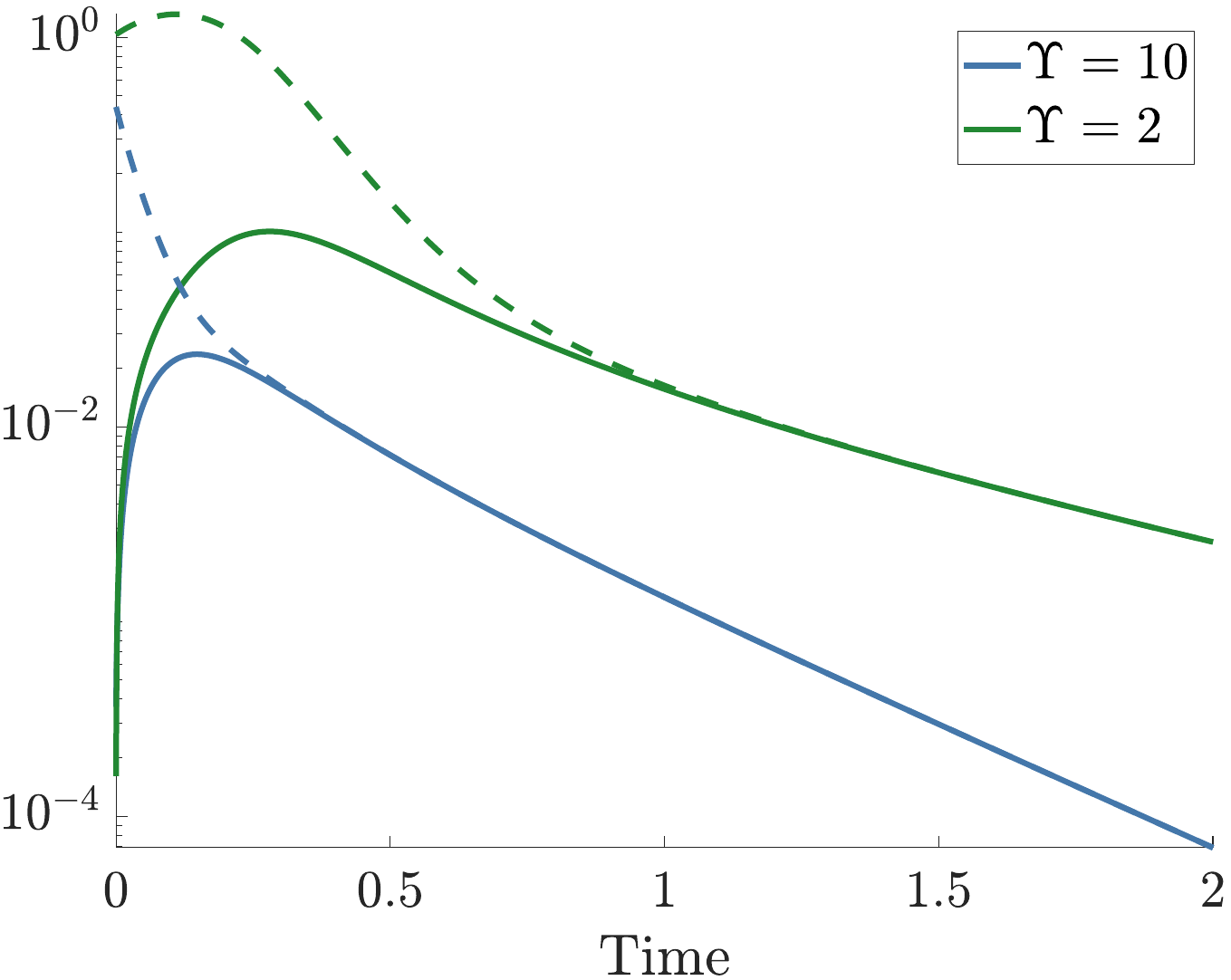}
	
	\caption{The error in $\left\langle x^2(t)\right\rangle$ for a finite truncation of the spectral expansion (\ref{eq:spec-2-point}) compared to the full F-P equation (\ref{eq: rescaled F-P}) initialised at $\chi(0) = 1$ for the polynomial potential. The dashed (solid) lines correspond to keeping the first two (50) non-zero terms. The bottom (blue) curve corresponds to $\Upsilon = 10$ and the top (green) curve to $\Upsilon = 2$.}
	\label{fig: Gtt_Schro_AllT_Gies_error}
\end{figure}

The above results for the change in the relative performance of LPA + WFR as temperature is lowered can be interpreted by resorting to the spectral expansion. In section \ref{sec:F-P} we recalled how all observables can be computed in a standard way from the Schr\"{o}dinger-like, Fokker-Planck equation (\ref{eq:FP1}) using an expansion in eigenfunctions and eigenenergies. It is straightforward to show that -- given our initial conditions considered above -- the average position and the two-point function can be expressed as:
\begin{eqnarray}
\left\langle x (t)\right\rangle = \chi_{eq} &+& \sum_{n=1}^{\infty}\Big[ e^{-E_n t}e^{V(x_0)/\Upsilon}p_n(x_0) \nonumber \\ 
 &\times &\int_{-\infty}^{\infty}\mathrm{d}x~x~e^{-V(x)/\Upsilon}p_n(x) \Big] \label{eq:spec-1-point}
\end{eqnarray} 
and
\begin{eqnarray}
\left\langle x^2(t)\right\rangle = \left\langle x^2\right\rangle_{eq} &+& \sum_{n=1}^{\infty}\Big[ e^{-E_n t}e^{V(x_0)/\Upsilon}p_n(x_0) \nonumber \\ 
 &\times &\int_{-\infty}^{\infty}\mathrm{d}x~x^2~e^{-V(x)/\Upsilon}p_n(x) \Big] \nonumber \\\label{eq:spec-2-point}
\end{eqnarray} 
where the subscript $eq$ indicates the equilibrium value, and $E_n$ and $p_n(x)$ are  respectively the eigenvalues and normalised eigenfunctions of the corresponding Schr\"{o}dinger-like problem. Obtaining the spectrum $E_n$ and $p_n(x)$ may be complicated  by the fact that the actual Schr\"{o}dinger potential $\bar{U}$ (\ref{eq: Ubar=}) can develop temperature dependent features as the temperature is decreased-- see Fig.~\ref{fig: multiT_Gies.png}. Even for the simple polynomial potential in the Langevin equation it is clear that at low temperatures $\bar{U}$ becomes non-trivial, developing highly asymmetrical trapping wells. The increasing energy gap between the two minima indicates that, for a fixed initial condition, higher order terms in the spectral expansion can become important as the temperature is lowered.

To illustrate the importance of these higher-order terms for the two-point function evolution in the polynomial potential we examine the accuracy of a finite truncation of the spectral expansion at two temperatures, $\Upsilon = 10$ and $\Upsilon = 2$, for the evolution of $\langle x^2(t)\rangle$, initializing trajectories at $x=1$: $P(x,t=0)=\delta(x-1)$. In Fig.~\ref{fig: Gtt_Schro_AllT_Gies_error} we plot the error associated with a finite truncation of the spectral expansion, keeping only the first two (dashed line) or fifty (solid line) terms, at two different temperatures $\Upsilon=2$ (green, top curve) or $\Upsilon=10$ (blue, bottom curve). This error is computed by comparing the truncated expansion to the numerical solution of the Fokker-Planck equation. At early times, the error associated with keeping only two terms is larger than when 50 terms are kept, as one would expect, the discrepancy being more pronounced at lower temperatures. As the system relaxes, the contribution form the higher order terms decreases and the errors of the two truncations converge, until they are essentially indistinguishable at later times, as expected. This decay of the contribution from the higher eigenvalues occurs faster for the higher temperature, making the two-term truncation more accurate earlier. This observation reinforces our inference from the previous paragraph that as temperature is lowered, higher order terms in the spectral expansion become more important for accurately describing the evolution, at least for a fixed initial condition. Crucially, this offers an explanation for why the the LPA + WFR offers poorer agreement as temperature is lowered since it would be expected to most accurately describe circumstance where the lowest order terms in a spectral expansion dominate. 

The relation between the spectral expansion and the range of validity of the effective action's derivative expansion is not entirely straightforward however, as it would also depend on the initial condition. The quantification of this relation would be an interesting undertaking which we leave for future work.

\section{Summary}\label{sec:Summary}

Collecting results scattered in the existing literature, we have recalled how Brownian motion can be formally described by a path integral involving a Euclidean Supersymmetric action and how an effective average action functional $\Gamma[\chi]$ of the average position $\chi$, incorporating the effects of the fluctuating force and encoding all statistical properties of the process, can be calculated using functional Renormalisation Group (fRG) methods. We  emphasised the importance of utilising the underlying symmetries of the problem, paying attention to the boundary terms which are often dropped, and showed how these can correctly incorporate any initial condition and, correspondingly, non-equilibrium evolution. The fRG flow equations were written down for the first two orders of the widely used derivative expansion of the effective action, referred to as the Local Potential Approximation (LPA) and Wavefunction Renormalisation (WFR). We used a particular type of regulator, the frequency independent Callan-Symanzik regulator, for which the flow equations take on a relatively simple form, and further recalled that obtaining flow equations within the supersymmetric framework is convenient for ensuring compatibility with the Boltzmann equilibrium distribution, something that is not a priori obvious or guaranteed if one starts with the Onsager-Machlup form of the action (\ref{eq:dim-action}) and considers it a Euclidean $N=1$ scalar theory in one dimension with the Schr\"{o}dinger potential $\bar{U} = \Upsilon/4 \, V'' - 1/4\,(V')^2$. We also reviewed how Brownian motion can be solved using a spectral expansion method of the Fokker-Planck (F-P) equation in a standard way. 

The Effective Action (EA) $\Gamma$ allows one to derive \textit{effective equations of motion} (EEOMs) for the average position $\chi(t)\equiv \langle x(t) \rangle$ and variance $\langle x^2(t) \rangle$ in an analogous manner to the classical equations of motion, by taking variational derivatives. We used the LPA and WFR to compute the elements entering the EEOM, for instance the \textit{dynamical effective potential}. We verified the accuracy of the equilibrium limit to these equations, further emphasising the physical significance of certain aspects of the effective potential $V_{k=0}$: namely how the minimum of $V_{k=0}$ corresponds to the equilibrium position and its second derivative evaluated at this point to the variance through equation (\ref{eq:equal2pt}). We noted here that while the LPA reproduces these equilibrium quantities, the accuracy of covariance's temporal evolution diminished as temperature was lowered.

Going beyond equilibrium, we examined how LPA+WFR handle relaxation towards it for the average position $\chi(t)$ in potentials such as a polynomial, a doublewell comprised of two LJ type interactions, or a bare $x^2$ plus gaussian bumps. The latter potential clearly demonstrates that the fRG is capable of capturing the effect of the non-trivial local features of this potential. In fact, the fRG could still offer reasonable approximations even in those cases where the Fokker-Planck numerics failed to converge. We have also shown how the fRG can closely match the relaxation of the variance $\langle x^2(t) \rangle$ to its equilibrium value: for both the unequal Lennard-Jones type potential and the $x^2$ plus gaussian bumps, the LPA variance has reasonable accuracy and still captures highly non-trivial behaviour such as the variance overshooting its equilibrium value before settling to it. This is in a system where numerically solving the Fokker-Planck equation failed to provide good results, at least using standard methods. Again, we find that accuracy decreases with decreasing temperature. 

A clear conclusion that can be drawn from the above investigations is that decreasing the temperature negatively impacts the accuracy of using the LPA + WFR derivative expansion for the fRG to describe Brownian motion in the potentials we examined. This appears to correlate with the increasing importance at lower temperatures of higher order terms in the spectral expansion; indeed, it is expected that the lowest order terms in the derivative expansion (LPA + WFR) are best placed to describe evolution dominated by the lowest non-zero eigenvalues of the Fokker-Planck spectral expansion. It would seem that the derivative expansion of the fRG for studying thermal fluctuations has utility in the range from moderate temperatures (roughly when the classical force is comparable to the noise), up to the very high temperature regime where the small local features of the potential become less relevant.  

Although our conclusions on the temperature dependent relation between the spectral expansion and the derivative expansion of the fRG are suggestive, and reasonable given the premise of the latter, a more precise quantitative comparison would be called for and should be addressed in future work. The insights gained from a more detailed, quantitative understanding of the fRG's range of validity when applied to the dynamical studies of thermally driven systems may lead to interesting and important applications. There are also issues relevant to the technicalities of applying the fRG programme and which may lead to better convergence properties. For example, the recent findings of \cite{Balog2019, DePolsi2020} suggest that an appropriately optimized regulator, which also excludes the regime $\omega>k$ from contributing to the flow, can ensure good convergence properties and a sizeable radius of convergence. Comparisons with \cite{Balog2019, DePolsi2020} are non-trivial because the supersymmetry of the Brownian motion problem makes the structure of the flow equations different to that of a simple scalar theory. This is an important question to be resolved however and we hope to return to it in future work. 


Future work could also examine if higher order approximations beyond the WFR offer any advantage, as these might better capture higher order terms in a spectral expansion. One could further investigate an ensemble of initial conditions and try to quantify better the computational time gain the fRG offers. It is worthwhile trying to see if there is a more concrete way to determine \textit{a priori} which systems will be well described by the fRG before having to compare to numerical simulations or doing a spectral expansion analysis. An interesting application that we didn't touch upon here might be thermal barrier escape, which in this formulation seems to be more akin to tunnelling in the corresponding euclidean quantum mechanics \cite{Strumia1999, Strumia1999b, Strumia1999a}; it would indeed be interesting to flesh out any analogies, if they exist. Most importantly, further understanding the application of fRG techniques to stochastically driven systems may allow extension to systems with more degrees of freedom, such as field theories and/or systems with a large number of particles. Advances in the above directions may lead to progress in theoretically tackling a broad range of physical phenomena with large separation between fundamental timescales of thermal fluctuations and long emergent timescales of macroscopic change, addressing what is now a major barrier for predictive simulations across scientific and engineering disciplines including materials science \cite{Yip2013}, drug design \cite{Zwier2010}, protein folding \cite{Freddolino2010}, and cosmology.

\section*{Acknowledgements}
AW would like to thank George Stagg for his numerical insight on solving the flow equations (\ref{eq:dV/dk}) and (\ref{eq:dzetax/dktilde}). AW is funded by the EPSRC under Project 2120421. GR would like to especially thank Nikos Tetradis for very useful discussions at the early stages of this project, Julien Serreau for providing much insight on fRG computations and Gabriel Moreau for sharing his PhD thesis, containing many new results on the application of the fRG to the Langevin equation. We would also like to thank the referees of an earlier partial version of this work for very useful criticisms, comments and suggestions which helped improve it substantially, and also for bringing to our attention the recent works \cite{Balog2019, DePolsi2020} and for prompting us to clarify the possible relation to \cite{Strumia1999, Strumia1999b, Strumia1999a}.   

\appendix

\section{\label{app:equil-flow} The equilibrium flow equation} 

In equilibrium, all equal-time expectation values can be obtained from the generating function \\
\begin{equation}\label{equil1}
	Z(J)=\int dx  \,  e^{-2 V(x)/\Upsilon + Jx} 
\end{equation}
in a manner directly analogous to that described in the text but with functional derivatives replaced by ordinary derivatives w.r.t. $J$. In a spirit identical to the renormalisation group but in the much simpler setting of one just degree of freedom, we can define a modified generating functional \cite{Guilleux2015}
\begin{equation}
	Z_k(J)=\int dx \, e^{-2 V(x)/\Upsilon - \frac{1}{2}R(k)x^2+Jx}
\end{equation}    
with an additional quadratic term controlled by an arbitrary function $R(k)$ of a parameter $k$, satisfying $\lim\limits_{k \rightarrow 0} R(k) = 0$, giving back the original $Z(J)$. Correlation functions are generated by $W_k(J)=\ln Z_k(J)$ via
\begin{equation}\label{app:correlations}
	\chi_k \equiv \langle x\rangle_k=\frac{\partial W_k(J)}{\partial J}\,,\quad \langle x^2\rangle_k - \chi^2_k=\frac{\partial^2 W_k(J)}{\partial J^2}
\end{equation}     
e.t.c. In the limit $k=0$ and after setting $J=0$ the usual predictions of the equilibrium Boltzmann distribution are recovered.   

The source $J$ has been considered as an external, independent variable controlling expectation values such as $\chi$ and higher correlators. One could also consider $\chi$ as the independent variable, solving $\chi = \partial W/\partial J$ for $J(\chi)$ and defining the effective potential $U(\chi)$ via a Legendre transform 
\begin{equation}
	\Gamma_k(\chi) + W_k(J) = J\chi -\frac{1}{2}R(k)\chi^2
\end{equation}  
with
\begin{equation}
	\Gamma(\chi) \equiv 2 U(\chi)/\Upsilon 
\end{equation}            
Note that 
\begin{equation}\label{Legendre}
	\frac{\partial \Gamma_k}{\partial \chi } = J_k - R(k)\chi
\end{equation}
implying that the minimum of the effective potential defines the equilibrium expectation value of $x$ (at $J=0$ and $k=0$).

The dependence of the generating function $W_k(J)$ on $k$ can be easily obtained as   
\begin{equation}
	\partial_k W_k(J)=-\frac{1}{2}\partial_kR\left[\frac{\partial^2 W_k(J)}{\partial J^2}+\left(\frac{\partial W_k(J)}{\partial J}\right)^2\right]
\end{equation}
which is an ``RG equation'' for $W_k(J)$. We can also obtain an equation determining how $\Gamma_k(\chi)$ runs with k. Reciprocally, taking $\chi$ as the independent variable, $J$ becomes a function of $\chi$ and $k$. Taking a $k$ derivative of (\ref{Legendre}) at fixed $\chi$ we obtain 
\begin{equation}
	\partial_k \Gamma_k(\chi) = \frac{1}{2}\partial_k R \frac{\partial^2 W_k}{\partial J^2} 
\end{equation}            
To express the rhs in terms of $\Gamma_k(\chi)$, consider the first relation of (\ref{app:correlations}). Taking a $\chi$ derivative we find 
\begin{equation}
	\left(\frac{\partial^2\Gamma_k}{\partial\chi^2}+R\right)\frac{\partial^2 W_k}{\partial J^2} =1
\end{equation}   
Hence, the ``RG flow'' of $\Gamma$ is determined by 
\begin{equation}\label{app:Gamma-flow}
	\partial_k\Gamma_k(\chi)=\frac{1}{2}\partial_kR\left(\frac{\partial^2\Gamma}{\partial\chi^2}+R \right)^{-1}
\end{equation} 
Note also that, at $k \rightarrow 0$ 
\begin{equation}
	\langle x^2 \rangle - \chi^2=\frac{\Upsilon}{2 \, \partial_\chi^2 U(\chi_{eq})} 
\end{equation}  
and hence the variance at equilibrium is determined by the curvature of the effective potential around its minimum. 

All the above manipulations can be generalized to many or even infinite degrees of freedom and continuum actions, leading to the Wetterich equation (\ref{Wetterich eqn}), which is directly equivalent to (\ref{app:Gamma-flow}), and the relations of section  (\ref{sec:1-point function}). For this work it is important to note that the equilibrium effective potential $U(\chi)$ discussed here obeys the LPA flow equation \emph{exactly} if we choose $R(k)=k$.         

\section{\label{app:2ptfuncderiv}Derivation of the two point function}
We start from equation (\ref{eq:2pointfuncdef}) repeated here for clarity:
\begin{eqnarray}
	\left(\dfrac{d^2}{dt^2} -  Q(t) \right)G(t,t') &=& -\dfrac{\Upsilon}{P(t)}\delta(t-t')  \label{eq:appendix2pointfuncdef} 
\end{eqnarray}
Where $Q(t) = \mathcal{U}(\chi(t))$ is given by (\ref{eq:U = }) and $P(t) = 1$ or $\zeta_{\chi}^2(\chi(t))$ for LPA and WFR respectively. If we now consider the homogeneous version of (\ref{eq:appendix2pointfuncdef}):
\begin{equation}
	\ddot{f}(t) -  Q(t)f(t) = 0 \label{eq:appendhomo2ptfunc}
\end{equation}
it will have a growing and a decaying solution since $Q(t)>0$, and its Wronskian $\mathcal{\mathcal{W}}$ will simply be a constant:
\begin{equation}
	\mathcal{W}(t) \equiv  Y_1(t)\dot{Y}_2(t) - \dot{Y}_1(t)Y_2(t) =\text{constant} \label{eq:appendWronskianconst}
\end{equation}
We take $Y_1(t)$ to be the growing solution and $Y_2(t)$ to be the decaying solution. Substituting the ansatz $G(t,t') = Y_1(t)F(t,t')$ where $F$ is some function to be determined into (\ref{eq:appendix2pointfuncdef}) we obtain:
\begin{eqnarray}
	\dot{F}(t,t') = \dfrac{1}{Y_{1}^{2}(t)}\left[-\Upsilon\dfrac{Y_1(t')}{P(t')}\theta (t-t') + C_1(t')\right] \nonumber \\
	\label{eq:append Fdot}
\end{eqnarray}
where $\theta (t-t')$ is the Heaviside step function and $C_1(t')$ is a `constant-of-integration' function of $t'$ to be determined. If we now integrate (\ref{eq:append Fdot}) we obtain the following expression for $G(t,t')$:
\begin{eqnarray}
	G(t,t') &=& -\Upsilon\dfrac{Y_1(t)}{P(t')}\left[\theta (t-t') \int_{t'}^{t} \dfrac{Y_1(t')}{Y_{1}^{2}(u)} du + C_2(t')\right] \nonumber \\
	& & \quad  +~C_1(t')Y_1(t)\int^{t} \dfrac{du}{Y_{1}^{2}(u)} \label{eq:appendGwithints}
\end{eqnarray}
where $C_2(t')$ is another `constant-of-integration' function of $t'$ to be determined. To compute the integrals in (\ref{eq:appendGwithints}) we note that by the definition of the Wronskian:
\begin{eqnarray}
	Y_1(t)\int^t \dfrac{\mathcal{W}(u)}{Y_{1}^{2}(u)}du = \mu Y_1(t) +  Y_2(t)
\end{eqnarray}
where $\mu$ is simply a constant of integration. As the Wronskian is constant here we can simply write:
\begin{eqnarray}
	Y_1(t)\int^t \dfrac{du}{Y_{1}^{2}(u)} = \dfrac{1}{\mathcal{W}}\left[\mu Y_1(t) + Y_2(t)\right]
\end{eqnarray}
such that (\ref{eq:appendGwithints}) becomes:
\begin{eqnarray}
	G(t,t') &=& \dfrac{\Upsilon}{\mathcal{W} P(t')}\Big\lbrace\bar{C}_1(t')Y_2(t) + \bar{C}_2(t')Y_1(t)  \nonumber \\
	& & \quad +~\theta (t-t')\left[Y_1(t)Y_2(t')-Y_1(t')Y_2(t)\right]\Big\rbrace \nonumber \\ \label{eq:append GCbar}
\end{eqnarray}
where $C_1$ and $C_2$ have been rescaled to $\bar{C}_1$ and $\bar{C}_2$ in order to absorb some irrelevant constant factors. We note that the functions $\bar{C}_i$ can only be linear combinations of $Y_1$ and $Y_2$:
\begin{eqnarray}
	\bar{C}_1(t') &\equiv &  \alpha ~Y_1(t') + \beta ~Y_2 (t') \\
	\bar{C}_2(t') &\equiv &  \gamma ~Y_1(t') + \delta ~Y_2 (t') 
\end{eqnarray}
where the constants $\alpha$, $\beta$, $\gamma$ and $\delta$ will be determined later\footnote{N.B. the $\delta$ here should not to be confused with the dirac delta function}. Combining all this together we obtain the most general solution:
\begin{eqnarray}
	G(t,t') &=& \dfrac{\Upsilon}{\mathcal{W} P(t')}\Big\lbrace \left[\alpha - \theta (t-t') \right] Y_1(t')Y_2(t) \nonumber \\
	&& \quad \quad \quad +~\beta ~Y_2(t')Y_2(t)  + \gamma ~Y_1(t')Y_1(t)\nonumber \\
	& & \quad  \quad  \quad+~\left[\delta + \theta (t-t')\right]Y_2(t')Y_1(t)\Big\rbrace \nonumber \\ \label{eq:append G general}
\end{eqnarray}
To obtain the values of the constants we must now impose physical conditions:
\begin{enumerate}
	\item $G(t,t)$ should remain finite as $t\rightarrow \infty$ i.e. an equilibrium distribution exists at late times \\
	$\Rightarrow \gamma = 0$	
	\item Covariance $G(t,0)$ should remain finite as $t\rightarrow \infty$  \\
	$\Rightarrow \delta = -1$
	\item  The equilibrium form of $G(t,t')$ should be symmetric under $t \leftrightarrow t'$ \\
	$\Rightarrow \alpha = 0$
	\item Setting the initial condition to be $G(0,0) \equiv G_{00}$ \\
	$\Rightarrow \dfrac{\beta\Upsilon}{\mathcal{W}} =  \dfrac{P(0)}{Y_2(0)Y_2(0)}\left[G_{00} + {Y_1(0)Y_2(0)}\dfrac{\Upsilon}{\mathcal{W} P(0)}\right]$
\end{enumerate}
These give us the two point function:
\begin{eqnarray}
	G(t,t') &=& \dfrac{\Upsilon}{2\lambda P(t')}\left[ \theta (t-t')\tilde{Y}_1(t')\tilde{Y}_2(t) + \theta (t'-t)\tilde{Y}_2(t')\tilde{Y}_1(t)\right] \nonumber \\
	&& \quad \quad +~ \dfrac{P(0)}{P(t')}\left[G_{00} - \dfrac{\Upsilon}{2\lambda P(0)}\right]\tilde{Y}_2(t')\tilde{Y}_2(t) \label{eq:append finalG}
\end{eqnarray}
where $\tilde{Y}_i(t) \equiv Y_i(t)/Y_i(0)$ and we have normalized the Wronskian as 
\begin{equation}
\mathcal{W}=-2\lambda Y_1(0)Y_2(0)
\end{equation}
which is the value at equilibrium when $Q(t)\rightarrow \lambda^2$,

Equation (\ref{eq:append finalG}) has two important limits: \\
The \textit{Variance} $t'\rightarrow t$:
\begin{eqnarray}
	\textbf{Var}(x) \equiv G(t,t) &=& \dfrac{\Upsilon}{2\lambda P(t)} \tilde{Y}_1(t)\tilde{Y}_2(t)  \nonumber \\
	&& +~ \dfrac{P(0)}{P(t)}\left[G_{00} - \dfrac{\Upsilon}{2\lambda P(0)}\right]\tilde{Y}_{2}^{2}(t)\nonumber \\
	\label{eq:append Variance}
\end{eqnarray}
and the \textit{Covariance} $t'\rightarrow 0$, $t > 0$:
\begin{eqnarray}
	\textbf{Cov}(x(0)x(t)) \equiv G(t,0) = G_{00}\tilde{Y}_2(t) \label{eq:append Covariance}
\end{eqnarray}
Equations (\ref{eq:append Variance}) \& (\ref{eq:append Covariance}) are the main results of this appendix.

\section{\label{app:det} Explicit computation of the determinant}
In this appendix we recall the computations of \cite{Marculescu:1990nf} explicitly showing that
\begin{eqnarray}
	\text{det } \textbf{M} &=& \int \mathcal{D}c\mathcal{D}\bar{c}\text{ exp}\left[\int\text{d}t \, \bar{c}\left( \partial_{t} +  V_{,xx} \right)c \right]\nonumber \\
	&\propto& \text{exp}\left[\dfrac{1}{2}\int \text{d}t~V_{,xx}\right] \label{eq: ccbar PI-Ap}
\end{eqnarray}
when the boundary conditions 
\begin{equation}  
	c(t_{\rm in})=0\,,\quad \bar{c}(t_{\rm f})=0
\end{equation}
are chosen. We first work with the more general condition 
\begin{equation}
	c(t_{\rm in}) = e^{-i\nu} c(t_{\rm f})\,,\quad 	e^{-i\nu}\bar{c}(t_{\rm in}) = \bar{c}(t_{\rm f})
\end{equation}
From here onwards we will be working in the time interval $t\in \left[0,T\right]$ to simplify notation. Note that such a boundary condition ensures that 
\begin{equation}
\int\limits_0^T dt \dfrac{d}{dt}\left(\bar{c}c\right) =0 
\end{equation}
allowing us to write 
\begin{equation}\label{eq:app-det-adjoint}
	\text{det } \textbf{M} = \int \mathcal{D}c\mathcal{D}\bar{c}\text{ exp}\left[\frac{1}{2}\int\limits_0^T\text{d}t \, \left(\begin{smallmatrix}c,\bar{c}\end{smallmatrix}\right)\left(\begin{smallmatrix}
		0 & -F^{\dagger}\\ F & 0\end{smallmatrix}\right)\left( \begin{smallmatrix} c\\ \bar{c}\end{smallmatrix} \right)\right]
\end{equation}
where
\begin{equation}
	F= \partial_{t} +  V_{,xx}\,,\quad F^\dagger= - \partial_{t} +  V_{,xx} 
\end{equation}
Therefore the operator in the exponent of (\ref{eq:app-det-adjoint}) is self-adjoint and the path integral is properly defined. The eigenfunctions and eigenvalues of $F$ and $F^{\dagger}$ 
\begin{equation}
	Fu_n=\alpha_n u_n \,,\quad  F^{\dagger}v_n=\alpha_n v_n
\end{equation}  
are
\begin{equation}\label{eq:eigenvalues}
	\alpha_n = \frac{i\left(2\pi n +\nu\right)}{T} + \frac{1}{T} \int\limits_0^T d\tau V_{,xx} 
\end{equation}
$n=0,\pm 1 , \pm 2 , \ldots$ and 
\begin{eqnarray}
	u_n(t) &=& \frac{1}{\sqrt{T}}\exp\left\{ \int\limits_0^t  d\tau \left(\alpha_n - V_{,xx} \right)\right\}\\
	v_n(t) &=& \frac{1}{\sqrt{T}}\exp\left\{ - \int\limits_0^t  d\tau \left(\alpha_n - V_{,xx} \right)\right\}
\end{eqnarray}
They form an orthonormal and complete set   
\begin{equation}
	\int\limits_0^T dt \,\, u_m(t)v_n(t)=\delta_{mn}
\end{equation}
\begin{equation}
	\sum\limits_{n=-\infty}^{\infty} u_n(t)v_n(t')=\delta(t-t')
\end{equation}
Expanding $c$ in $u_n$ and $\bar{c}$ in $v_n$ with (Grassmann) coefficients $b_n$ and $\bar{b}_n$ respectively, we can represent the determinant as 
\begin{eqnarray}
	\text{det } \textbf{M} &=& \int \prod\limits_{n=-\infty}^{\infty} db_n d\bar{b}_n \exp\left\{ \sum\limits_{n=-\infty}^{\infty} \alpha_n \bar{b}_n b_n \right\}\nonumber \\
	 &=& \prod\limits_{n=-\infty}^{\infty}  \alpha_n
\end{eqnarray}
Using (\ref{eq:eigenvalues}) we have 
\begin{eqnarray}
	\prod\limits_{n=-\infty}^{\infty}  \alpha_n &=& \left\{\prod\limits_{n=-\infty}^{\infty}  \left(\frac{2\pi n}{T} i \right)\right\} \frac{i\nu+\int\limits_0^T \! dt\, V_{,xx}}{T}\nonumber \\
	&&\times\prod\limits_{n=1}^{\infty}\left[1+\frac{\left(i\nu+\int\limits_0^T \! dt \, V_{,xx}\right)^2}{4\pi^2n^2}\right]
\end{eqnarray}
which gives 
\begin{equation}
	\prod\limits_{n=-\infty}^{\infty}  \alpha_n = \frac{2\prod\limits_{n=-\infty}^{\infty}  \left(\frac{2\pi n}{T} i \right)}{T} \sinh\left[\frac{1}{2}\left(i\nu+\int\limits_0^T \! dt \, V_{,xx}\right)\right]
\end{equation}
With the infinite constant absorbed in the definition of the path integral measure, the choice $\nu\rightarrow -i\infty$ gives the required result $\ref{eq:det-Alg}$. Note that other, perhaps more obvious choices, e.g. periodic ($\nu=0$) or anti-periodic ($\nu=\pi$) boundary conditions do not reproduce the determinant which corresponds to the causal stochastic problem.

\bibliography{library}

\begin{thebibliography}{55}%
\makeatletter
\providecommand \@ifxundefined [1]{%
 \@ifx{#1\undefined}
}%
\providecommand \@ifnum [1]{%
 \ifnum #1\expandafter \@firstoftwo
 \else \expandafter \@secondoftwo
 \fi
}%
\providecommand \@ifx [1]{%
 \ifx #1\expandafter \@firstoftwo
 \else \expandafter \@secondoftwo
 \fi
}%
\providecommand \natexlab [1]{#1}%
\providecommand \enquote  [1]{``#1''}%
\providecommand \bibnamefont  [1]{#1}%
\providecommand \bibfnamefont [1]{#1}%
\providecommand \citenamefont [1]{#1}%
\providecommand \href@noop [0]{\@secondoftwo}%
\providecommand \href [0]{\begingroup \@sanitize@url \@href}%
\providecommand \@href[1]{\@@startlink{#1}\@@href}%
\providecommand \@@href[1]{\endgroup#1\@@endlink}%
\providecommand \@sanitize@url [0]{\catcode `\\12\catcode `\$12\catcode
  `\&12\catcode `\#12\catcode `\^12\catcode `\_12\catcode `\%12\relax}%
\providecommand \@@startlink[1]{}%
\providecommand \@@endlink[0]{}%
\providecommand \url  [0]{\begingroup\@sanitize@url \@url }%
\providecommand \@url [1]{\endgroup\@href {#1}{\urlprefix }}%
\providecommand \urlprefix  [0]{URL }%
\providecommand \Eprint [0]{\href }%
\providecommand \doibase [0]{https://doi.org/}%
\providecommand \selectlanguage [0]{\@gobble}%
\providecommand \bibinfo  [0]{\@secondoftwo}%
\providecommand \bibfield  [0]{\@secondoftwo}%
\providecommand \translation [1]{[#1]}%
\providecommand \BibitemOpen [0]{}%
\providecommand \bibitemStop [0]{}%
\providecommand \bibitemNoStop [0]{.\EOS\space}%
\providecommand \EOS [0]{\spacefactor3000\relax}%
\providecommand \BibitemShut  [1]{\csname bibitem#1\endcsname}%
\let\auto@bib@innerbib\@empty
\bibitem [{\citenamefont {{Van Kampen}}(2007)}]{VanKampen2007}%
  \BibitemOpen
  \bibfield  {author} {\bibinfo {author} {\bibfnamefont {N.~G.}\ \bibnamefont
  {{Van Kampen}}},\ }\href {https://doi.org/10.1016/B978-0-444-52965-7.X5000-4}
  {\emph {\bibinfo {title} {{Stochastic Processes in Physics and Chemistry}}}}\
  (\bibinfo  {publisher} {Elsevier},\ \bibinfo {year} {2007})\BibitemShut
  {NoStop}%
\bibitem [{\citenamefont {Gardiner}(2009)}]{gardiner2009stochastic}%
  \BibitemOpen
  \bibfield  {author} {\bibinfo {author} {\bibfnamefont {C.}~\bibnamefont
  {Gardiner}},\ }\href@noop {} {\emph {\bibinfo {title} {{Stochastic
  methods}}}},\ Vol.~\bibinfo {volume} {4}\ (\bibinfo {year}
  {2009})\BibitemShut {NoStop}%
\bibitem [{\citenamefont {Starobinsky}\ and\ \citenamefont
  {Yokoyama}(1994)}]{Starobinsky1994}%
  \BibitemOpen
  \bibfield  {author} {\bibinfo {author} {\bibfnamefont {A.~A.}\ \bibnamefont
  {Starobinsky}}\ and\ \bibinfo {author} {\bibfnamefont {J.}~\bibnamefont
  {Yokoyama}},\ }\bibfield  {title} {\bibinfo {title} {{Equilibrium state of a
  self-interacting scalar field in the de Sitter background}},\ }\href
  {https://doi.org/10.1103/PhysRevD.50.6357} {\bibfield  {journal} {\bibinfo
  {journal} {Physical Review D}\ }\textbf {\bibinfo {volume} {50}},\ \bibinfo
  {pages} {6357} (\bibinfo {year} {1994})}\BibitemShut {NoStop}%
\bibitem [{\citenamefont {Wetterich}(1993)}]{Wetterich1993}%
  \BibitemOpen
  \bibfield  {author} {\bibinfo {author} {\bibfnamefont {C.}~\bibnamefont
  {Wetterich}},\ }\bibfield  {title} {\bibinfo {title} {{Exact evolution
  equation for the effective potential}},\ }\href
  {https://doi.org/10.1016/0370-2693(93)90726-X} {\bibfield  {journal}
  {\bibinfo  {journal} {Physics Letters B}\ }\textbf {\bibinfo {volume}
  {301}},\ \bibinfo {pages} {90} (\bibinfo {year} {1993})},\ \Eprint
  {https://arxiv.org/abs/1710.05815v1} {arXiv:1710.05815v1} \BibitemShut
  {NoStop}%
\bibitem [{\citenamefont {Morris}(1994)}]{Morris1994}%
  \BibitemOpen
  \bibfield  {author} {\bibinfo {author} {\bibfnamefont {T.~R.}\ \bibnamefont
  {Morris}},\ }\bibfield  {title} {\bibinfo {title} {{The exact Renormalization
  group and approximate solutions}},\ }\href
  {https://doi.org/10.1142/S0217751X94000972} {\bibfield  {journal} {\bibinfo
  {journal} {International Journal of Modern Physics A}\ }\textbf {\bibinfo
  {volume} {09}},\ \bibinfo {pages} {2411} (\bibinfo {year}
  {1994})}\BibitemShut {NoStop}%
\bibitem [{\citenamefont {Berges}\ \emph {et~al.}(2002)\citenamefont {Berges},
  \citenamefont {Tetradis},\ and\ \citenamefont {Wetterich}}]{Berges2002}%
  \BibitemOpen
  \bibfield  {author} {\bibinfo {author} {\bibfnamefont {J.}~\bibnamefont
  {Berges}}, \bibinfo {author} {\bibfnamefont {N.}~\bibnamefont {Tetradis}},\
  and\ \bibinfo {author} {\bibfnamefont {C.}~\bibnamefont {Wetterich}},\
  }\bibfield  {title} {\bibinfo {title} {{Non-perturbative renormalization flow
  in quantum field theory and statistical physics}},\ }\href
  {https://doi.org/10.1016/S0370-1573(01)00098-9} {\bibfield  {journal}
  {\bibinfo  {journal} {Physics Report}\ }\textbf {\bibinfo {volume} {363}},\
  \bibinfo {pages} {223} (\bibinfo {year} {2002})},\ \Eprint
  {https://arxiv.org/abs/0005122} {arXiv:0005122 [hep-ph]} \BibitemShut
  {NoStop}%
\bibitem [{\citenamefont {Dupuis}\ \emph {et~al.}(2020)\citenamefont {Dupuis},
  \citenamefont {Canet}, \citenamefont {Eichhorn}, \citenamefont {Metzner},
  \citenamefont {Pawlowski}, \citenamefont {Tissier},\ and\ \citenamefont
  {Wschebor}}]{Dupuis2020}%
  \BibitemOpen
  \bibfield  {author} {\bibinfo {author} {\bibfnamefont {N.}~\bibnamefont
  {Dupuis}}, \bibinfo {author} {\bibfnamefont {L.}~\bibnamefont {Canet}},
  \bibinfo {author} {\bibfnamefont {A.}~\bibnamefont {Eichhorn}}, \bibinfo
  {author} {\bibfnamefont {W.}~\bibnamefont {Metzner}}, \bibinfo {author}
  {\bibfnamefont {J.~M.}\ \bibnamefont {Pawlowski}}, \bibinfo {author}
  {\bibfnamefont {M.}~\bibnamefont {Tissier}},\ and\ \bibinfo {author}
  {\bibfnamefont {N.}~\bibnamefont {Wschebor}},\ }\bibfield  {title} {\bibinfo
  {title} {{The nonperturbative functional renormalization group and its
  applications}},\ }\href {http://arxiv.org/abs/2006.04853} {\bibfield
  {journal} {\bibinfo  {journal} {arXiv}\ ,\ \bibinfo {pages} {1}} (\bibinfo
  {year} {2020})},\ \Eprint {https://arxiv.org/abs/2006.04853}
  {arXiv:2006.04853} \BibitemShut {NoStop}%
\bibitem [{\citenamefont {Gies}(2012)}]{Gies2012}%
  \BibitemOpen
  \bibfield  {author} {\bibinfo {author} {\bibfnamefont {H.}~\bibnamefont
  {Gies}},\ }\bibfield  {title} {\bibinfo {title} {{Introduction to the
  functional RG and applications to gauge theories}},\ }\href
  {https://doi.org/10.1007/978-3-642-27320-9_6} {\bibfield  {journal} {\bibinfo
   {journal} {Lecture Notes in Physics}\ }\textbf {\bibinfo {volume} {852}},\
  \bibinfo {pages} {287} (\bibinfo {year} {2012})},\ \Eprint
  {https://arxiv.org/abs/0611146} {arXiv:0611146 [hep-ph]} \BibitemShut
  {NoStop}%
\bibitem [{\citenamefont {Delamotte}(2012)}]{Delamotte2012}%
  \BibitemOpen
  \bibfield  {author} {\bibinfo {author} {\bibfnamefont {B.}~\bibnamefont
  {Delamotte}},\ }\bibfield  {title} {\bibinfo {title} {{An introduction to the
  nonperturbative renormalization group}},\ }\href
  {https://doi.org/10.1007/978-3-642-27320-9_2} {\bibfield  {journal} {\bibinfo
   {journal} {Lecture Notes in Physics}\ }\textbf {\bibinfo {volume} {852}},\
  \bibinfo {pages} {49} (\bibinfo {year} {2012})},\ \Eprint
  {https://arxiv.org/abs/0702365} {arXiv:0702365 [cond-mat]} \BibitemShut
  {NoStop}%
\bibitem [{\citenamefont {Wilson}(1983)}]{Wilson1983}%
  \BibitemOpen
  \bibfield  {author} {\bibinfo {author} {\bibfnamefont {K.~G.}\ \bibnamefont
  {Wilson}},\ }\bibfield  {title} {\bibinfo {title} {{The renormalization group
  and critical phenomena}},\ }\href {https://doi.org/10.1103/RevModPhys.55.583}
  {\bibfield  {journal} {\bibinfo  {journal} {Reviews of Modern Physics}\
  }\textbf {\bibinfo {volume} {55}},\ \bibinfo {pages} {583} (\bibinfo {year}
  {1983})}\BibitemShut {NoStop}%
\bibitem [{\citenamefont {Peskin}\ and\ \citenamefont
  {Schroeder}(1995)}]{Peskin:1995ev}%
  \BibitemOpen
  \bibfield  {author} {\bibinfo {author} {\bibfnamefont {M.~E.}\ \bibnamefont
  {Peskin}}\ and\ \bibinfo {author} {\bibfnamefont {D.~V.}\ \bibnamefont
  {Schroeder}},\ }\href@noop {} {\emph {\bibinfo {title} {{An Introduction to
  quantum field theory}}}}\ (\bibinfo  {publisher} {Addison-Wesley},\ \bibinfo
  {address} {Reading, USA},\ \bibinfo {year} {1995})\BibitemShut {NoStop}%
\bibitem [{\citenamefont {Chaikin}\ and\ \citenamefont
  {Lubensky}(1995)}]{Chaikin1995}%
  \BibitemOpen
  \bibfield  {author} {\bibinfo {author} {\bibfnamefont {P.~M.}\ \bibnamefont
  {Chaikin}}\ and\ \bibinfo {author} {\bibfnamefont {T.~C.}\ \bibnamefont
  {Lubensky}},\ }\href {https://doi.org/10.1017/CBO9780511813467} {\emph
  {\bibinfo {title} {{Principles of Condensed Matter Physics}}}}\ (\bibinfo
  {publisher} {Cambridge University Press},\ \bibinfo {year}
  {1995})\BibitemShut {NoStop}%
\bibitem [{\citenamefont {Vasil'ev}(2004)}]{Vasiliev2004}%
  \BibitemOpen
  \bibfield  {author} {\bibinfo {author} {\bibfnamefont {A.}~\bibnamefont
  {Vasil'ev}},\ }\href@noop {} {\emph {\bibinfo {title} {{The Field Theoretic
  Renormalization Group in Critical Behavior Theory and Stochastic
  Dynamics}}}}\ (\bibinfo  {publisher} {Chapman and Hall/CRC},\ \bibinfo {year}
  {2004})\BibitemShut {NoStop}%
\bibitem [{\citenamefont {Parisi}\ and\ \citenamefont
  {Sourlas}(1982)}]{Parisi:1982ud}%
  \BibitemOpen
  \bibfield  {author} {\bibinfo {author} {\bibfnamefont {G.}~\bibnamefont
  {Parisi}}\ and\ \bibinfo {author} {\bibfnamefont {N.}~\bibnamefont
  {Sourlas}},\ }\bibfield  {title} {\bibinfo {title} {{Supersymmetric Field
  Theories and Stochastic Differential Equations}},\ }\href
  {https://doi.org/10.1016/0550-3213(82)90538-7} {\bibfield  {journal}
  {\bibinfo  {journal} {Nucl. Phys. B}\ }\textbf {\bibinfo {volume} {206}},\
  \bibinfo {pages} {321} (\bibinfo {year} {1982})}\BibitemShut {NoStop}%
\bibitem [{\citenamefont {Zinn-Justin}(2002)}]{zinn2002quantum}%
  \BibitemOpen
  \bibfield  {author} {\bibinfo {author} {\bibfnamefont {J.}~\bibnamefont
  {Zinn-Justin}},\ }\href {https://books.google.co.uk/books?id=N8DBpTzBCJYC}
  {\emph {\bibinfo {title} {{Quantum Field Theory and Critical Phenomena}}}},\
  International series of monographs on physics\ (\bibinfo  {publisher}
  {Clarendon Press},\ \bibinfo {year} {2002})\BibitemShut {NoStop}%
\bibitem [{\citenamefont {Synatschke}\ \emph {et~al.}(2009)\citenamefont
  {Synatschke}, \citenamefont {Bergner}, \citenamefont {Gies},\ and\
  \citenamefont {Wipf}}]{Synatschke2009}%
  \BibitemOpen
  \bibfield  {author} {\bibinfo {author} {\bibfnamefont {F.}~\bibnamefont
  {Synatschke}}, \bibinfo {author} {\bibfnamefont {G.}~\bibnamefont {Bergner}},
  \bibinfo {author} {\bibfnamefont {H.}~\bibnamefont {Gies}},\ and\ \bibinfo
  {author} {\bibfnamefont {A.}~\bibnamefont {Wipf}},\ }\bibfield  {title}
  {\bibinfo {title} {{Flow equation for supersymmetric quantum mechanics}},\
  }\href {https://doi.org/10.1088/1126-6708/2009/03/028} {\bibfield  {journal}
  {\bibinfo  {journal} {Journal of High Energy Physics}\ }\textbf {\bibinfo
  {volume} {2009}},\ \bibinfo {pages} {028} (\bibinfo {year} {2009})},\ \Eprint
  {https://arxiv.org/abs/0809.4396} {arXiv:0809.4396} \BibitemShut {NoStop}%
\bibitem [{\citenamefont {Vasiliev}(1998)}]{Vasiliev1998}%
  \BibitemOpen
  \bibfield  {author} {\bibinfo {author} {\bibfnamefont {A.}~\bibnamefont
  {Vasiliev}},\ }\href@noop {} {\emph {\bibinfo {title} {{Functional Methods in
  Quantum Field Theory and Statistical Physics}}}}\ (\bibinfo  {publisher} {CRC
  Press},\ \bibinfo {year} {1998})\BibitemShut {NoStop}%
\bibitem [{\citenamefont {{De Dominicis}}(1976)}]{DEDOMINICIS1976}%
  \BibitemOpen
  \bibfield  {author} {\bibinfo {author} {\bibfnamefont {C.}~\bibnamefont {{De
  Dominicis}}},\ }\bibfield  {title} {\bibinfo {title} {{Techniques de
  Renormalisation de la Th{\'{e}}orie des Champs Et Dynamique Des
  Ph{\'{e}}nom{\`{e}}nes Critiques}},\ }\href
  {https://doi.org/10.1051/jphyscol:1976138} {\bibfield  {journal} {\bibinfo
  {journal} {Le Journal de Physique Colloques}\ }\textbf {\bibinfo {volume}
  {37}},\ \bibinfo {pages} {C1} (\bibinfo {year} {1976})}\BibitemShut {NoStop}%
\bibitem [{\citenamefont {Janssen}(1976)}]{Janssen1976}%
  \BibitemOpen
  \bibfield  {author} {\bibinfo {author} {\bibfnamefont {H.-K.}\ \bibnamefont
  {Janssen}},\ }\bibfield  {title} {\bibinfo {title} {{On a Lagrangean for
  classical field dynamics and renormalization group calculations of dynamical
  critical properties}},\ }\href {https://doi.org/10.1007/BF01316547}
  {\bibfield  {journal} {\bibinfo  {journal} {Zeitschrift f\"{u}r Physik B
  Condensed Matter and Quanta}\ }\textbf {\bibinfo {volume} {23}},\ \bibinfo
  {pages} {377} (\bibinfo {year} {1976})}\BibitemShut {NoStop}%
\bibitem [{\citenamefont {{De Dominicis}}\ and\ \citenamefont
  {Peliti}(1978)}]{DeDominicis1978}%
  \BibitemOpen
  \bibfield  {author} {\bibinfo {author} {\bibfnamefont {C.}~\bibnamefont {{De
  Dominicis}}}\ and\ \bibinfo {author} {\bibfnamefont {L.}~\bibnamefont
  {Peliti}},\ }\bibfield  {title} {\bibinfo {title} {{Field-theory
  renormalization and critical dynamics above $T_{c}$ : Helium,
  antiferromagnets, and liquid-gas systems}},\ }\href
  {https://doi.org/10.1103/PhysRevB.18.353} {\bibfield  {journal} {\bibinfo
  {journal} {Physical Review B}\ }\textbf {\bibinfo {volume} {18}},\ \bibinfo
  {pages} {353} (\bibinfo {year} {1978})}\BibitemShut {NoStop}%
\bibitem [{\citenamefont {Martin}\ \emph {et~al.}(1973)\citenamefont {Martin},
  \citenamefont {Siggia},\ and\ \citenamefont {Rose}}]{Martin1973}%
  \BibitemOpen
  \bibfield  {author} {\bibinfo {author} {\bibfnamefont {P.~C.}\ \bibnamefont
  {Martin}}, \bibinfo {author} {\bibfnamefont {E.~D.}\ \bibnamefont {Siggia}},\
  and\ \bibinfo {author} {\bibfnamefont {H.~A.}\ \bibnamefont {Rose}},\
  }\bibfield  {title} {\bibinfo {title} {{Statistical Dynamics of Classical
  Systems}},\ }\href {https://doi.org/10.1103/PhysRevA.8.423} {\bibfield
  {journal} {\bibinfo  {journal} {Physical Review A}\ }\textbf {\bibinfo
  {volume} {8}},\ \bibinfo {pages} {423} (\bibinfo {year} {1973})}\BibitemShut
  {NoStop}%
\bibitem [{\citenamefont {Lau}\ and\ \citenamefont {Lubensky}(2007)}]{Lau2007}%
  \BibitemOpen
  \bibfield  {author} {\bibinfo {author} {\bibfnamefont {A.~W.}\ \bibnamefont
  {Lau}}\ and\ \bibinfo {author} {\bibfnamefont {T.~C.}\ \bibnamefont
  {Lubensky}},\ }\bibfield  {title} {\bibinfo {title} {{State-dependent
  diffusion: Thermodynamic consistency and its path integral formulation}},\
  }\bibfield  {journal} {\bibinfo  {journal} {Physical Review E - Statistical,
  Nonlinear, and Soft Matter Physics}\ }\textbf {\bibinfo {volume} {76}},\
  \href {https://doi.org/10.1103/PhysRevE.76.011123}
  {10.1103/PhysRevE.76.011123} (\bibinfo {year} {2007}),\ \Eprint
  {https://arxiv.org/abs/0707.2234} {arXiv:0707.2234} \BibitemShut {NoStop}%
\bibitem [{\citenamefont {Hertz}\ \emph {et~al.}(2017)\citenamefont {Hertz},
  \citenamefont {Roudi},\ and\ \citenamefont {Sollich}}]{Hertz2017}%
  \BibitemOpen
  \bibfield  {author} {\bibinfo {author} {\bibfnamefont {J.~A.}\ \bibnamefont
  {Hertz}}, \bibinfo {author} {\bibfnamefont {Y.}~\bibnamefont {Roudi}},\ and\
  \bibinfo {author} {\bibfnamefont {P.}~\bibnamefont {Sollich}},\ }\bibfield
  {title} {\bibinfo {title} {{Path integral methods for the dynamics of
  stochastic and disordered systems}},\ }\bibfield  {journal} {\bibinfo
  {journal} {Journal of Physics A: Mathematical and Theoretical}\ }\textbf
  {\bibinfo {volume} {50}},\ \href
  {https://doi.org/10.1088/1751-8121/50/3/033001}
  {10.1088/1751-8121/50/3/033001} (\bibinfo {year} {2017}),\ \Eprint
  {https://arxiv.org/abs/1604.05775} {arXiv:1604.05775} \BibitemShut {NoStop}%
\bibitem [{\citenamefont {Markkanen}\ \emph {et~al.}(2019)\citenamefont
  {Markkanen}, \citenamefont {Rajantie}, \citenamefont {Stopyra},\ and\
  \citenamefont {Tenkanen}}]{Markkanen:2019kpv}%
  \BibitemOpen
  \bibfield  {author} {\bibinfo {author} {\bibfnamefont {T.}~\bibnamefont
  {Markkanen}}, \bibinfo {author} {\bibfnamefont {A.}~\bibnamefont {Rajantie}},
  \bibinfo {author} {\bibfnamefont {S.}~\bibnamefont {Stopyra}},\ and\ \bibinfo
  {author} {\bibfnamefont {T.}~\bibnamefont {Tenkanen}},\ }\bibfield  {title}
  {\bibinfo {title} {{Scalar correlation functions in de Sitter space from the
  stochastic spectral expansion}},\ }\href
  {https://doi.org/10.1088/1475-7516/2019/08/001} {\bibfield  {journal}
  {\bibinfo  {journal} {JCAP}\ }\textbf {\bibinfo {volume} {08}},\ \bibinfo
  {pages} {001}},\ \Eprint {https://arxiv.org/abs/1904.11917} {arXiv:1904.11917
  [gr-qc]} \BibitemShut {NoStop}%
\bibitem [{\citenamefont {Leschke}\ and\ \citenamefont
  {Schmutz}(1977)}]{Leschke1977}%
  \BibitemOpen
  \bibfield  {author} {\bibinfo {author} {\bibfnamefont {H.}~\bibnamefont
  {Leschke}}\ and\ \bibinfo {author} {\bibfnamefont {M.}~\bibnamefont
  {Schmutz}},\ }\bibfield  {title} {\bibinfo {title} {{Operator orderings and
  functional formulations of quantum and stochastic dynamics}},\ }\href
  {https://doi.org/10.1007/BF01315509} {\bibfield  {journal} {\bibinfo
  {journal} {Zeitschrift f\"{u}r Physik B: Condensed Matter and Quanta}\
  }\textbf {\bibinfo {volume} {27}},\ \bibinfo {pages} {85} (\bibinfo {year}
  {1977})}\BibitemShut {NoStop}%
\bibitem [{\citenamefont {Mallick}\ \emph {et~al.}(2011)\citenamefont
  {Mallick}, \citenamefont {Moshe},\ and\ \citenamefont
  {Orland}}]{Mallick2011}%
  \BibitemOpen
  \bibfield  {author} {\bibinfo {author} {\bibfnamefont {K.}~\bibnamefont
  {Mallick}}, \bibinfo {author} {\bibfnamefont {M.}~\bibnamefont {Moshe}},\
  and\ \bibinfo {author} {\bibfnamefont {H.}~\bibnamefont {Orland}},\
  }\bibfield  {title} {\bibinfo {title} {{A field-theoretic approach to
  non-equilibrium work identities}},\ }\bibfield  {journal} {\bibinfo
  {journal} {Journal of Physics A: Mathematical and Theoretical}\ }\textbf
  {\bibinfo {volume} {44}},\ \href
  {https://doi.org/10.1088/1751-8113/44/9/095002}
  {10.1088/1751-8113/44/9/095002} (\bibinfo {year} {2011}),\ \Eprint
  {https://arxiv.org/abs/1009.4800} {arXiv:1009.4800} \BibitemShut {NoStop}%
\bibitem [{\citenamefont {Damgaard}\ and\ \citenamefont
  {H{\"{u}}ffel}(1987)}]{Damgaard1987}%
  \BibitemOpen
  \bibfield  {author} {\bibinfo {author} {\bibfnamefont {P.~H.}\ \bibnamefont
  {Damgaard}}\ and\ \bibinfo {author} {\bibfnamefont {H.}~\bibnamefont
  {H{\"{u}}ffel}},\ }\bibfield  {title} {\bibinfo {title} {{Stochastic
  quantization}},\ }\href {https://doi.org/10.1016/0370-1573(87)90144-X}
  {\bibfield  {journal} {\bibinfo  {journal} {Physics Reports}\ }\textbf
  {\bibinfo {volume} {152}},\ \bibinfo {pages} {227} (\bibinfo {year}
  {1987})}\BibitemShut {NoStop}%
\bibitem [{\citenamefont {Canet}\ \emph {et~al.}(2004)\citenamefont {Canet},
  \citenamefont {Delamotte}, \citenamefont {Deloubri{\`{e}}re},\ and\
  \citenamefont {Wschebor}}]{Canet2004}%
  \BibitemOpen
  \bibfield  {author} {\bibinfo {author} {\bibfnamefont {L.}~\bibnamefont
  {Canet}}, \bibinfo {author} {\bibfnamefont {B.}~\bibnamefont {Delamotte}},
  \bibinfo {author} {\bibfnamefont {O.}~\bibnamefont {Deloubri{\`{e}}re}},\
  and\ \bibinfo {author} {\bibfnamefont {N.}~\bibnamefont {Wschebor}},\
  }\bibfield  {title} {\bibinfo {title} {{Nonperturbative Renormalization-Group
  Study of Reaction-Diffusion Processes}},\ }\href
  {https://doi.org/10.1103/PhysRevLett.92.195703} {\bibfield  {journal}
  {\bibinfo  {journal} {Physical Review Letters}\ }\textbf {\bibinfo {volume}
  {92}},\ \bibinfo {pages} {195703} (\bibinfo {year} {2004})}\BibitemShut
  {NoStop}%
\bibitem [{\citenamefont {Gezzi}\ \emph {et~al.}(2007)\citenamefont {Gezzi},
  \citenamefont {Pruschke},\ and\ \citenamefont {Meden}}]{Gezzi2007}%
  \BibitemOpen
  \bibfield  {author} {\bibinfo {author} {\bibfnamefont {R.}~\bibnamefont
  {Gezzi}}, \bibinfo {author} {\bibfnamefont {T.}~\bibnamefont {Pruschke}},\
  and\ \bibinfo {author} {\bibfnamefont {V.}~\bibnamefont {Meden}},\ }\bibfield
   {title} {\bibinfo {title} {{Functional renormalization group for
  nonequilibrium quantum many-body problems}},\ }\href
  {https://doi.org/10.1103/PhysRevB.75.045324} {\bibfield  {journal} {\bibinfo
  {journal} {Physical Review B}\ }\textbf {\bibinfo {volume} {75}},\ \bibinfo
  {pages} {045324} (\bibinfo {year} {2007})}\BibitemShut {NoStop}%
\bibitem [{\citenamefont {Jakobs}\ \emph {et~al.}(2007)\citenamefont {Jakobs},
  \citenamefont {Meden},\ and\ \citenamefont {Schoeller}}]{Jakobs2007}%
  \BibitemOpen
  \bibfield  {author} {\bibinfo {author} {\bibfnamefont {S.~G.}\ \bibnamefont
  {Jakobs}}, \bibinfo {author} {\bibfnamefont {V.}~\bibnamefont {Meden}},\ and\
  \bibinfo {author} {\bibfnamefont {H.}~\bibnamefont {Schoeller}},\ }\bibfield
  {title} {\bibinfo {title} {{Nonequilibrium Functional Renormalization Group
  for Interacting Quantum Systems}},\ }\href
  {https://doi.org/10.1103/PhysRevLett.99.150603} {\bibfield  {journal}
  {\bibinfo  {journal} {Physical Review Letters}\ }\textbf {\bibinfo {volume}
  {99}},\ \bibinfo {pages} {150603} (\bibinfo {year} {2007})}\BibitemShut
  {NoStop}%
\bibitem [{\citenamefont {Gasenzer}\ and\ \citenamefont
  {Pawlowski}(2008)}]{Gasenzer2008}%
  \BibitemOpen
  \bibfield  {author} {\bibinfo {author} {\bibfnamefont {T.}~\bibnamefont
  {Gasenzer}}\ and\ \bibinfo {author} {\bibfnamefont {J.~M.}\ \bibnamefont
  {Pawlowski}},\ }\bibfield  {title} {\bibinfo {title} {{Towards
  far-from-equilibrium quantum field dynamics: A functional
  renormalisation-group approach}},\ }\href
  {https://doi.org/10.1016/j.physletb.2008.10.049} {\bibfield  {journal}
  {\bibinfo  {journal} {Physics Letters B}\ }\textbf {\bibinfo {volume}
  {670}},\ \bibinfo {pages} {135} (\bibinfo {year} {2008})}\BibitemShut
  {NoStop}%
\bibitem [{\citenamefont {Berges}\ and\ \citenamefont
  {Hoffmeister}(2009)}]{Berges2009}%
  \BibitemOpen
  \bibfield  {author} {\bibinfo {author} {\bibfnamefont {J.}~\bibnamefont
  {Berges}}\ and\ \bibinfo {author} {\bibfnamefont {G.}~\bibnamefont
  {Hoffmeister}},\ }\bibfield  {title} {\bibinfo {title} {{Nonthermal fixed
  points and the functional renormalization group}},\ }\href
  {https://doi.org/10.1016/j.nuclphysb.2008.12.017} {\bibfield  {journal}
  {\bibinfo  {journal} {Nuclear Physics B}\ }\textbf {\bibinfo {volume}
  {813}},\ \bibinfo {pages} {383} (\bibinfo {year} {2009})}\BibitemShut
  {NoStop}%
\bibitem [{\citenamefont {Gasenzer}\ \emph {et~al.}(2010)\citenamefont
  {Gasenzer}, \citenamefont {Ke{\ss}ler},\ and\ \citenamefont
  {Pawlowski}}]{Gasenzer2010}%
  \BibitemOpen
  \bibfield  {author} {\bibinfo {author} {\bibfnamefont {T.}~\bibnamefont
  {Gasenzer}}, \bibinfo {author} {\bibfnamefont {S.}~\bibnamefont
  {Ke{\ss}ler}},\ and\ \bibinfo {author} {\bibfnamefont {J.~M.}\ \bibnamefont
  {Pawlowski}},\ }\bibfield  {title} {\bibinfo {title} {{Far-from-equilibrium
  quantum many-body dynamics}},\ }\href
  {https://doi.org/10.1140/epjc/s10052-010-1430-3} {\bibfield  {journal}
  {\bibinfo  {journal} {The European Physical Journal C}\ }\textbf {\bibinfo
  {volume} {70}},\ \bibinfo {pages} {423} (\bibinfo {year} {2010})}\BibitemShut
  {NoStop}%
\bibitem [{\citenamefont {Kloss}\ and\ \citenamefont
  {Kopietz}(2011)}]{Kloss2011}%
  \BibitemOpen
  \bibfield  {author} {\bibinfo {author} {\bibfnamefont {T.}~\bibnamefont
  {Kloss}}\ and\ \bibinfo {author} {\bibfnamefont {P.}~\bibnamefont
  {Kopietz}},\ }\bibfield  {title} {\bibinfo {title} {{Nonequilibrium time
  evolution of bosons from the functional renormalization group}},\ }\href
  {https://doi.org/10.1103/PhysRevB.83.205118} {\bibfield  {journal} {\bibinfo
  {journal} {Physical Review B}\ }\textbf {\bibinfo {volume} {83}},\ \bibinfo
  {pages} {205118} (\bibinfo {year} {2011})}\BibitemShut {NoStop}%
\bibitem [{\citenamefont {Canet}\ \emph {et~al.}(2011)\citenamefont {Canet},
  \citenamefont {Chat{\'{e}}},\ and\ \citenamefont {Delamotte}}]{Canet2011}%
  \BibitemOpen
  \bibfield  {author} {\bibinfo {author} {\bibfnamefont {L.}~\bibnamefont
  {Canet}}, \bibinfo {author} {\bibfnamefont {H.}~\bibnamefont {Chat{\'{e}}}},\
  and\ \bibinfo {author} {\bibfnamefont {B.}~\bibnamefont {Delamotte}},\
  }\bibfield  {title} {\bibinfo {title} {{General framework of the
  non-perturbative renormalization group for non-equilibrium steady states}},\
  }\href {https://doi.org/10.1088/1751-8113/44/49/495001} {\bibfield  {journal}
  {\bibinfo  {journal} {Journal of Physics A: Mathematical and Theoretical}\
  }\textbf {\bibinfo {volume} {44}},\ \bibinfo {pages} {495001} (\bibinfo
  {year} {2011})},\ \Eprint {https://arxiv.org/abs/1106.4129} {arXiv:1106.4129}
  \BibitemShut {NoStop}%
\bibitem [{\citenamefont {Sieberer}\ \emph {et~al.}(2014)\citenamefont
  {Sieberer}, \citenamefont {Huber}, \citenamefont {Altman},\ and\
  \citenamefont {Diehl}}]{Sieberer2014}%
  \BibitemOpen
  \bibfield  {author} {\bibinfo {author} {\bibfnamefont {L.~M.}\ \bibnamefont
  {Sieberer}}, \bibinfo {author} {\bibfnamefont {S.~D.}\ \bibnamefont {Huber}},
  \bibinfo {author} {\bibfnamefont {E.}~\bibnamefont {Altman}},\ and\ \bibinfo
  {author} {\bibfnamefont {S.}~\bibnamefont {Diehl}},\ }\bibfield  {title}
  {\bibinfo {title} {{Nonequilibrium functional renormalization for
  driven-dissipative Bose-Einstein condensation}},\ }\href
  {https://doi.org/10.1103/PhysRevB.89.134310} {\bibfield  {journal} {\bibinfo
  {journal} {Physical Review B}\ }\textbf {\bibinfo {volume} {89}},\ \bibinfo
  {pages} {134310} (\bibinfo {year} {2014})}\BibitemShut {NoStop}%
\bibitem [{\citenamefont {Mukherjee}\ \emph {et~al.}(2015)\citenamefont
  {Mukherjee}, \citenamefont {Venugopalan},\ and\ \citenamefont
  {Yin}}]{Mukherjee2015}%
  \BibitemOpen
  \bibfield  {author} {\bibinfo {author} {\bibfnamefont {S.}~\bibnamefont
  {Mukherjee}}, \bibinfo {author} {\bibfnamefont {R.}~\bibnamefont
  {Venugopalan}},\ and\ \bibinfo {author} {\bibfnamefont {Y.}~\bibnamefont
  {Yin}},\ }\bibfield  {title} {\bibinfo {title} {{Real-time evolution of
  non-Gaussian cumulants in the QCD critical regime}},\ }\href
  {https://doi.org/10.1103/PhysRevC.92.034912} {\bibfield  {journal} {\bibinfo
  {journal} {Physical Review C}\ }\textbf {\bibinfo {volume} {92}},\ \bibinfo
  {pages} {034912} (\bibinfo {year} {2015})}\BibitemShut {NoStop}%
\bibitem [{\citenamefont {Pawlowski}\ and\ \citenamefont
  {Strodthoff}(2015)}]{Pawlowski2015}%
  \BibitemOpen
  \bibfield  {author} {\bibinfo {author} {\bibfnamefont {J.~M.}\ \bibnamefont
  {Pawlowski}}\ and\ \bibinfo {author} {\bibfnamefont {N.}~\bibnamefont
  {Strodthoff}},\ }\bibfield  {title} {\bibinfo {title} {{Real time correlation
  functions and the functional renormalization group}},\ }\href
  {https://doi.org/10.1103/PhysRevD.92.094009} {\bibfield  {journal} {\bibinfo
  {journal} {Physical Review D}\ }\textbf {\bibinfo {volume} {92}},\ \bibinfo
  {pages} {094009} (\bibinfo {year} {2015})}\BibitemShut {NoStop}%
\bibitem [{\citenamefont {Corell}\ \emph {et~al.}(2019)\citenamefont {Corell},
  \citenamefont {Cyrol}, \citenamefont {Heller},\ and\ \citenamefont
  {Pawlowski}}]{Corell2019}%
  \BibitemOpen
  \bibfield  {author} {\bibinfo {author} {\bibfnamefont {L.}~\bibnamefont
  {Corell}}, \bibinfo {author} {\bibfnamefont {A.~K.}\ \bibnamefont {Cyrol}},
  \bibinfo {author} {\bibfnamefont {M.}~\bibnamefont {Heller}},\ and\ \bibinfo
  {author} {\bibfnamefont {J.~M.}\ \bibnamefont {Pawlowski}},\ }\bibfield
  {title} {\bibinfo {title} {{Flowing with the Temporal Renormalisation
  Group}},\ }\href {http://arxiv.org/abs/1910.09369} {\  (\bibinfo {year}
  {2019})},\ \Eprint {https://arxiv.org/abs/1910.09369} {arXiv:1910.09369}
  \BibitemShut {NoStop}%
\bibitem [{\citenamefont {Duclut}\ and\ \citenamefont
  {Delamotte}(2017)}]{Duclut2017}%
  \BibitemOpen
  \bibfield  {author} {\bibinfo {author} {\bibfnamefont {C.}~\bibnamefont
  {Duclut}}\ and\ \bibinfo {author} {\bibfnamefont {B.}~\bibnamefont
  {Delamotte}},\ }\bibfield  {title} {\bibinfo {title} {{Frequency regulators
  for the nonperturbative renormalization group: A general study and the model
  A as a benchmark}},\ }\href {https://doi.org/10.1103/PhysRevE.95.012107}
  {\bibfield  {journal} {\bibinfo  {journal} {Physical Review E}\ }\textbf
  {\bibinfo {volume} {95}},\ \bibinfo {pages} {012107} (\bibinfo {year}
  {2017})},\ \Eprint {https://arxiv.org/abs/1611.07301v3} {arXiv:1611.07301v3}
  \BibitemShut {NoStop}%
\bibitem [{\citenamefont {Moreau}\ and\ \citenamefont
  {Serreau}(2020)}]{Moreau2020}%
  \BibitemOpen
  \bibfield  {author} {\bibinfo {author} {\bibfnamefont {G.}~\bibnamefont
  {Moreau}}\ and\ \bibinfo {author} {\bibfnamefont {J.}~\bibnamefont
  {Serreau}},\ }\bibfield  {title} {\bibinfo {title} {{Unequal time correlators
  of stochastic scalar fields in de Sitter space}},\ }\href
  {https://doi.org/10.1103/physrevd.101.045015} {\bibfield  {journal} {\bibinfo
   {journal} {Physical Review D}\ }\textbf {\bibinfo {volume} {101}},\ \bibinfo
  {pages} {1} (\bibinfo {year} {2020})},\ \Eprint
  {https://arxiv.org/abs/1912.05358} {arXiv:1912.05358} \BibitemShut {NoStop}%
\bibitem [{\citenamefont {Prokopec}\ and\ \citenamefont
  {Rigopoulos}(2018)}]{Prokopec2018}%
  \BibitemOpen
  \bibfield  {author} {\bibinfo {author} {\bibfnamefont {T.}~\bibnamefont
  {Prokopec}}\ and\ \bibinfo {author} {\bibfnamefont {G.}~\bibnamefont
  {Rigopoulos}},\ }\bibfield  {title} {\bibinfo {title} {{Functional
  renormalization group for stochastic inflation}},\ }\href
  {https://doi.org/10.1088/1475-7516/2018/08/013} {\bibfield  {journal}
  {\bibinfo  {journal} {Journal of Cosmology and Astroparticle Physics}\
  }\textbf {\bibinfo {volume} {2018}}\bibfield  {number} {\bibinfo  {number} {
  (8)}},\ }\Eprint {https://arxiv.org/abs/1710.07333} {arXiv:1710.07333}
  \BibitemShut {NoStop}%
\bibitem [{\citenamefont {Moreau}(2020)}]{Moreau2020a}%
  \BibitemOpen
  \bibfield  {author} {\bibinfo {author} {\bibfnamefont {G.}~\bibnamefont
  {Moreau}},\ }\emph {\bibinfo {title} {{Nonperturbative dynamics of quantum
  fields in de Sitter spacetime}}},\ \href@noop {} {Ph.D. thesis},\ \bibinfo
  {school} {APC Paris} (\bibinfo {year} {2020})\BibitemShut {NoStop}%
\bibitem [{\citenamefont {Guilleux}\ and\ \citenamefont
  {Serreau}(2015)}]{Guilleux2015}%
  \BibitemOpen
  \bibfield  {author} {\bibinfo {author} {\bibfnamefont {M.}~\bibnamefont
  {Guilleux}}\ and\ \bibinfo {author} {\bibfnamefont {J.}~\bibnamefont
  {Serreau}},\ }\bibfield  {title} {\bibinfo {title} {{Quantum scalar fields in
  de Sitter space from the nonperturbative renormalization group}},\ }\href
  {https://doi.org/10.1103/PhysRevD.92.084010} {\bibfield  {journal} {\bibinfo
  {journal} {Physical Review D}\ }\textbf {\bibinfo {volume} {92}},\ \bibinfo
  {pages} {084010} (\bibinfo {year} {2015})},\ \Eprint
  {https://arxiv.org/abs/1506.06183} {arXiv:1506.06183} \BibitemShut {NoStop}%
\bibitem [{\citenamefont {Guilleux}\ and\ \citenamefont
  {Serreau}(2017)}]{Guilleux2017}%
  \BibitemOpen
  \bibfield  {author} {\bibinfo {author} {\bibfnamefont {M.}~\bibnamefont
  {Guilleux}}\ and\ \bibinfo {author} {\bibfnamefont {J.}~\bibnamefont
  {Serreau}},\ }\bibfield  {title} {\bibinfo {title} {{Nonperturbative
  renormalization group for scalar fields in de Sitter space: Beyond the local
  potential approximation}},\ }\href
  {https://doi.org/10.1103/PhysRevD.95.045003} {\bibfield  {journal} {\bibinfo
  {journal} {Physical Review D}\ }\textbf {\bibinfo {volume} {95}},\ \bibinfo
  {pages} {045003} (\bibinfo {year} {2017})},\ \Eprint
  {https://arxiv.org/abs/1611.08106} {arXiv:1611.08106} \BibitemShut {NoStop}%
\bibitem [{\citenamefont {Markkanen}\ and\ \citenamefont
  {Rajantie}(2020)}]{Markkanen2020}%
  \BibitemOpen
  \bibfield  {author} {\bibinfo {author} {\bibfnamefont {T.}~\bibnamefont
  {Markkanen}}\ and\ \bibinfo {author} {\bibfnamefont {A.}~\bibnamefont
  {Rajantie}},\ }\bibfield  {title} {\bibinfo {title} {{Scalar correlation
  functions for a double-well potential in de Sitter space}},\ }\href
  {https://doi.org/10.1088/1475-7516/2020/03/049} {\bibfield  {journal}
  {\bibinfo  {journal} {Journal of Cosmology and Astroparticle Physics}\
  }\textbf {\bibinfo {volume} {2020}}\bibfield  {number} {\bibinfo  {number} {
  (03)},\ \bibinfo {pages} {049}},\ }\Eprint {https://arxiv.org/abs/2001.04494}
  {arXiv:2001.04494} \BibitemShut {NoStop}%
\bibitem [{\citenamefont {Balog}\ \emph {et~al.}(2019)\citenamefont {Balog},
  \citenamefont {Chat{\'{e}}}, \citenamefont {Delamotte}, \citenamefont
  {Marohni{\'{c}}},\ and\ \citenamefont {Wschebor}}]{Balog2019}%
  \BibitemOpen
  \bibfield  {author} {\bibinfo {author} {\bibfnamefont {I.}~\bibnamefont
  {Balog}}, \bibinfo {author} {\bibfnamefont {H.}~\bibnamefont {Chat{\'{e}}}},
  \bibinfo {author} {\bibfnamefont {B.}~\bibnamefont {Delamotte}}, \bibinfo
  {author} {\bibfnamefont {M.}~\bibnamefont {Marohni{\'{c}}}},\ and\ \bibinfo
  {author} {\bibfnamefont {N.}~\bibnamefont {Wschebor}},\ }\bibfield  {title}
  {\bibinfo {title} {{Convergence of Nonperturbative Approximations to the
  Renormalization Group}},\ }\href
  {https://doi.org/10.1103/PhysRevLett.123.240604} {\bibfield  {journal}
  {\bibinfo  {journal} {Physical Review Letters}\ }\textbf {\bibinfo {volume}
  {123}},\ \bibinfo {pages} {1} (\bibinfo {year} {2019})},\ \Eprint
  {https://arxiv.org/abs/1907.01829} {arXiv:1907.01829} \BibitemShut {NoStop}%
\bibitem [{\citenamefont {{De Polsi}}\ \emph {et~al.}(2020)\citenamefont {{De
  Polsi}}, \citenamefont {Balog}, \citenamefont {Tissier},\ and\ \citenamefont
  {Wschebor}}]{DePolsi2020}%
  \BibitemOpen
  \bibfield  {author} {\bibinfo {author} {\bibfnamefont {G.}~\bibnamefont {{De
  Polsi}}}, \bibinfo {author} {\bibfnamefont {I.}~\bibnamefont {Balog}},
  \bibinfo {author} {\bibfnamefont {M.}~\bibnamefont {Tissier}},\ and\ \bibinfo
  {author} {\bibfnamefont {N.}~\bibnamefont {Wschebor}},\ }\bibfield  {title}
  {\bibinfo {title} {{Precision calculation of critical exponents in the O(N)
  universality classes with the nonperturbative renormalization group}},\
  }\href {https://doi.org/10.1103/PhysRevE.101.042113} {\bibfield  {journal}
  {\bibinfo  {journal} {Physical Review E}\ }\textbf {\bibinfo {volume}
  {101}},\ \bibinfo {pages} {042113} (\bibinfo {year} {2020})}\BibitemShut
  {NoStop}%
\bibitem [{\citenamefont {Strumia}\ and\ \citenamefont
  {Tetradis}(1999{\natexlab{a}})}]{Strumia1999}%
  \BibitemOpen
  \bibfield  {author} {\bibinfo {author} {\bibfnamefont {A.}~\bibnamefont
  {Strumia}}\ and\ \bibinfo {author} {\bibfnamefont {N.}~\bibnamefont
  {Tetradis}},\ }\bibfield  {title} {\bibinfo {title} {{A consistent
  calculation of bubble-nucleation rates}},\ }\href
  {https://doi.org/10.1016/S0550-3213(98)00804-9} {\bibfield  {journal}
  {\bibinfo  {journal} {Nuclear Physics B}\ }\textbf {\bibinfo {volume}
  {542}},\ \bibinfo {pages} {719} (\bibinfo {year}
  {1999}{\natexlab{a}})}\BibitemShut {NoStop}%
\bibitem [{\citenamefont {Strumia}\ and\ \citenamefont
  {Tetradis}(1999{\natexlab{b}})}]{Strumia1999b}%
  \BibitemOpen
  \bibfield  {author} {\bibinfo {author} {\bibfnamefont {A.}~\bibnamefont
  {Strumia}}\ and\ \bibinfo {author} {\bibfnamefont {N.}~\bibnamefont
  {Tetradis}},\ }\bibfield  {title} {\bibinfo {title} {{Bubble-nucleation rates
  for radiatively induced first-order phase transitions}},\ }\href
  {https://doi.org/10.1016/S0550-3213(99)00285-0} {\bibfield  {journal}
  {\bibinfo  {journal} {Nuclear Physics B}\ }\textbf {\bibinfo {volume}
  {554}},\ \bibinfo {pages} {697} (\bibinfo {year}
  {1999}{\natexlab{b}})}\BibitemShut {NoStop}%
\bibitem [{\citenamefont {Strumia}\ \emph {et~al.}(1999)\citenamefont
  {Strumia}, \citenamefont {Tetradis},\ and\ \citenamefont
  {Wetterich}}]{Strumia1999a}%
  \BibitemOpen
  \bibfield  {author} {\bibinfo {author} {\bibfnamefont {A.}~\bibnamefont
  {Strumia}}, \bibinfo {author} {\bibfnamefont {N.}~\bibnamefont {Tetradis}},\
  and\ \bibinfo {author} {\bibfnamefont {C.}~\bibnamefont {Wetterich}},\
  }\bibfield  {title} {\bibinfo {title} {{The region of validity of homogeneous
  nucleation theory}},\ }\href {https://doi.org/10.1016/S0370-2693(99)01158-2}
  {\bibfield  {journal} {\bibinfo  {journal} {Physics Letters B}\ }\textbf
  {\bibinfo {volume} {467}},\ \bibinfo {pages} {279} (\bibinfo {year}
  {1999})}\BibitemShut {NoStop}%
\bibitem [{\citenamefont {Yip}\ and\ \citenamefont {Short}(2013)}]{Yip2013}%
  \BibitemOpen
  \bibfield  {author} {\bibinfo {author} {\bibfnamefont {S.}~\bibnamefont
  {Yip}}\ and\ \bibinfo {author} {\bibfnamefont {M.~P.}\ \bibnamefont
  {Short}},\ }\bibfield  {title} {\bibinfo {title} {{Multiscale materials
  modelling at the mesoscale}},\ }\href {https://doi.org/10.1038/nmat3746}
  {\bibfield  {journal} {\bibinfo  {journal} {Nature Materials}\ }\textbf
  {\bibinfo {volume} {12}},\ \bibinfo {pages} {774} (\bibinfo {year}
  {2013})}\BibitemShut {NoStop}%
\bibitem [{\citenamefont {Zwier}\ and\ \citenamefont
  {Chong}(2010)}]{Zwier2010}%
  \BibitemOpen
  \bibfield  {author} {\bibinfo {author} {\bibfnamefont {M.~C.}\ \bibnamefont
  {Zwier}}\ and\ \bibinfo {author} {\bibfnamefont {L.~T.}\ \bibnamefont
  {Chong}},\ }\bibfield  {title} {\bibinfo {title} {{Reaching biological
  timescales with all-atom molecular dynamics simulations}},\ }\href
  {https://doi.org/10.1016/j.coph.2010.09.008} {\bibfield  {journal} {\bibinfo
  {journal} {Current Opinion in Pharmacology}\ }\textbf {\bibinfo {volume}
  {10}},\ \bibinfo {pages} {745} (\bibinfo {year} {2010})}\BibitemShut
  {NoStop}%
\bibitem [{\citenamefont {Freddolino}\ \emph {et~al.}(2010)\citenamefont
  {Freddolino}, \citenamefont {Harrison}, \citenamefont {Liu},\ and\
  \citenamefont {Schulten}}]{Freddolino2010}%
  \BibitemOpen
  \bibfield  {author} {\bibinfo {author} {\bibfnamefont {P.~L.}\ \bibnamefont
  {Freddolino}}, \bibinfo {author} {\bibfnamefont {C.~B.}\ \bibnamefont
  {Harrison}}, \bibinfo {author} {\bibfnamefont {Y.}~\bibnamefont {Liu}},\ and\
  \bibinfo {author} {\bibfnamefont {K.}~\bibnamefont {Schulten}},\ }\bibfield
  {title} {\bibinfo {title} {{Challenges in protein-folding simulations}},\
  }\href {https://doi.org/10.1038/nphys1713} {\bibfield  {journal} {\bibinfo
  {journal} {Nature Physics}\ }\textbf {\bibinfo {volume} {6}},\ \bibinfo
  {pages} {751} (\bibinfo {year} {2010})}\BibitemShut {NoStop}%
\bibitem [{\citenamefont {Marculescu}\ \emph {et~al.}(1991)\citenamefont
  {Marculescu}, \citenamefont {Okano},\ and\ \citenamefont
  {Schulke}}]{Marculescu:1990nf}%
  \BibitemOpen
  \bibfield  {author} {\bibinfo {author} {\bibfnamefont {S.}~\bibnamefont
  {Marculescu}}, \bibinfo {author} {\bibfnamefont {K.}~\bibnamefont {Okano}},\
  and\ \bibinfo {author} {\bibfnamefont {L.}~\bibnamefont {Schulke}},\
  }\bibfield  {title} {\bibinfo {title} {{Superspace Renormalization and
  Stochastic Quantization}},\ }\href
  {https://doi.org/10.1016/0550-3213(91)90333-S} {\bibfield  {journal}
  {\bibinfo  {journal} {Nucl. Phys. B}\ }\textbf {\bibinfo {volume} {349}},\
  \bibinfo {pages} {463} (\bibinfo {year} {1991})}\BibitemShut {NoStop}%
\end{thebibliography}%
\end{document}